\documentclass{article}

\usepackage{PRIMEarxiv}
\usepackage{authblk} 
\usepackage[utf8]{inputenc} 
\usepackage[T1]{fontenc}    
\usepackage{hyperref}       
\usepackage{url}            
\usepackage{booktabs}       
\usepackage{amsfonts}       
\usepackage{nicefrac}       
\usepackage{microtype}      
\usepackage{lipsum}
\usepackage{float}
\usepackage{multirow}%
\usepackage{graphicx}
\usepackage{siunitx}
\usepackage{lineno}
\usepackage{booktabs}
\usepackage{flafter} 
\usepackage{tcolorbox}
\usepackage{xspace}
\usepackage{adjustbox}
\usepackage{standalone}
\usepackage{tikz}
\usepackage{tabularx}
\usepackage{placeins}
\usepackage{amssymb}
\usepackage{amsmath}
\usepackage{subcaption}
\usepackage{esdiff}
\usepackage[title]{appendix}%
\usepackage[backend=biber, style=numeric, sorting=none]{biblatex}

\usepackage{algorithm}
\usepackage{algpseudocode}   
\graphicspath{{Images/}}     

\makeatletter
\newcommand{\resetAuthors}{%
  \renewcommand\AB@authlist{}%
  \renewcommand\AB@affillist{}%
  \renewcommand\AB@authors{}%
  \renewcommand\AB@affilsep{\protect\Affilfont}%
  \renewcommand\AB@blk@and{\protect\Authfont\protect\AB@setsep}%
}
\makeatother

\title{Shower Shapes and Data-Driven Simulation Studies using a Highly Granular Crystal Electromagnetic Calorimeter}

\author[a,b]{J.\,Rolph\thanks{corresponding author},}
\author[a,b]{Y.\,Liu,}
\author[a,b]{B.\,Qi\thanks{formerly at IHEP/UCAS}}

\affil[a]{Institute of High Energy Physics, Chinese Academy of Sciences (IHEP, CAS),\\
  19B Yuquan Road, Shijingshan District, Beijing 100049, China}

\affil[b]{University of Chinese Academy of Sciences (UCAS),\\
  19A Yuquan Road, Shijingshan District, Beijing 100049, China}

\author[d,c,e]{S.\,Li,}
\author[d,e]{Z.\,Zhao,}
\author[c,d,e]{H.\,Yang}

\affil[c]{School of Physics and Astronomy, Shanghai Jiao Tong University,\\
  800 Dongchuan Road, Shanghai 200240, China}

\affil[d]{Tsung-Dao Lee Institute, Shanghai Jiao Tong University,\\
  1 Lisuo Road, Shanghai 201210, China}

\affil[e]{National Key Laboratory of Dark Matter Physics; Shanghai Key Laboratory for Particle Physics and Cosmology;\\
  Key Laboratory for Particle Astrophysics and Cosmology (MoE), Shanghai 200240, China}

\bibliography{main}

\begin{document}
\maketitle

\begin{abstract}
Electron shower shapes were measured using standard shower profile models for events observed with a highly granular crystal electromagnetic calorimeter prototype, with data taken from a 2025 test beam at the CERN Proton Synchrotron T9 beamline. The shower maximum, longitudinal shape parameter, and lateral variance were found to be well approximated by log-normal distributions. From these distributions, properties of the absorber were measured to be in agreement with existing literature. Energy-dependent correlations were observed between all parameters and most strongly between the shower maximum and the shower shape parameter. The distributions of energy and event-by-event pairwise log ratios of energy fractions in individual active volumes of the calorimeter conditional on the shower shape were found to be consistent with shifted and scaled Gamma and Fisher's-$z$ distributions, respectively. Generative methods were developed to simulate the logarithm of shower shape variables using Gaussian copulas. Copula sampling for shower shape parameters was found to produce superior modeling of heavy tails observed for the shower maximum and longitudinal shape parameter distributions when compared to direct sampling from a correlated multivariate normal distribution. A copula method was also devised to sample full shower events conditional on the shower shape parameters by exploiting the observation that the vector of energy fractions follows a logistic Log-Gamma distribution, the joint relation implied by the observed Fisher's-$z$ form of the pairwise log-ratios. The generative model was evaluated using standard manifold and classifier metrics and was found to approach the data-versus-data ceiling. The fastest generation time was found to be around \qty{30}{\micro \second} per particle shower.
\end{abstract}



\section{Introduction}

Future Higgs factories such as the Circular Electron-Positron Collider (CEPC) \cite{group_cepc_2025} and Future Circular Electron-Positron Collider (FCC-ee) \cite{benedikt_future_2025} will precisely measure the properties of the Higgs boson, weak vector bosons, and the top quark, while also performing critical validations of the Standard Model. A typical figure of merit for the required performance is that a particle flow algorithm reconstructs the mass of the Higgs boson with \qtyrange{3}{4}{\percent} resolution. This approach necessitates accurate tracking of charged particles within a jet, sophisticated event reconstruction techniques, and highly granular sampling calorimeters. Furthermore, the electromagnetic calorimeter (ECAL) must have sufficient resolution to reconstruct photon energy while also being granular enough to resolve individual particles with excellent energy and time resolution. 

The overall properties of electromagnetic showers are typically defined in terms of observables. Shower shapes measure the properties of the distribution of energy in a particle shower. Examples include the shower maximum, which is the depth at which the multiplication of an initiated particle shower stops, and after which the shower particles are more likely to undergo scattering or absorption; and the Molière radius, which is the radius of the cylinder around the shower development axis that contains \qty{90}{\percent} of the shower energy on average \cite{groom_passage_2000}. These quantities are related to material properties of the absorber, such as the radiation length and critical energy, and can be measured using shower shapes. Experimental measurement of these properties from shower shapes is critical input into future collider experiments, as the lateral and longitudinal spread of particle showers are fundamental to how well the showers can be separated in a highly granular calorimeter. CALICE has performed extensive studies of the mean shower shapes of hadron and electron particle showers in highly granular calorimeters using the CALICE Fe-Sc AHCAL in Ref.~\cite{pinto_shower_2022} for electrons and pions in the energy range \qtyrange{10}{100}{\giga \electronvolt};  in Ref.~\cite{chadeeva_parametrisation_2015} for pions in the energy range \qtyrange{10}{80}{\giga \electronvolt}; and for a Sc-W HCAL in Ref.~\cite{adloff_shower_2014} for pions and protons at \qty{4}{\giga \electronvolt} and \qty{10}{\giga \electronvolt}. Additionally, the shower shapes have been studied for electrons in Ref.~\cite{alme_performance_2023} in the energy range \qtyrange{1}{5.8}{\giga \electronvolt} using the EPICAL-2 digital ECAL. It is noted that there is a dearth of literature regarding experimental studies of the event-by-event shapes of particle showers and of showers at sub-\unit{\giga \electronvolt} energies. 

Secondly, the generation of simulations will be one of the most computationally intensive aspects of future collider experiments. For example, given conservative research and development, Monte Carlo (MC) simulation is predicted to take up around \qty{31}{\percent} of the CPU resources for ATLAS by 2031 \cite{noauthor_atlas_2022}. Of the simulation, most will be attributed to the calorimeters, as they require the tracking of many individual secondary particles. The event-by-event fluctuations of electromagnetic shower shapes have been explored in the simulation study of Ref.~ \cite{grindhammer_parameterized_2000}, which parameterized the shower shapes and forms the basis for the \texttt{Geant4} \texttt{GFlash} fast EM shower generation program \cite{agostinelli_geant4simulation_2003, barberio_gflashhomoshowerparameterisationcc_2004}. More recently, generative machine learning techniques have been developed to sample full shower events for a fixed detector geometry. Various methods have been explored for fast generation of particle showers, reviewed in Ref.~\cite{krause_calochallenge_2025}: generative adversarial networks \cite{paganini_accelerating_2018}, conditional flow matching \cite{favaro_calodream_2025}, diffusion-based models \cite{amram_denoising_2023, mikuni_score_based_2022, mikuni_caloscore_2024}, normalizing flows \cite{buss_convolutional_2024} have all been used to generate particle showers with high fidelity. However, these models typically aim for a single model to describe all aspects of a particle shower: the uncertainty on the energy due to the calorimeter resolution, the fluctuations in the shower shape, and the individual fluctuations per unit active volume in a highly granular calorimeter. By contrast, the \texttt{GFlash} model scales a deterministic energy density profile with fluctuating shower-shape parameters, and distributes the energy as spots within a modeled active volume. The fluctuations within individual active regions of the calorimeter are not modeled; only the shower shape fluctuates event by event. The prospect of modeling the shower fluctuations conditional on the shower shape parameters remains unexplored in the literature. 

This study performs a complete treatment of shower shapes and data-driven generative modeling of particle showers. The paper is structured as follows. Methods and tools are introduced in Sec.~\ref{sec:methods_and_tools}. Sec.~\ref{sec:hgccal} introduces the highly granular crystal calorimeter and describes the experimental conditions of data taken in the 2025 testbeam experiment at the PS-T9 beamline at CERN, as well as the reconstruction and selection criteria for the data used in the study. Sec.~\ref{sec:statistical_methods} introduces the required statistical methods. Sec.~\ref{sec:analysis_and_results} presents the analysis and results, ordered such that each study builds upon the results of the preceding one. Sec.~\ref{sec:shower_shape_distributions} discusses the analysis of the shower shapes from fits, their distribution, and measurement of the key physics observables of the absorber: the radiation length, the critical energy, and the Molière radius. Parameterizations are also provided, as well as studies of the correlations between the shower shape variables. Finally, the event-by-event fluctuations are studied conditional on the shower shape parameters. Sec.~\ref{sec:generative_modelling} builds on the results of the preceding section by proposing two classes of generative models. In Sec.~\ref{sec:shower_shape_modelling}, two methods of shower shape sampling, using correlated log-normal sampling and sampling by a Gaussian copula, are studied and validated. Finally, Sec.~\ref{sec:full_event_modelling} builds a model for full shower events conditioned on shower shape parameters and is assessed using a similar evaluation approach to that presented in Ref.~\cite{krause_calochallenge_2025}.

As a caveat, the generative methods studied in this paper seek to illustrate the distributional structure of electron showers and make no claim to produce superior or higher-quality simulated shower profiles or events than any other model. All generative models presented are pedagogical and serve as corroborations of the observations made in the analysis.

\section{Methods and Tools} \label{sec:methods_and_tools}

The key methods and tools of the study are presented in this section. It is noted that a number of non-standard statistical methods are applied in the study, particularly for Sec.~\ref{sec:event_to_event_fluctuations}, Sec.~\ref{sec:shower_shape_modelling} and  Sec.~\ref{sec:full_event_modelling}, and are available in the Appendix. Empirical cumulative distribution functions are introduced in Appendix Sec.~\ref{sec:ecdf}. Gaussian Copulas, Factor Analysis, and Tobit censored fits, which are necessary to understand the generative modeling of particle showers, are introduced in the Appendix. Sec.~\ref{sec:gaussian_copula}, Sec.~\ref{sec:factor_analysis} and Sec.~\ref{sec:censored_fits}. Similarity measures like the Kolmogorov-Smirnov (KS) statistic, Probability Integral Transform KS statistic (PIT-KS), and Wasserstein-1 statistic (W1) are presented in Sec.~\ref{sec:similarity_measures}. The Crystal Ball distribution (DSCB) and Fisher's z-distribution are introduced in Sec.~\ref{sec:non_standard_distributions}. Finally, multi-layer perceptrons (MLPs) are presented in Sec.~\ref{sec:mlps}.

\subsection{Highly-Granular Crystal ECAL Prototype} \label{sec:hgccal}

The High-Granularity Crystal Calorimeter (HGCCAL) is the primary choice for the CEPC ECAL system \cite{group_cepc_2025}. A prototype has been developed to test the performance. Each layer of the crystal ECAL prototype consists of 14 scintillating crystal bars, each with a volume of \qtyproduct[product-units = single]{12x12x2}{\centi \meter \cubed}. Each subsequent layer is rotated orthogonally with respect to the previous one. This ’cross-hatch’ structure enables reconstruction of the energy density of deposits in the crystal by combining information from two orthogonal, adjacent layers and transverse granularity, while minimizing dead volume and the number of readout channels compared to a design with single crystals. Scintillation light from the crystal is read out from each end by two SiPMs. There are 14 layers of the calorimeter. The first 12 layers are composed of beryllium germanium oxide (BGO). The last 2 layers are composed of bismuth silicate (BSO), which acts as a tail-catcher. The total depth of the calorimeter is \qty{24.9}{X_{0}}.

\begin{figure}
    \centering
    \includegraphics[width=\linewidth]{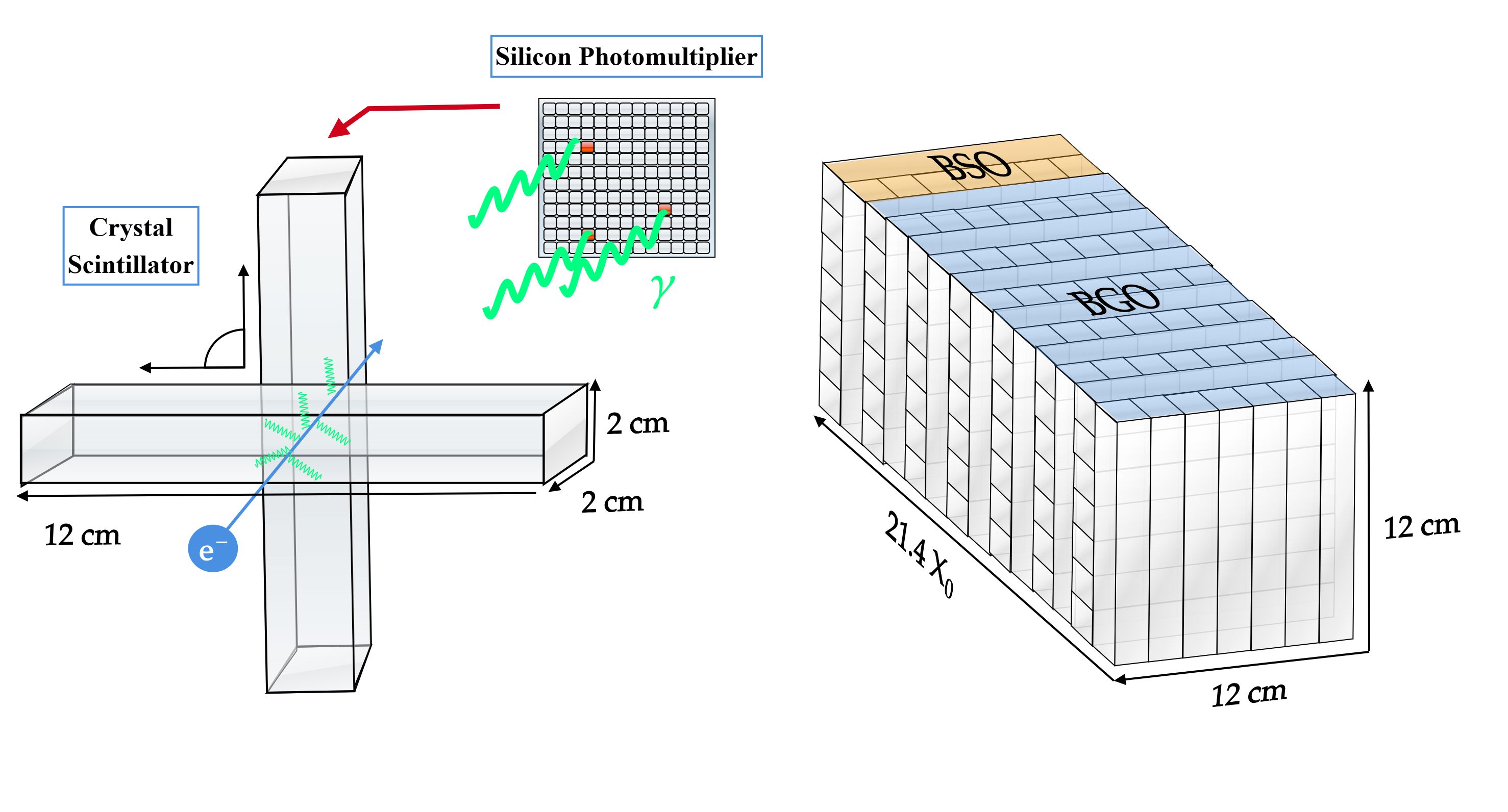}
    \caption{Diagram illustrating the design of the HGCCAL calorimeter.}
    \label{fig:fig_gravnet}
\end{figure}

During 2025, the calorimeter was used in a testbeam campaign at the Proton Synchrotron (PS) and Super Proton Synchrotron (SPS) at CERN in Geneva, Switzerland. The entire system was configured in a low-energy beamline setup with all upstream beam instrumentation removed to minimize particle scattering and momentum spread. A total of 20.8 million events were collected across an energy range of \qtyrange{0.1}{10}{\giga \electronvolt}. However, low beam purity for certain energies meant that electron samples were heavily contaminated by (anti)muons and charged pions. A boosted-decision tree-based PID was therefore used to select electrons \cite{rolph_particle_2026}.  

\begin{figure}
    \centering
    \includegraphics[width=0.8\linewidth]{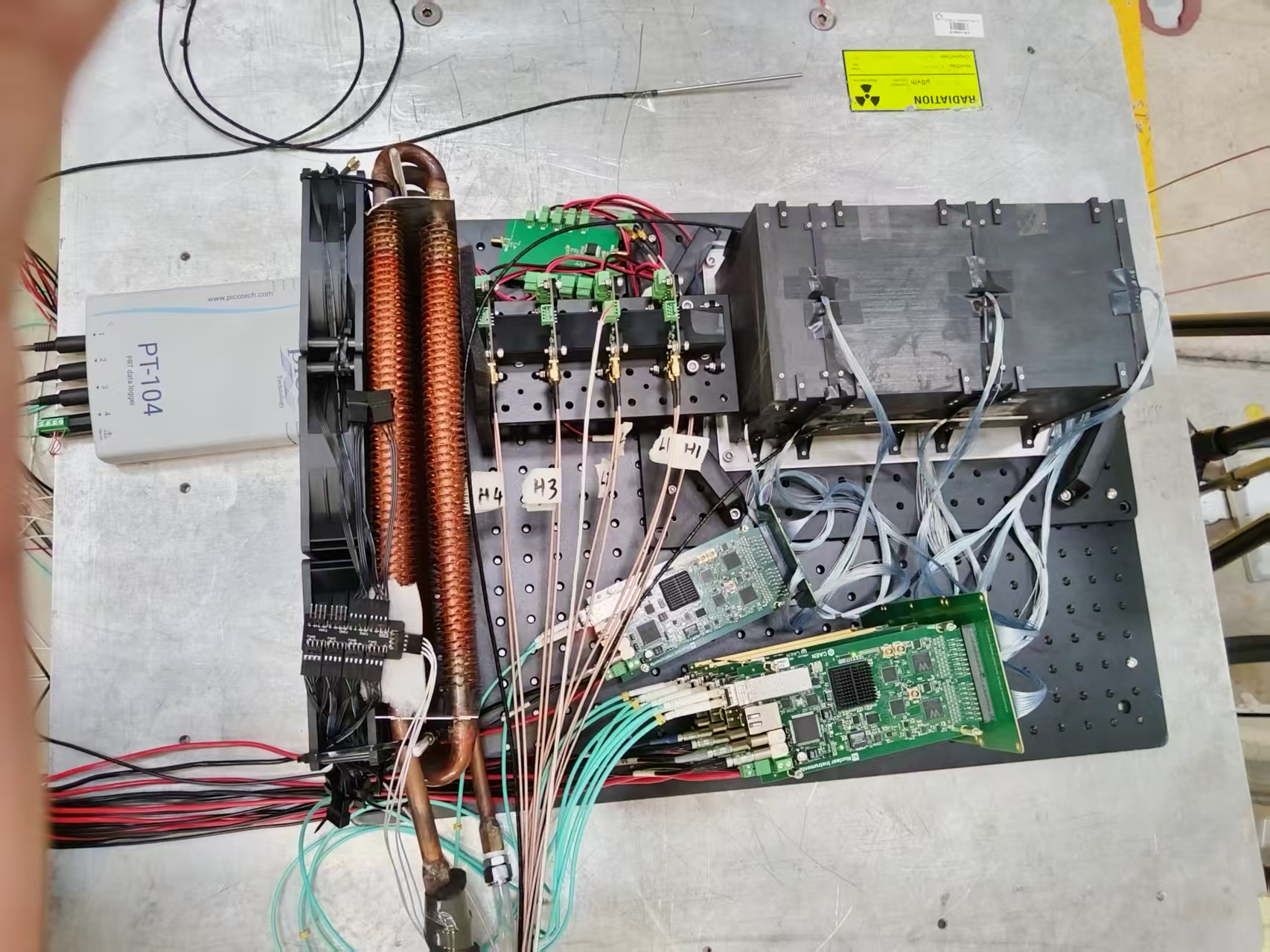}
    \caption{Photograph of the HGCCAL at the test beam at PS in 2025.}
    \label{fig:fig_hgcal_tb}
\end{figure}

A photograph of the calorimeter at testbeam is shown in Fig.~\ref{fig:fig_hgcal_tb}. 

\subsubsection{Datasets}

During 2025, experimental data were taken at the CERN PS-T9 beamline using the crystal ECAL prototype. During this test beam, upstream instrumentation, such as SiFi trackers and Cerenkov detectors, was removed to minimize the beam momentum spread and improve resolution measurements. Energies from \qtyrange{0.1}{6}{\giga \electronvolt} of the campaign are studied henceforth, which were chosen as they were of high electron purity.

The data were reconstructed using a standard analysis procedure for calorimeters with SiPM-based readout. The SiPM response is measured in 'analog-to-digital' counts (ADC counts). Two ASIC operation modes were measured simultaneously during the test beam: high-gain and low-gain ADC values. Measurements of the pedestal, $\mathrm{ADC}_\mathrm{ped}$  and MIP-scale $\mathrm{ADC}_\mathrm{MPV}$, are required for energy reconstruction. The pedestal was measured for each readout channel using a DSCB fit of the SiPM pedestal for both high- and low-gain modes. The mode of the fit yields the pedestal value, $\mathrm{ADC}_{\mathrm{ped}}$. A normal distribution was also found to be consistent. The MIP-scale was measured using $\mu^{+}$ events. Ionizing energy deposits were selected by rejecting pedestal energy deposits based on the fitted distribution of the pedestal, $\mathcal{L}(\operatorname{DSCB}(\mathrm{ADC} | \theta_{\mathrm{ped}})) < -10$ and $\mathrm{ADC} > \mathrm{ADC}_{\mathrm{ped}}$, where $\theta_{\mathrm{ped}}$ are the parameters from the pedestal fit. All subsequent layers were required to have a single energy  $\pm 1$ bar index (lateral/horizontal), to ensure a well-defined, straight track per event. A Landau-Gaussian convolution was fitted to each channel to measure the MPV for the high-gain mode, $\mathrm{ADC}_{\mathrm{MPV, HG}}$.

 The high- and low-gain modes are non-linear and must be corrected. Events are selected from a high-purity \qty{5}{\giga \electronvolt} dataset containing predominantly $e^{-}$ events,  and the joint distribution of the high- and low-gain ADC values is examined. Firstly, a locally weighted scatterplot smoothing (LOWESS) spline \cite{cleveland_smoothing_1996, josef_perktold_statsmodelsstatsmodels_2024} was fitted to the joint distribution to extract the overall profile of the non-linearity. Then, a reference linear relationship between high- and low-gain modes was reconstructed by selecting the spline region with the least curvature and requiring the line to pass through the pedestals of both modes. The low-gain MPV value, $\mathrm{ADC}_{\mathrm{MPV, LG}}$, is then estimated from the line based on the value of $\mathrm{ADC}_{\mathrm{MPV, HG}}$. Finally, the correction for the low-gain mode is obtained using a translation with a spline, $C_{\mathrm{corr}}(\mathrm{ADC}_\mathrm{HG},  i_{\mathrm{channel}})$, which maps the spline to the reference line. An additive constant is used, as it does not affect the variance of the low-gain ADC value around the new line.

 The energy is then reconstructed according to Eq. ~\ref{eq:energy_reco}
 
\begin{equation}
E_{\mathrm{MIP}} =
\begin{cases}
    \dfrac{\mathrm{ADC}_{\mathrm{HG}} - \mathrm{ADC}_{\mathrm{Ped, HG}}}{\mathrm{ADC}_{\mathrm{MPV, HG}}} 
        & \text{if } \mathrm{ADC}_{\mathrm{HG}} < 7900~\text{counts} \\[1em]
    \dfrac{\mathrm{ADC}_{\mathrm{LG}} + C_{\mathrm{corr}}(\mathrm{ADC}_{\mathrm{HG}}, i_{\mathrm{channel}}) - \mathrm{ADC}_{\mathrm{Ped, LG}}}{\mathrm{ADC}_{\mathrm{MPV, LG}}} 
        & \text{otherwise}
\end{cases}
\label{eq:energy_reco}
\end{equation}

A selection cut was made using a boosted decision tree PID method to select electrons \cite{rolph_particle_2026}. In addition, events were rejected that did not match the expectations from the reconstructed energy distribution of a single shower,  $\mathcal{L}(\operatorname{DSCB}(E_{\mathrm{sum}} | \theta_{\mathrm{E_{sum}}})) < -10$. This method excludes rare, high-energy events where multiple particles deposit energy. 

Eq.~\ref{eq:lateral_profile_main} was fitted to the total energy deposited in each horizontal $y$ and vertical $x$ crystal bar, per event. This produced two fit parameters: $\gamma_{x}$ and $\gamma_{y}$. Eq.~\ref{eq:longitudinal_profile_main} was fitted to the total energy deposited per layer, yielding two fit parameters: $\alpha_{z}$ and $T_{z}$. The estimated value of the lateral scale parameter, $\gamma$, was obtained using the geometric mean of Eq.~\ref{eq:geom_mean_gammaxy}. Events were selected for study to require at least three horizontal, three vertical bars, and three layers to be simultaneously active during the event, and for which the fitted $T_{z} > \qty{1}{\milli \meter}$ to ensure physically meaningful results. Energies from \qtyrange{0.2}{6}{\giga \electronvolt} were used in this study. A total of around \num{2.85e6} events were used for this study. A table of events per particle energy is shown in Table~\ref{tab:event_counts}.

\subsection{Shower Models}

Two phenomenological shower-profile models are used in this study to model longitudinal and lateral development. The chosen shower models were empirically identified as good estimates of the overall profile of EM showers. The longitudinal and lateral empirical models for hadron shower development used were introduced in  Refs.~\cite{bock_parametrization_1981, shih_parametrization_1986, longo_monte_1975}. The study of Ref.~\cite{grindhammer_parameterized_2000} also provides a parameterization of electron showers in terms of event-by-event fluctuations in the shower profiles. More recently, the CALICE Collaboration has performed studies on the shower shapes with highly granular calorimeters. Studies on hadron showers in the Sc-Fe CALICE AHCAL \cite{chadeeva_parametrisation_2015} and, most notably, on both electron and hadron showers for the same detector in Ref.~\cite{pinto_shower_2022}, the latter of which demonstrated its validity for both simulation and experimental data. These studies serve as baselines for the work that follows.

\subsubsection{Lateral Development}

The lateral development of the particle shower can be modeled according to the differential energy loss per unit area of infinitesimal rings, $S$, around the core of the EM shower, by Eq.~\ref{eq:lateral_shower_pdf_r}, derived in Appendix Sec.~\ref{sec:lateral_profile_pdf}:

\begin{equation}
\diff{E}{S}(r; \gamma) = E_{T} \cdot \frac{e^{-\frac{r}{\gamma}}}{2\pi \gamma^{2}}
\label{eq:lateral_shower_pdf_r}
\end{equation}

where $r$ is the radial distance from the shower axis, $\gamma$ is the scale parameter of the shower, related to the Molière radius according to the relationship $\rho_{M} \simeq 3.89 \gamma$, $E_{T}$ is the total energy of the shower, and $\rho_{M}$ is hereby defined as the 'effective' Molière radius, which depends on the geometry and shower shape model used to describe the shower shape.

Several caveats and observations are presented. Firstly, the energy density is typically modeled as a mixture of a dense core and a diffuse halo. Sufficient containment volume is required to observe the halo. It was estimated that the halo component could only be isolated in the outermost bars of the calorimeter, and no advantage in fit quality was observed by including a halo. For this analysis, it was omitted. 

Secondly, an alternative model was proposed in Ref.~\cite{grindhammer_parameterized_2000}, which includes a shower-maximum-dependent scaling to parameterize the lateral development of an EM shower with core and halo components. While an increasing variance in the lateral energy profile with layer was observed, the parameters describing a core and halo component according to this alternative model were found to be only marginally identifiable at the highest energy point studied, \qty{6}{\giga \electronvolt}. Even where the fit converged, the model did not improve on the proposed single-component exponential model of Eq.~\ref{eq:lateral_shower_pdf_r}. The failure of the model is most likely attributable to the coarse granularity of the calorimeter in this case.

Thirdly, energy is deposited at a particular depth in either $x$ or $y$ depending on whether the layer was even or odd, not both simultaneously. This means the only way to reconstruct the lateral development of the shower is to marginalize the distribution on its $x$- or $y$-development. Fortunately, this is possible since the average lateral shower development is approximately isotropic. Appendix Sec~\ref{sec:lateral_profile_pdf} derives the average lateral profile projected to Cartesian coordinates to be the Bessel-K distribution: 

\begin{equation}
    \diff{E}{x}(x; c_{x}, \gamma_{x}) = E_{T} \cdot \frac{|x-c_{x}|}{\pi \gamma_{x}^{2}} \cdot K_{1} \left(\frac{|x-c_{x}|}{\gamma_{x}} \right)
    \label{eq:lateral_profile_main}
\end{equation}

Where $K_{1}$ is the modified Bessel function of the second kind with $\nu=1$, $\gamma_{x}$, and $c_{x}$ are the scale parameter and core position, measured in the odd layers. The equivalent variables $\gamma_{y}$ and $c_{y}$ are obtained for the $y$-plane in the even layers. The shower scale estimator is reconstructed using the independent measurements with the geometric mean:  

\begin{equation}
    \gamma_{xy} = \sqrt{\gamma_{x}\gamma_{y}}
    \label{eq:geom_mean_gammaxy}
\end{equation}

Reasoning for this choice and its validity is given in Sec.~\ref{sec:shower_shape_distributions}.

\subsubsection{Longitudinal Development}

The longitudinal development of the particle shower is well-described by the Gamma distribution: 

\begin{equation}
    \diff{E}{z}(z; \alpha_{z}, \beta_{z}, \Delta_{z})=\frac{\beta_{z}^{\alpha_{z}} \cdot  (z - \Delta_{z})^{\alpha_{z}-1} \cdot  e^{-\beta_{z} (z - \Delta_{z})}}{\Gamma(\alpha_{z})}
    \label{eq:longitudinal_profile_main}
\end{equation}

where $z$ is longitudinal depth, $\beta = (\alpha-1)/T_{z}$, $T_{z}$ is the shower's maximum position, $\alpha_{z}$ is a shape parameter that characterizes the shower's symmetry, and $\Delta_{z}$ is an offset. Fits were performed with the conditions $T_z{z} >  \qty{1}{\milli \meter}$ and $\alpha_{z} > 1$. This model is consistent with many EM-shower parameterization studies, such as in Ref.~\cite{grindhammer_parameterized_2000} and Ref.~\cite{pinto_shower_2022}.

The minimum case where the shower maximum would be inside the calorimeter, $\alpha_{z} = 1$ is an exponential distribution, with the shower maximum position at the very front of the calorimeter. The peakedness of the distribution decreases as $\alpha_{z}>1$.

\section{Analysis and Results}\label{sec:analysis_and_results}

\subsection{Shower Shape Distributions} \label{sec:shower_shape_distributions}

\subsubsection{Goodness-of-Fit and Example Fits}

An extended binned log-likelihood was used to fit the distributions using \texttt{iminuit}, under the simplifying assumption that the bin fluctuations are Poisson-distributed. Weighting methods such as the Bohm-Zech transformation, which uses an effective Poisson count $n_{\mathrm{eff}} = (\Sigma E_{i})^2 / \Sigma E_{i}^2$, were considered. However, in low-occupancy bins where only a couple of high-energy hits dominate, $n_{\mathrm{eff}}$ becomes small, and the fit becomes unstable. Low-energy bins must nonetheless be included due to the limited number of bins available. However, the unweighted Poisson assumption is incorrect and underestimates the uncertainty in each bin, resulting in cases where reduced-$\chi^{2}$ can be less than 1. Therefore, to interpret the results, the KS and Wasserstein-1 ($W_{1}$) statistics are used to assess the fit between the empirical cumulative distribution function (ECDF) of the data and the binned CDF.

\begin{figure}
    \centering
    \includegraphics[width=\linewidth]{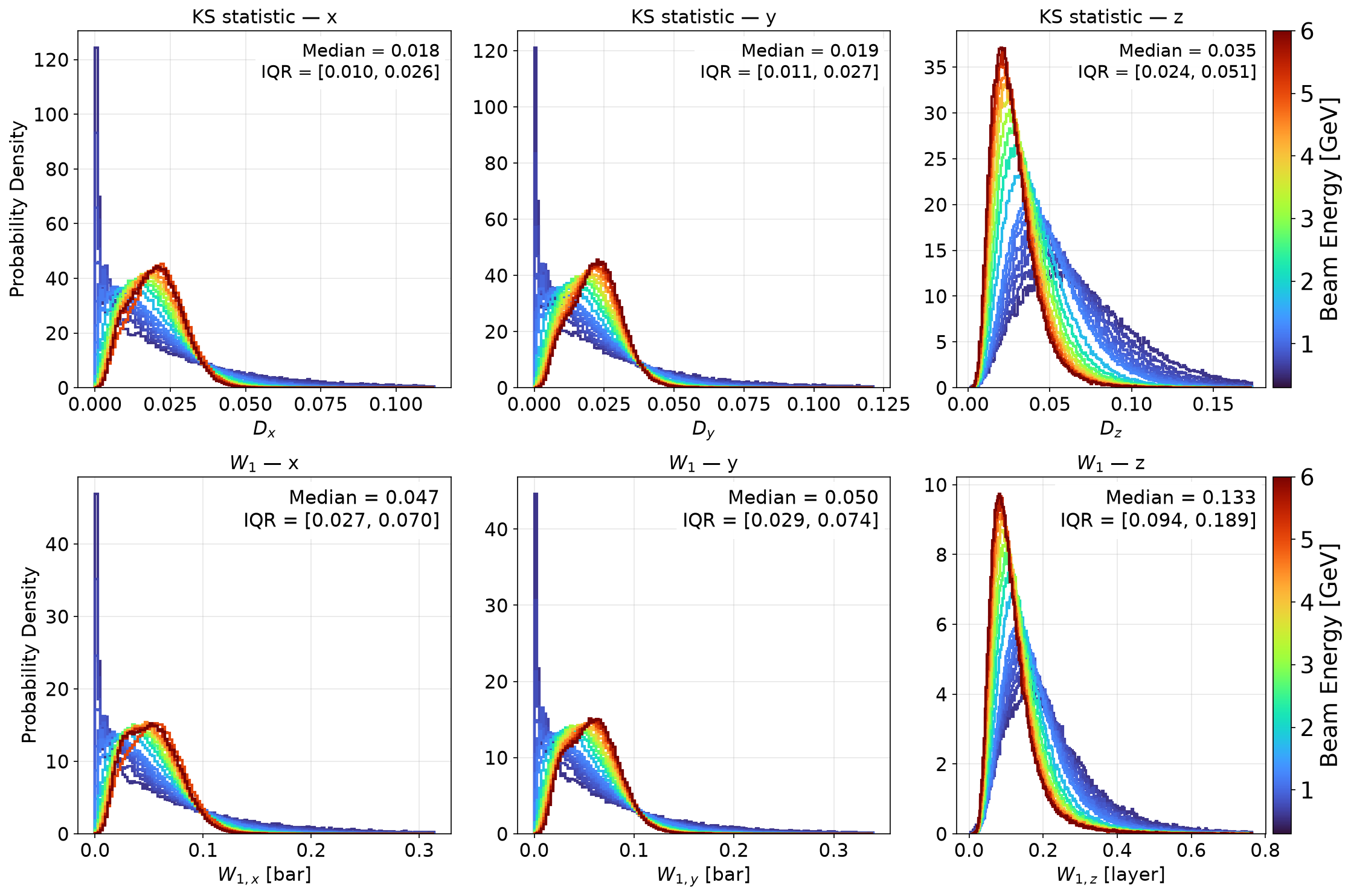}
    \caption{Distributions of the goodness-of-fit parameters used to describe the profile fits. Top row: KS-statistic, Bottom row: Wasserstein-1 (Earth Mover's Distance) statistic. Blue to red color is low to high energy.}
    \label{fig:fig_W1_KS}
\end{figure}

The results are shown for the $x$, $y$ and $z$ directions in Fig.~\ref{fig:fig_W1_KS}. 
The top row of Fig.~\ref{fig:fig_W1_KS} show that the median KS-statistic of $D_{x},~ D_{y} \simeq 0.02$ for $x$ and $y$ profiles and $D \simeq 0.035$. the largest vertical distance between the empirical and fitted CDFs is a median \qty{2}{\percent} (\qty{3.5}{\percent}) for lateral (longitudinal) profiles. The sharp bin at low energy in the lateral profiles indicates events where all the energy is deposited in the central two bins, and the calorimeter is not sufficiently granular to resolve the shower shape. However, this is only the case for the lowest-energy bins and is not relevant above about \qty{0.3}{\giga \electronvolt}. Similarly, the bottom row shows that the $W_{1}$ statistic indicates that the distribution would need to be displaced by less than one bin in all cases, meaning the model precision exceeds the detector granularity. The median required displacement is around 
\qty{5}{\percent} and \qty{13}{\percent}, or \qty{1}{\milli \meter} and \qty{2.6}{\milli \meter}, for the lateral and longitudinal profile fits, both of which are considerably smaller than the uncertainty on a single bin. In all cases, the skewness decreases with energy. This indicates shower fluctuations are more extreme at low energy. In summary, both results indicate good agreement with the overall shape of the shower, particularly as reflected in $W_{1}$.

\begin{figure}
    \centering
    \includegraphics[width=\linewidth]{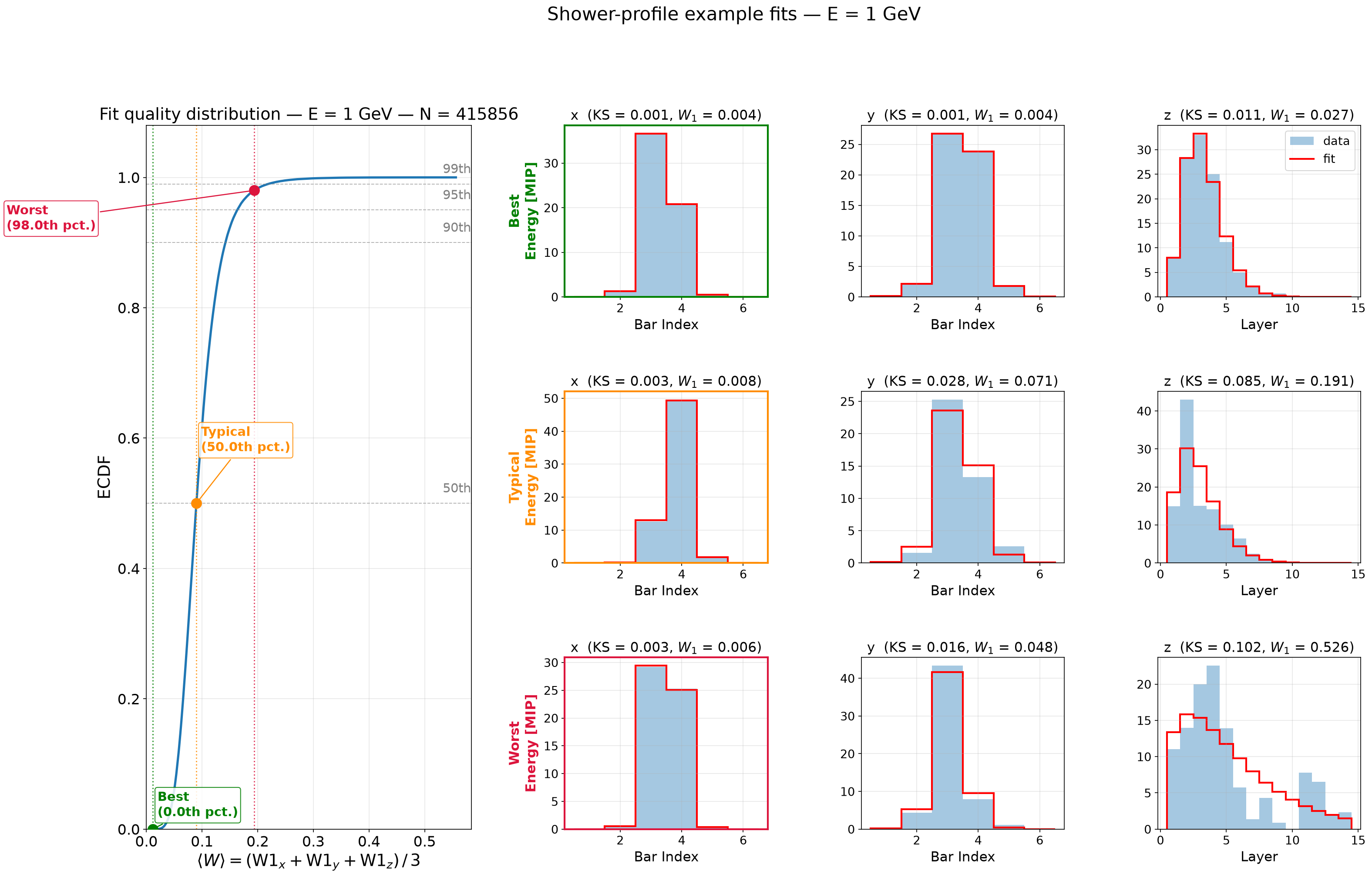}
    \caption{Examples of the best, median and worst fits for \qty{1}{\giga \electronvolt} showers. The left subfigure shows the empirical cumulative distribution function (ECDF) vs. the average Wasserstein-1 values, $\langle W_1 \rangle$, over the entire set of profiles ($x$, $y$, $z$), with the percentiles shown on the right in grey text. The green, blue, and red text indicate the best, typical (median), and worst case for the fits. The blue energy-weighted histograms are shown on the right, with each row corresponding to the coloured axis label. The $y$-axis of the histograms shows the energy deposited per bar index or layer. The red line indicates the fitted profile for that event.}
    \label{fig:fig_example_fits_profile_1GeV}
\end{figure}

Example fits for \qty{1}{\giga \electronvolt} showers are shown in Fig.~\ref{fig:fig_example_fits_profile_1GeV}, with \qty{0.2}{\giga \electronvolt}, \qty{0.5}{\giga \electronvolt} and \qty{5}{\giga \electronvolt} samples shown in Appendix Figs.~\ref{fig:fig_example_fits_profile_0p2GeV}-\ref{fig:fig_example_fits_profile_5GeV}. The median average $W_{1}$ statistic was found to be only around 0.1 bar widths (\qty{2}{\milli \meter}) over all dimensions, indicating the agreement is well within the spatial measurement precision of the calorimeter. The plots illustrate the same conclusion as the goodness-of-fit: that the fits generally describe the shape of the shower, and that this holds even in the worst cases. At and above the \qty{99}{\percent} percentile of average $W_{1}$, pathological fits are observed where there are small deposits of energy before the main shower, which is not representative of a true fit failure. As it would introduce a bias to remove the events, they are retained in the analysis.

\subsubsection{Reconstruction of $\gamma_{xy}$}\label{sec:gammaxy_reconstruction}

The reconstruction of $\gamma_{xy}$ from independent measurements is discussed henceforth. 

The distribution of the log-ratio of $\gamma_{x}$ and $\gamma_{y}$ measured per event in the dataset was fitted with the DSCB distribution. Normality can be assessed by the quality of the fit and the fraction of events that belong to the Gaussian core at each energy. If the mean is close to 0, there is negligible bias, and the variables measure the same quantity.

\begin{figure}
\centering

\begin{subfigure}{0.49\linewidth}
\centering
\includegraphics[width=\linewidth]{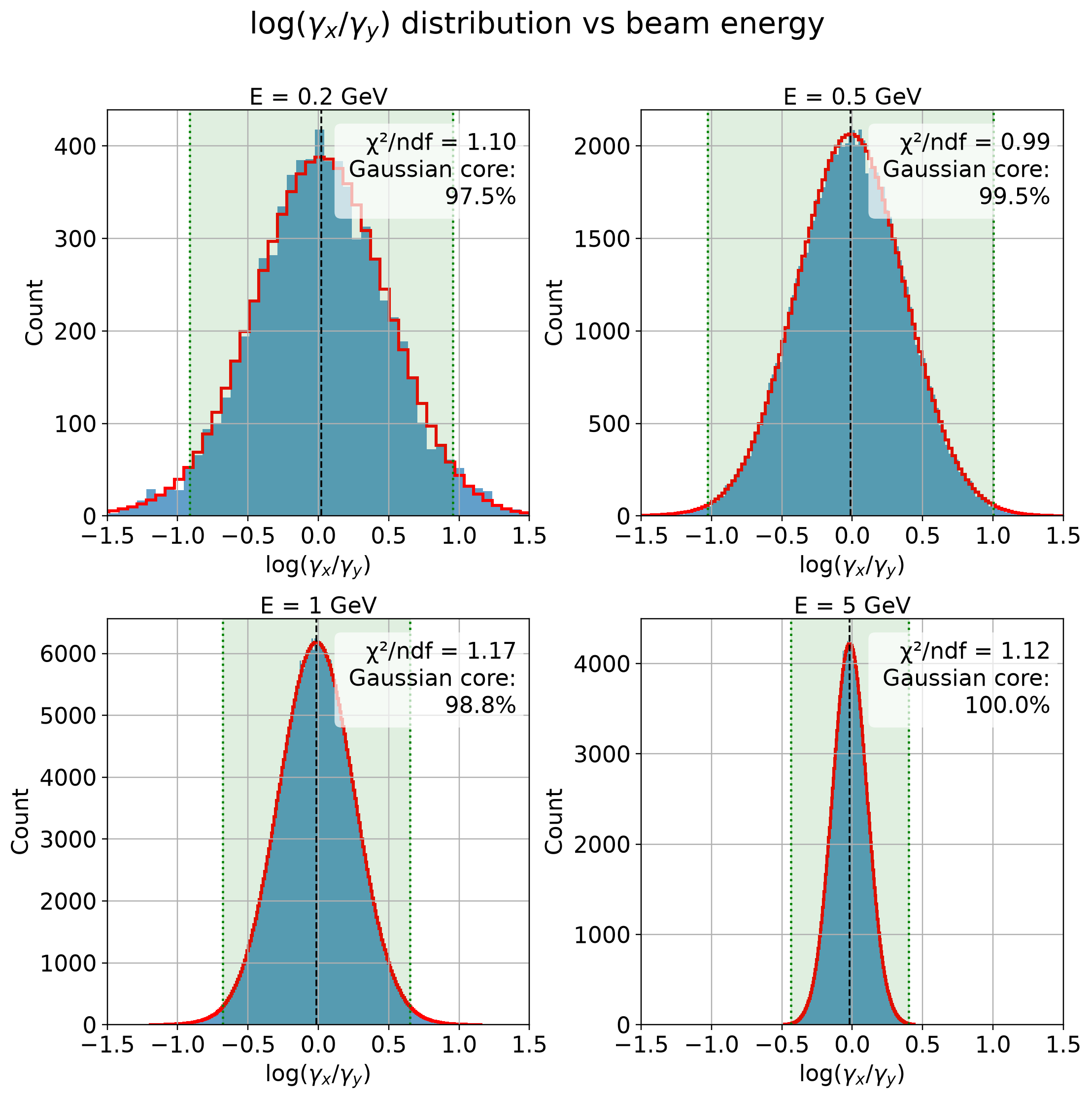}
\caption{}
\label{fig:fig_log_ratio_gamma_dists}
\end{subfigure}
\hfill
\begin{subfigure}{0.49\linewidth}
\centering
\includegraphics[width=\linewidth]{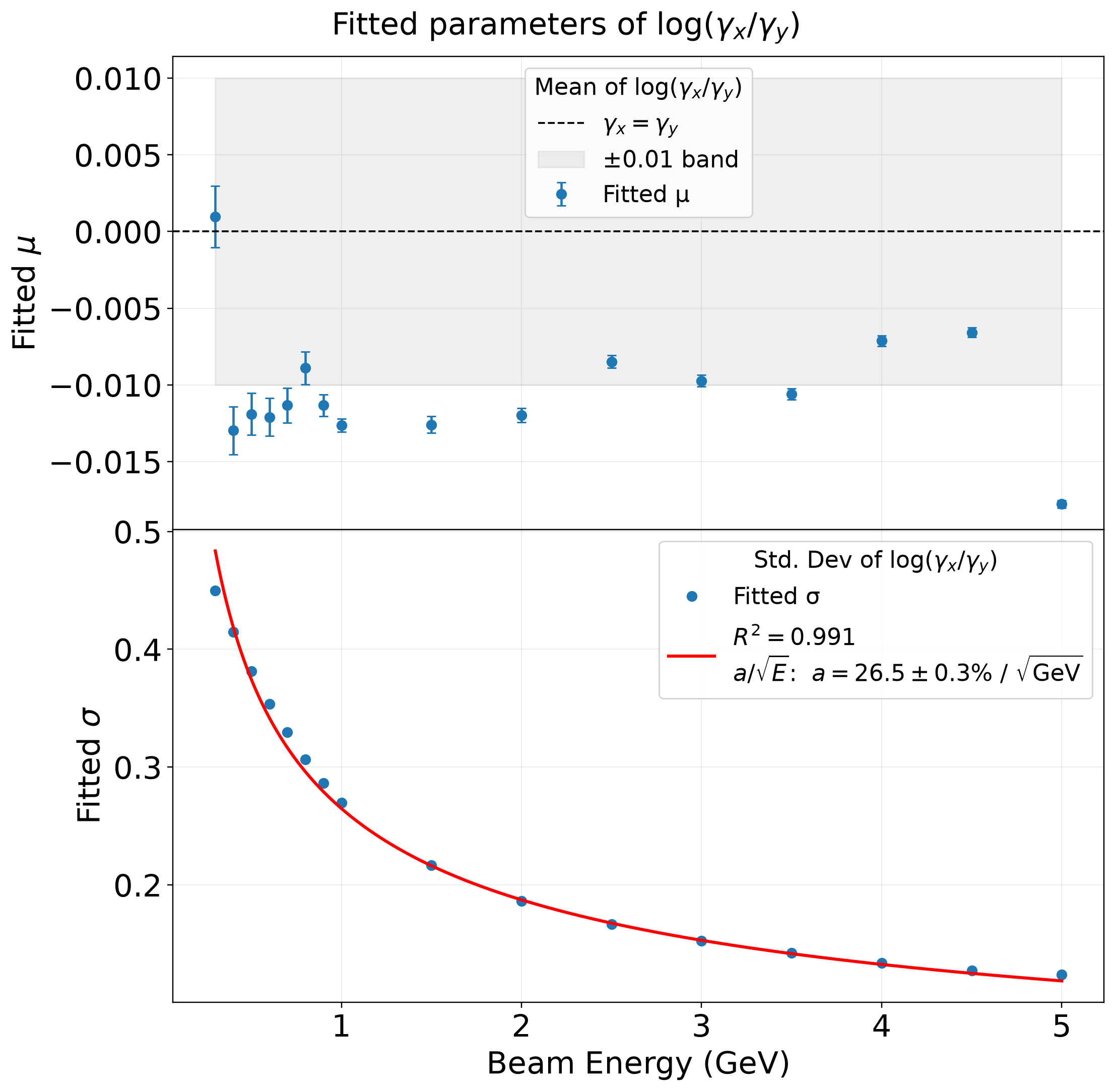}
\caption{}
\label{fig:fig_log_ratio_gamma_stats}
\end{subfigure}

\caption{Fig.~\ref{fig:fig_log_ratio_gamma_dists} shows the distributions of the \qty{0.2}{\giga \electronvolt}, \qty{0.5}{\giga \electronvolt}, \qty{1}{\giga \electronvolt} and \qty{5}{\giga \electronvolt} are shown in data in blue. A DSCB is shown as a red line, and the normal fraction of the DSCB is indicated by the green-shaded area. The reduced $\chi^{2}$ and core fraction are indicated in the legend. Fig~\ref{fig:fig_log_ratio_gamma_stats} shows the fitted parameters of the normal distribution. The top subplot shows the fitted $\mu$ as blue dots. The black dashed line indicates perfect equality between $\gamma_{x}$ and $\gamma_{y}$, The shaded $\pm$ \num{0.01} band indicates an approximately \qty{1}{\percent} deviation in the ratio of $\gamma_{x}$ and $\gamma_{y}$. The bottom subplot shows the fitted $\sigma$ as blue dots. The red dashed line indicates an empirical fit of the relation $\sigma = a/\sqrt{E_{\mathrm{particle}}}$. The coefficient of determination is shown for the fit, and its value is in the legend. }
\label{fig:fig_log_ratio_gamma}
\end{figure}

The results are shown in Fig.~\ref{fig:fig_log_ratio_gamma}. Fig.~\ref{fig:fig_log_ratio_gamma_dists} shows that the distributions are approximately normal, with \qty{95}{\percent} or more of the events observed within the Gaussian core. The observed log-normal distribution implies multiplicative noise sources in the measurement. This also motivates the use of the mean in log space, or the geometric mean, as the estimator for the true scale parameter of the event. 

Fig.~\ref{fig:fig_log_ratio_gamma_stats} shows the mean and standard deviation of the log ratios. The top plot shows a small bias of \qty{1}{\percent}, which is marginal and indicates that $\gamma_{x}$ and $\gamma_{y}$ measure the same quantity on average.  The final observation is that the standard deviation of the ratio increases as $a/\sqrt{E_{\mathrm{particle}}}$.

\subsubsection{Lateral Distributions}\label{sec:lateral_distributions}

\begin{figure}
    \centering
    \includegraphics[width=0.5\linewidth]{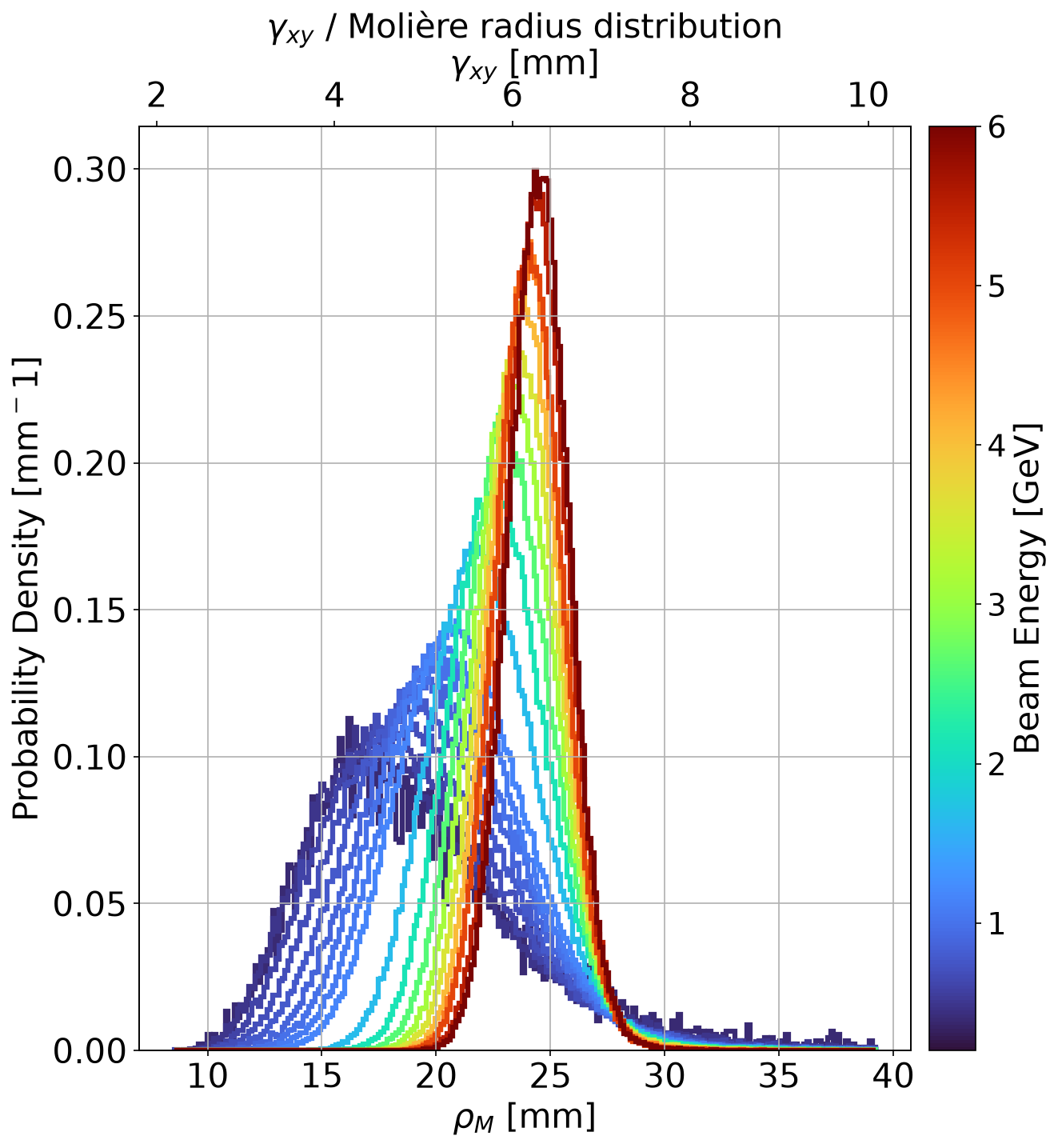}
    \caption{Distributions of the $\gamma_{xy}$/$\rho_{M}$ variables. The color scale blue-red indicates lower to higher energies. The dual $x$-axes indicate Molière radius ($\rho_{M}$) and $\gamma_{xy}$, respectively}
    \label{fig:fig_rho_m_gamma_xy}
\end{figure}

The lateral shower distribution of $\gamma_{xy}$ and the corresponding estimate of the Molière radius are shown in Fig.~\ref{fig:fig_rho_m_gamma_xy}. The distributions show that at low energy, the shower profiles have a wider and more skewed range of spreads. As the energy increases, so too does the mode saturate at a particular value, and the variance of the shower shapes decreases. The distribution was found to be approximately log-normal but with heavy, asymmetrical tails.

\begin{figure}
\centering

\begin{subfigure}{0.49\linewidth}
\centering
\includegraphics[width=\linewidth]{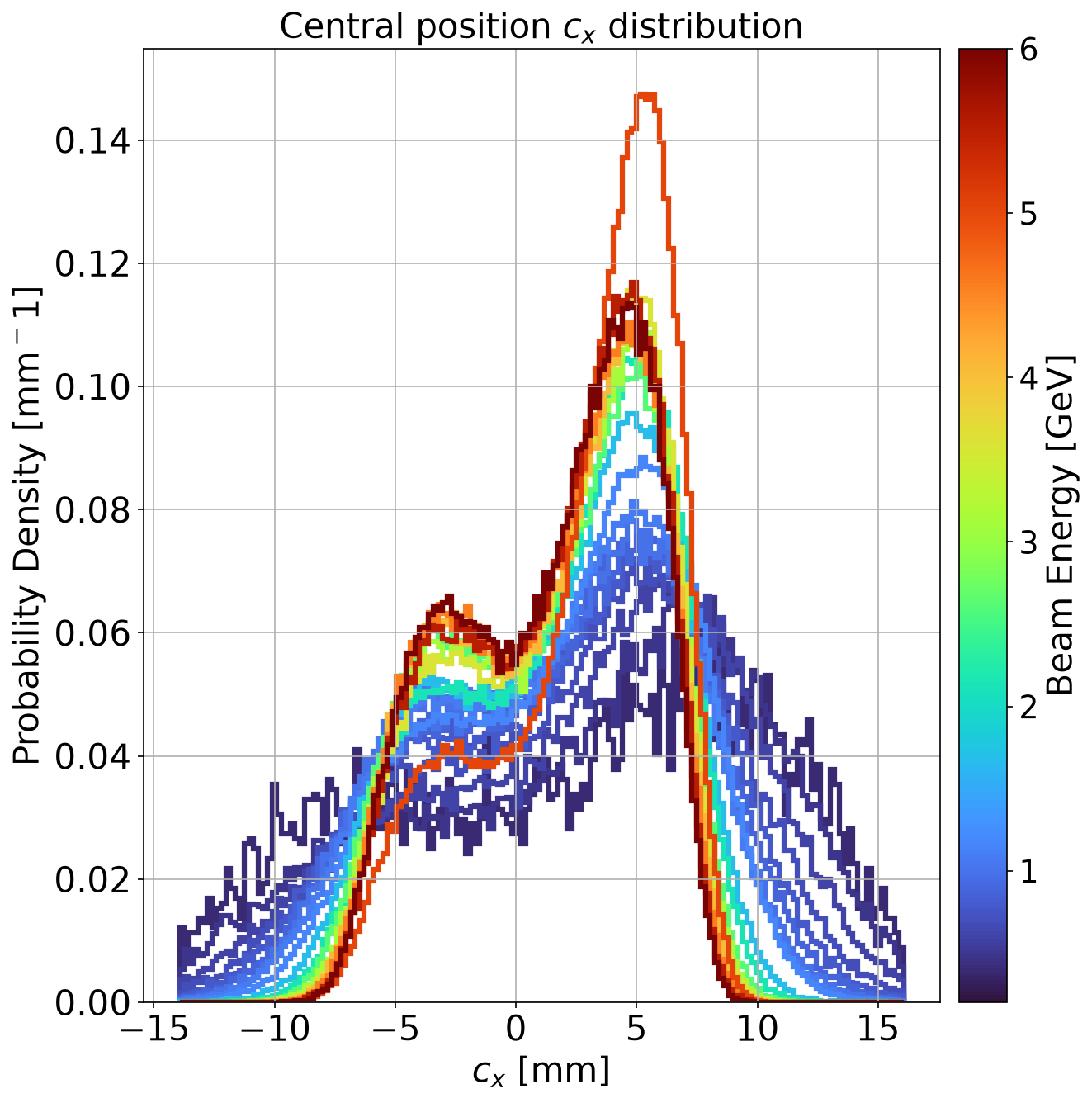}
\caption{}
\label{fig:fig_central_x}
\end{subfigure}
\hfill
\begin{subfigure}{0.49\linewidth}
\centering
\includegraphics[width=\linewidth]{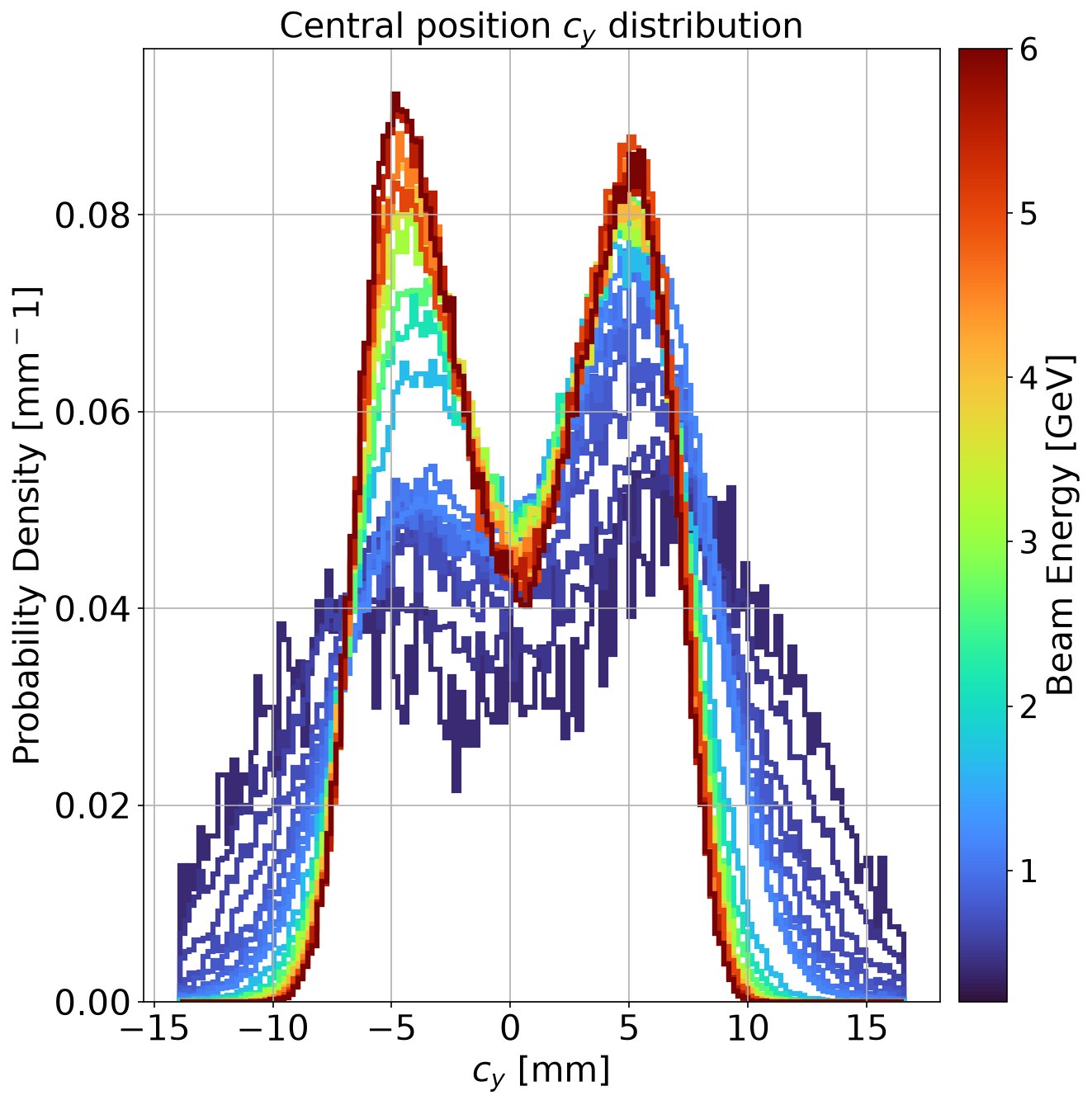}
\caption{}
\label{fig:fig_central_y}
\end{subfigure}

\caption{Distributions of the fitted $c_{x}$ (left) and $c_{y}$ (right) variables, respectively. The variables have been offset to the center of the calorimeter, $C_{\mathrm{calo}} = \left[\qty{70}{\milli \meter}, \qty{70}{\milli \meter} \right]$. Else, as in Fig.~\ref{fig:fig_rho_m_gamma_xy}}
\label{fig:fig_central}
\end{figure}

The distributions of $c_{x}$ and $c_{y}$ are shown in Fig.~\ref{fig:fig_central}. The bimodal structure is a geometrical effect of the beam being fired between crystal bars. It is noted as a minor effect: the spread of the reconstructed position decreases with energy and is again most likely a result of the greater shower fluctuations at low energy, and it indicates that the model recovers expected geometrical dependencies.

Ref.~\cite{grindhammer_parameterized_2000} indicates that the expected shower fluctuations follow a log-normal distribution. The mean of the distribution was estimated under the assumption of a log-normal distribution by performing a DSCB fit. The reduced $\chi^{2}$ was less than 5 for each fit. For each energy, at least \qty{90}{\percent} of the central region could be described by the normal region of the DSCB. However, skewed and heavy tails were observed. 

The mean and variance of the Molière radius, $\langle \rho_{M} \rangle$ and $\operatorname{Var}(\rho_{M})$ were determined using the log-normal mean and variance definitions, $\langle \rho_{M} \rangle = e^{\mu + \sigma^{2}}$. The variance, $\operatorname{Var}(\rho_{M}) = e^{\sigma ^{2}-1} \cdot e^{2\mu +\sigma ^{2}}$.

\begin{figure}
\centering
    \begin{subfigure}{0.5\linewidth}
    \includegraphics[width=\linewidth]{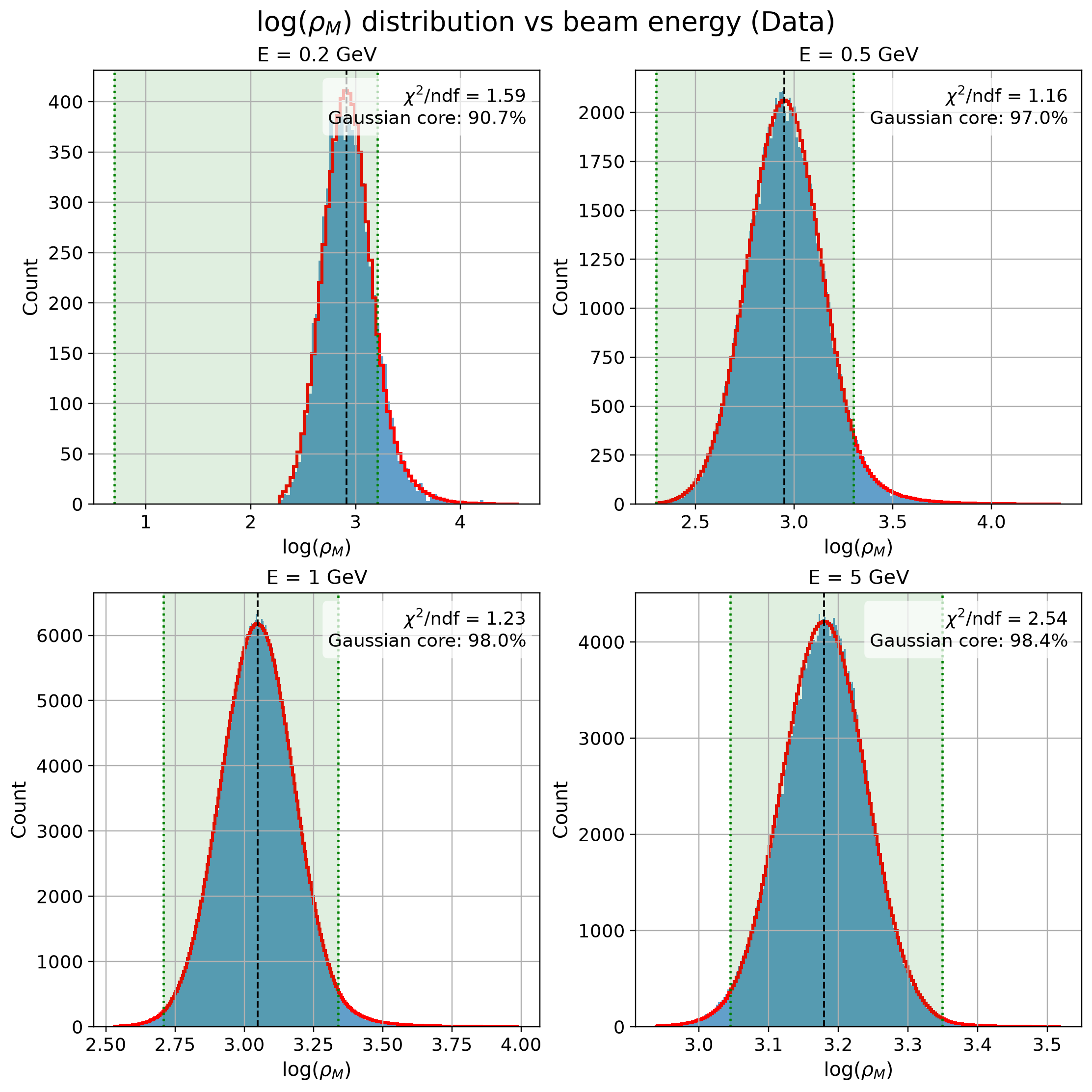}
    \caption{Distributions of $\log{\rho_{M}}$ variables. Else, as in Fig.~\ref{fig:fig_log_ratio_gamma_dists}}
    \label{fig:fig_log_rho_m_dist}
    \end{subfigure}
    \vfill
    \begin{subfigure}{0.45\linewidth}
    \includegraphics[width=\linewidth]{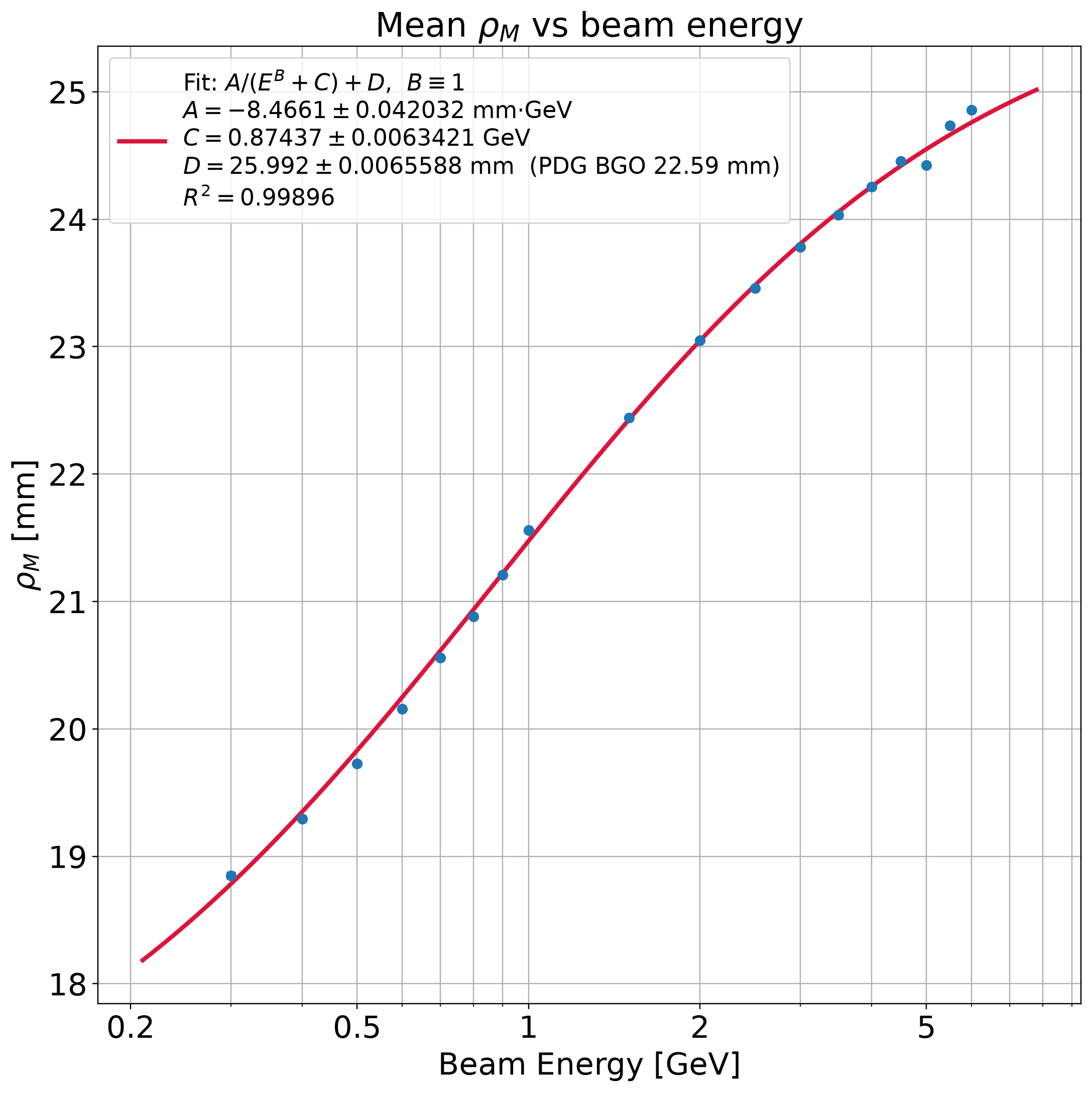}
    \caption{$\langle \rho_{M} \rangle$ as a function of energy. The blue dots indicate the mean and the red line indicates a fit of the model Eq.~\ref{eq:gen_rational}.}
    \label{fig:fig_rho_m_mean}
    \end{subfigure}
    \hfill
    \begin{subfigure}{0.45\linewidth}
    \includegraphics[width=\linewidth]{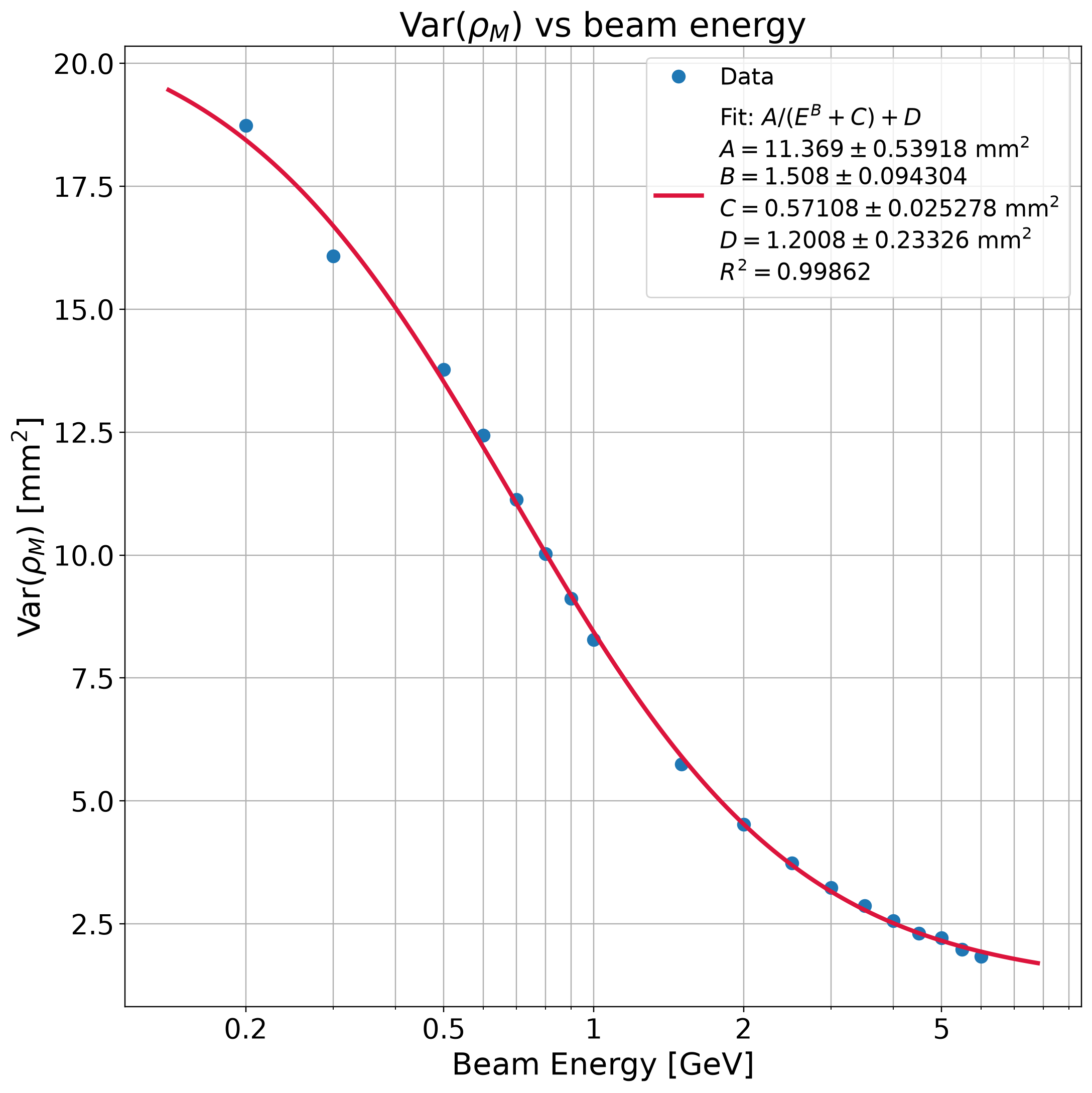}
    \caption{$\operatorname{Var}(\rho_{M})$ as a function of energy. The blue dots indicate the variance. Else, as in Fig.~\ref{fig:fig_rho_m_mean}.}
    \label{fig:fig_rho_m_variance}
    \end{subfigure}
    \caption{Distribution of the logarithm of $\rho_{M}$ and derived mean and variance of $\rho_{M}$.}
    \label{fig:rho_M_results}
\end{figure}

Fits of the DSCB to the distribution are shown in Fig.~\ref{fig:fig_log_ratio_gamma_dists}. These plots show that the log-normal distribution is a good fit to the fitted values of $\log{\gamma_{xy}}$, with more than \qty{95}{\percent} of the distribution covered by the DSCB distribution.  

Fig.~\ref{fig:fig_rho_m_mean} and Fig.~\ref{fig:fig_rho_m_variance} show the mean and variance of the distributions. Since no physical model is expected, the best-fitting functions were determined via symbolic regression. Unless otherwise stated, the best-fitting models for both the log-fit values and the reconstructed mean and variance were found by symbolic regression to take general rational or log-rational forms, given in Eq.~\ref{eq:gen_rational}:
\begin{equation}
    f(E) = \frac{A}{E^{B} + C} + D 
    \label{eq:gen_rational}
\end{equation}

As an exception, $\langle \log{\alpha_{z}} \rangle$ and $\operatorname{Var} (\log{\alpha_{z}})$ 

\begin{equation}
    f(E) = \frac{A}{\log{E} + C} + D 
    \label{eq:log_rational}
\end{equation}

With $B=1$ for the $\langle \rho_{M} \rangle$ fit, Fig.~\ref{fig:fig_rho_m_mean} and Fig.~\ref{fig:fig_rho_m_variance} show similar and inverted behavior. Both models reveal saturation behavior for the Molière radius. The mean value increases to a saturation value of \qty{25.992(0.007)}{\milli \meter}. The value for $\rho_{M}$ quoted by the PDG is \qty{22.59}{\milli \meter} \cite{pdg_review_2026, pdg_review_2026_url}. Therefore, a larger Molière radius was measured of around \qty{3.4}{\milli \meter}, which is attributed to the granularity of the detector. By contrast, the variance of $\rho_{M}$ decreases to a constant of around \qty{1.2}{\milli \meter}. Consistent behavior was observed in an independent \texttt{Geant4} simulation of the calorimeter, following the same form factors with similar levels of agreement, though this is not shown.  
These results may be interpreted as follows. The definition of the Molière radius requires a fully developed particle shower, which is not the case at energies below \qty{1}{\giga \electronvolt}. The fitted function tends toward a constant value, in agreement with expectations for a constant at high energy. The fit model yields an asymptote in both the mean and the variance, indicating that the expected constant behavior is restored at high energy with low variance. In other words, the multiple scattering is truncated before it can develop to its maximum extent, after which it saturates. Consequently, particle showers tend to have less energy in the core on average but exhibit greater variability.

 Fits to the mean and variance of $\log{\gamma_{xy}}$ are shown in Appendix Fig.~\ref{fig:fig_log_gamma_xy_mean} and Fig.~\ref{fig:fig_log_gamma_xy_variance}, and are well described by Eq.~\ref{eq:gen_rational}.

\subsubsection{Longitudinal Distributions}\label{sec:longitudinal_distributions}

\begin{figure}
\centering

\begin{subfigure}{0.49\linewidth}
\centering
\includegraphics[width=\linewidth]{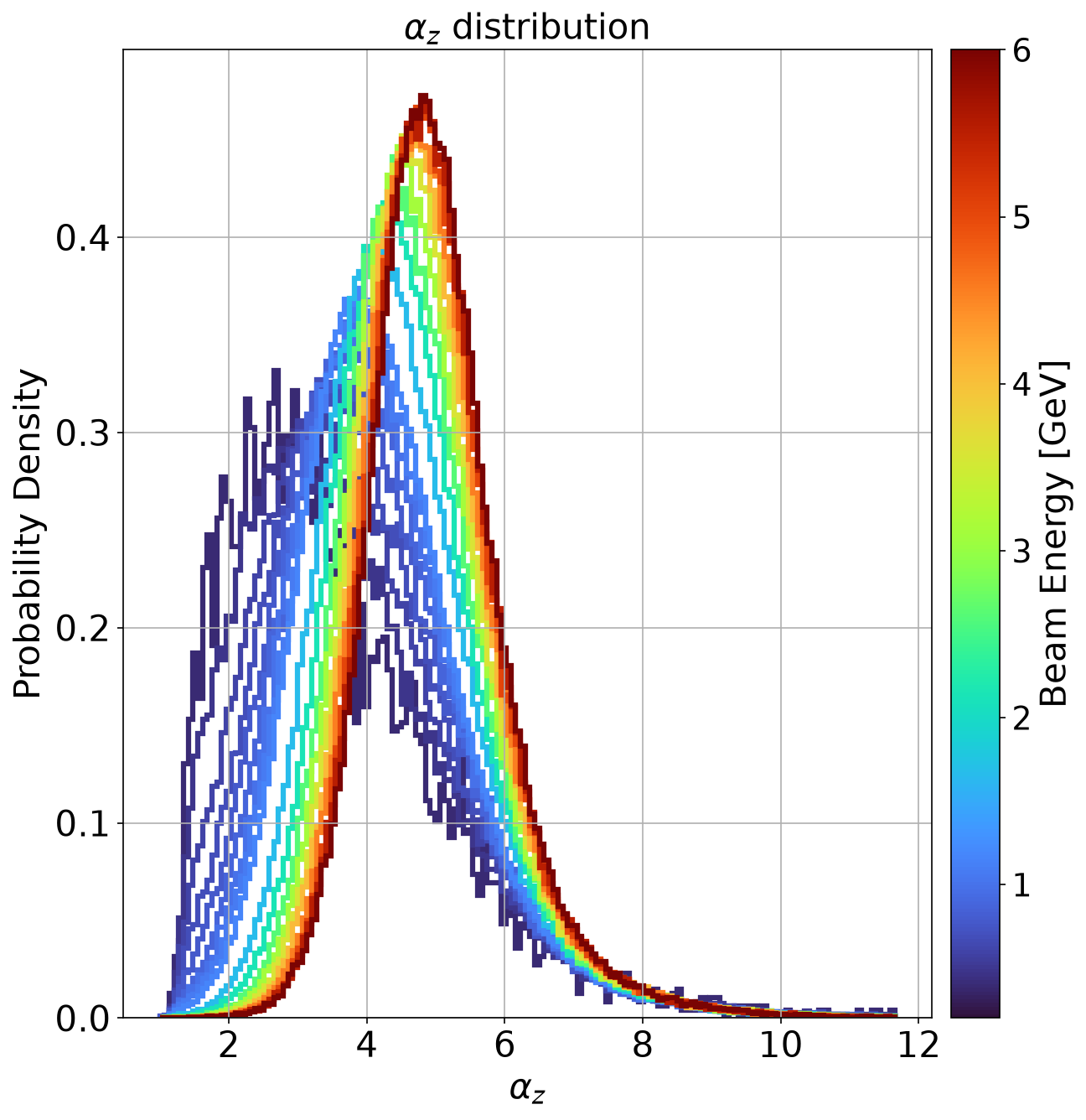}
\caption{}
\label{fig:fig_alpha_z}
\end{subfigure}
\hfill
\begin{subfigure}{0.49\linewidth}
\centering
\includegraphics[width=\linewidth]{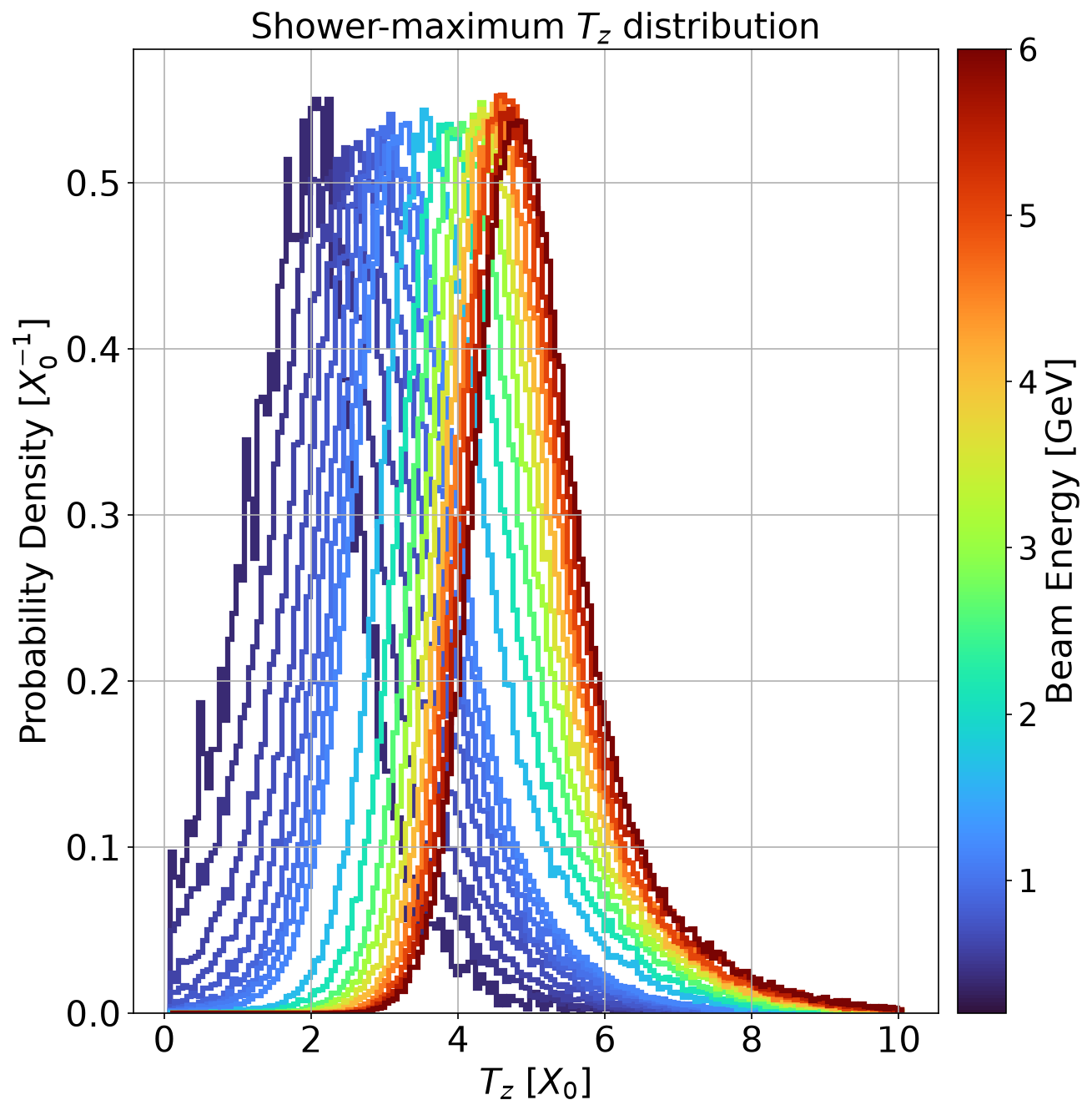}
\caption{}
\label{fig:fig_T_z}
\end{subfigure}

\caption{Distributions of the fitted $\alpha_{z}$ (left) and $T_{z}$ (right) variables, respectively. Else, as in Fig.~\ref{fig:fig_rho_m_gamma_xy}.}
\label{fig:fig_T_z_alpha_Z}
\end{figure}

The longitudinal distributions of $\alpha_{z}$ and $T_{z}$ are shown in Fig.~\ref{fig:fig_alpha_z} and Fig ~\ref{fig:fig_T_z}. The distributions are approximately log-normal and are both observed to increase with energy.

\begin{figure}
\centering
    \begin{subfigure}{0.5\linewidth}
    \includegraphics[width=\linewidth]{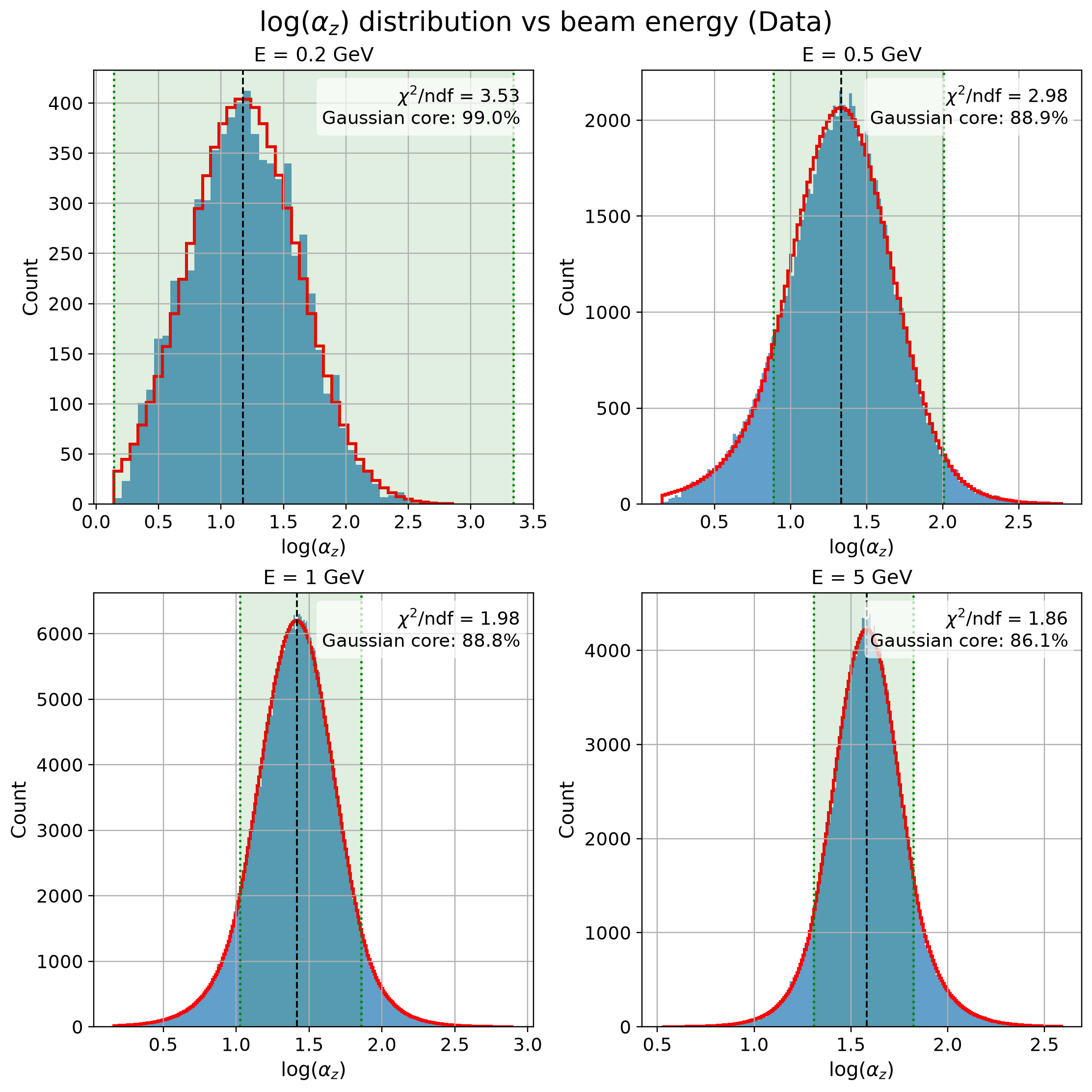}
    \caption{Distributions of $\log{\alpha_{z}}$ variables. Else, as in Fig.~\ref{fig:fig_log_gamma_alpha_z_dist}}
    \label{fig:fig_log_gamma_alpha_z_dist}
    \end{subfigure}
    \vfill
    \begin{subfigure}{0.45\linewidth}
    \includegraphics[width=\linewidth]{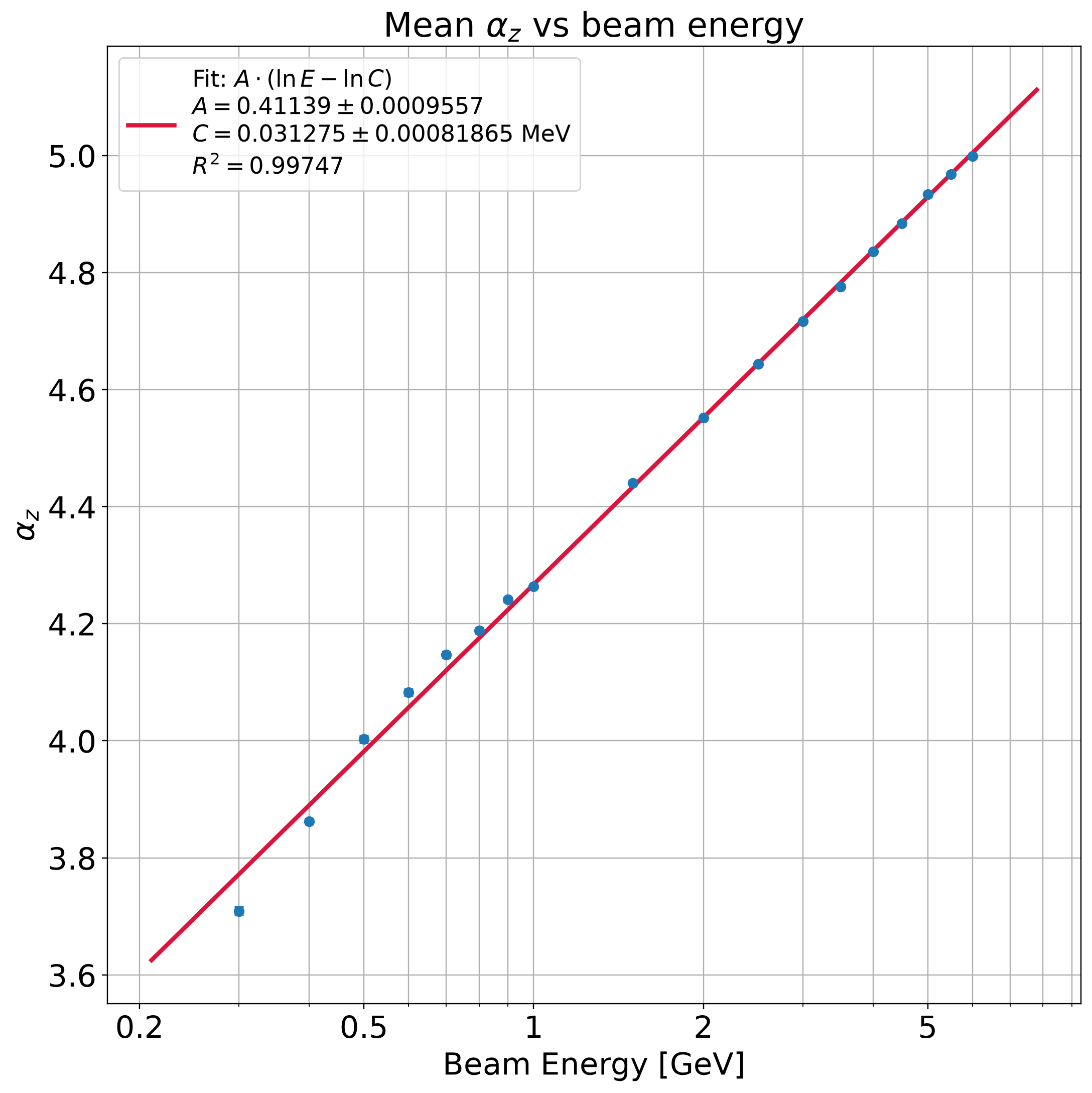}
    \caption{$\langle \alpha_{z} \rangle$ as a function of energy. The blue dots indicate the mean, and the red line indicates a fit of the empirical model of Eq.~\ref{eq:log_linear}.}
    \label{fig:fig_alpha_z_mean}
    \end{subfigure}
    \hfill
    \begin{subfigure}{0.45\linewidth}
    \includegraphics[width=\linewidth]{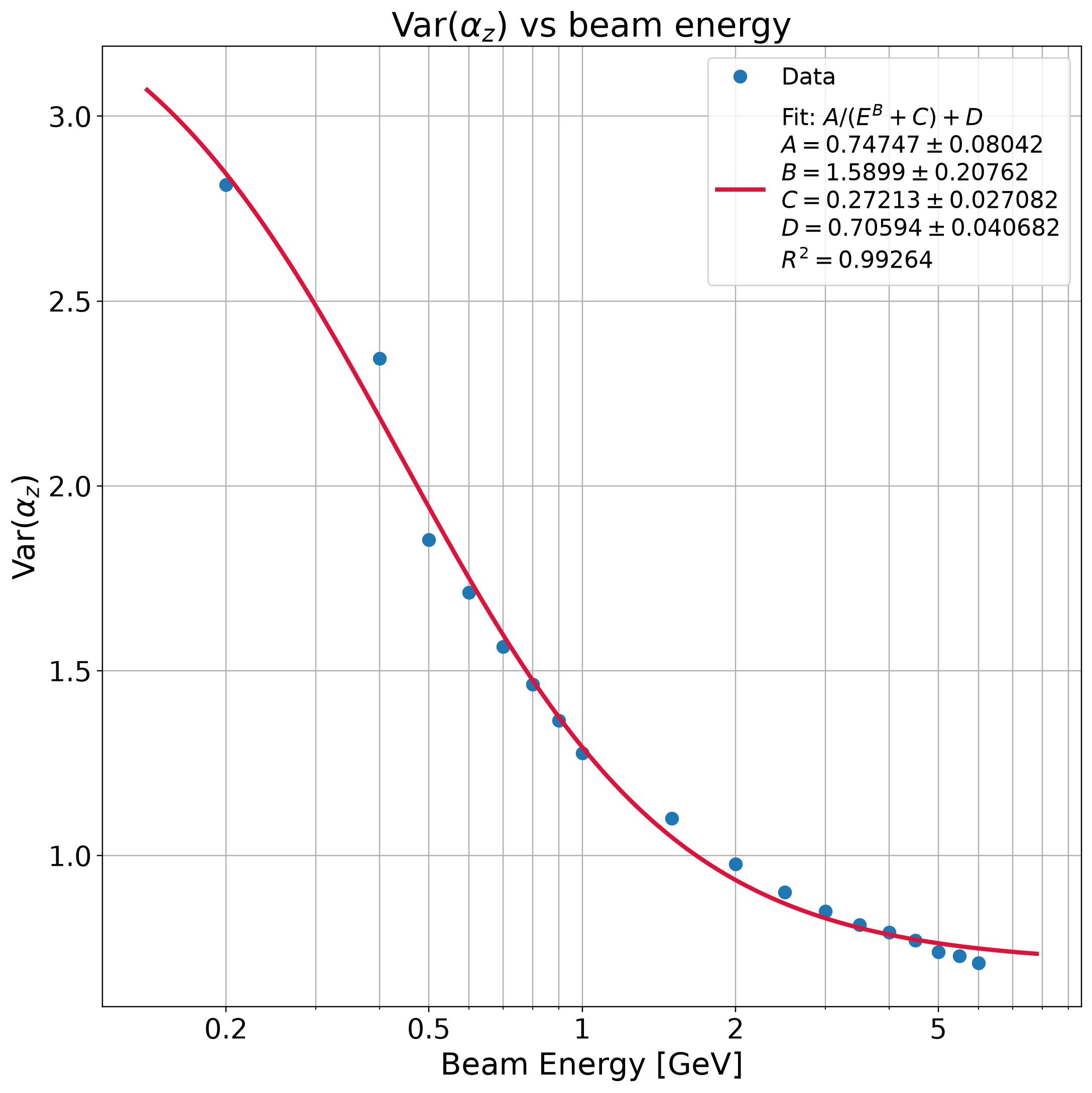}
    \caption{$\operatorname{Var}(\alpha_{z})$ as a function of energy. The blue dots indicate the variance. Else as in Fig.~\ref{fig:fig_rho_m_mean}}
    \label{fig:fig_alpha_z_variance}
    \end{subfigure}
    \caption{Distribution of the logarithm of $\alpha_{z}$ and derived mean and variance of $\alpha_{z}$.}
    \label{fig:alpha_z_results}
\end{figure}

\begin{figure}
\centering
    \begin{subfigure}{0.5\linewidth}
    \includegraphics[width=\linewidth]{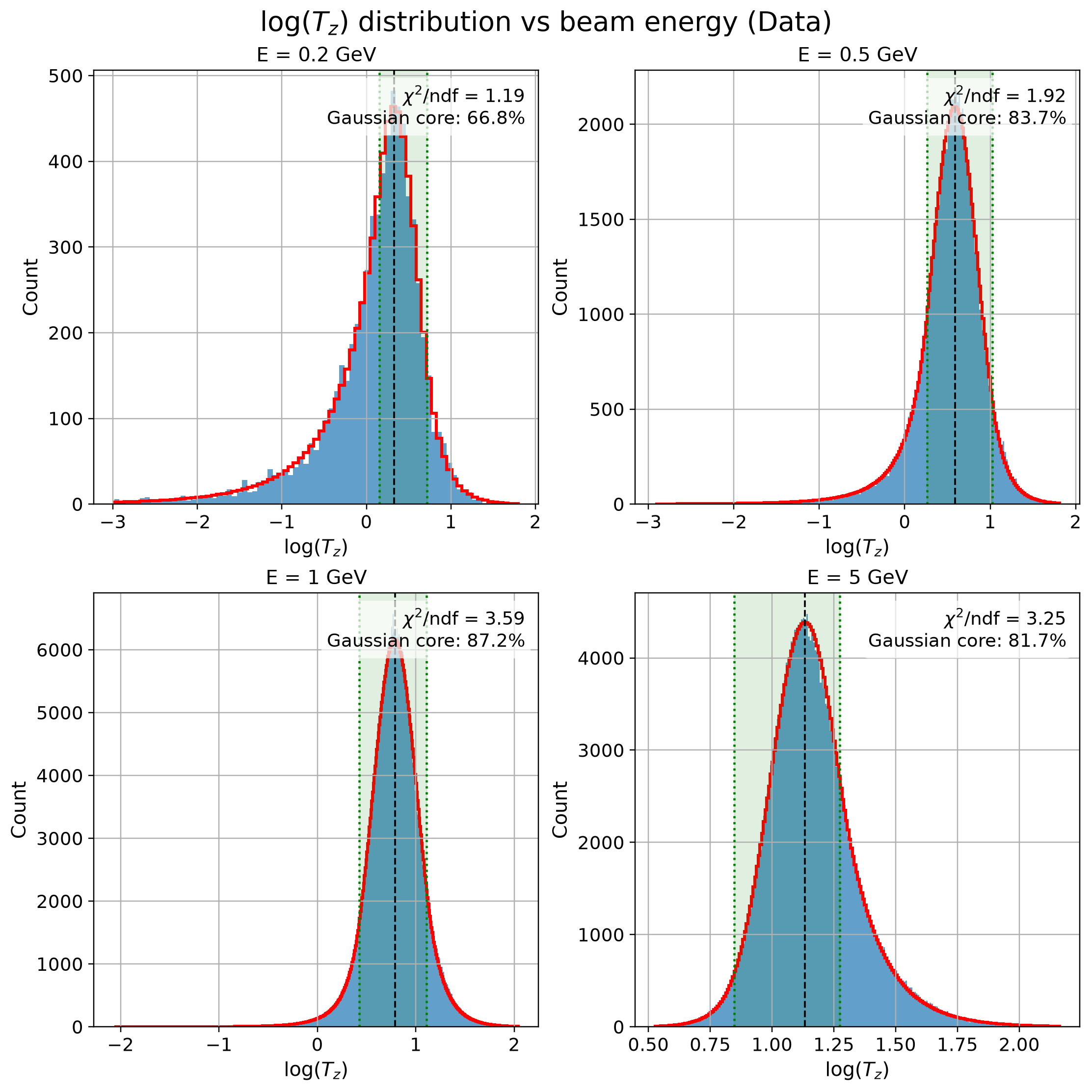}
    \caption{Distributions of $\log{T_{z}}$ variables. Else, as in Fig.~\ref{fig:fig_log_T_z_dist}}
    \label{fig:fig_log_T_z_dist}
    \end{subfigure}
    \vfill
    \begin{subfigure}{0.45\linewidth}
    \includegraphics[width=\linewidth]{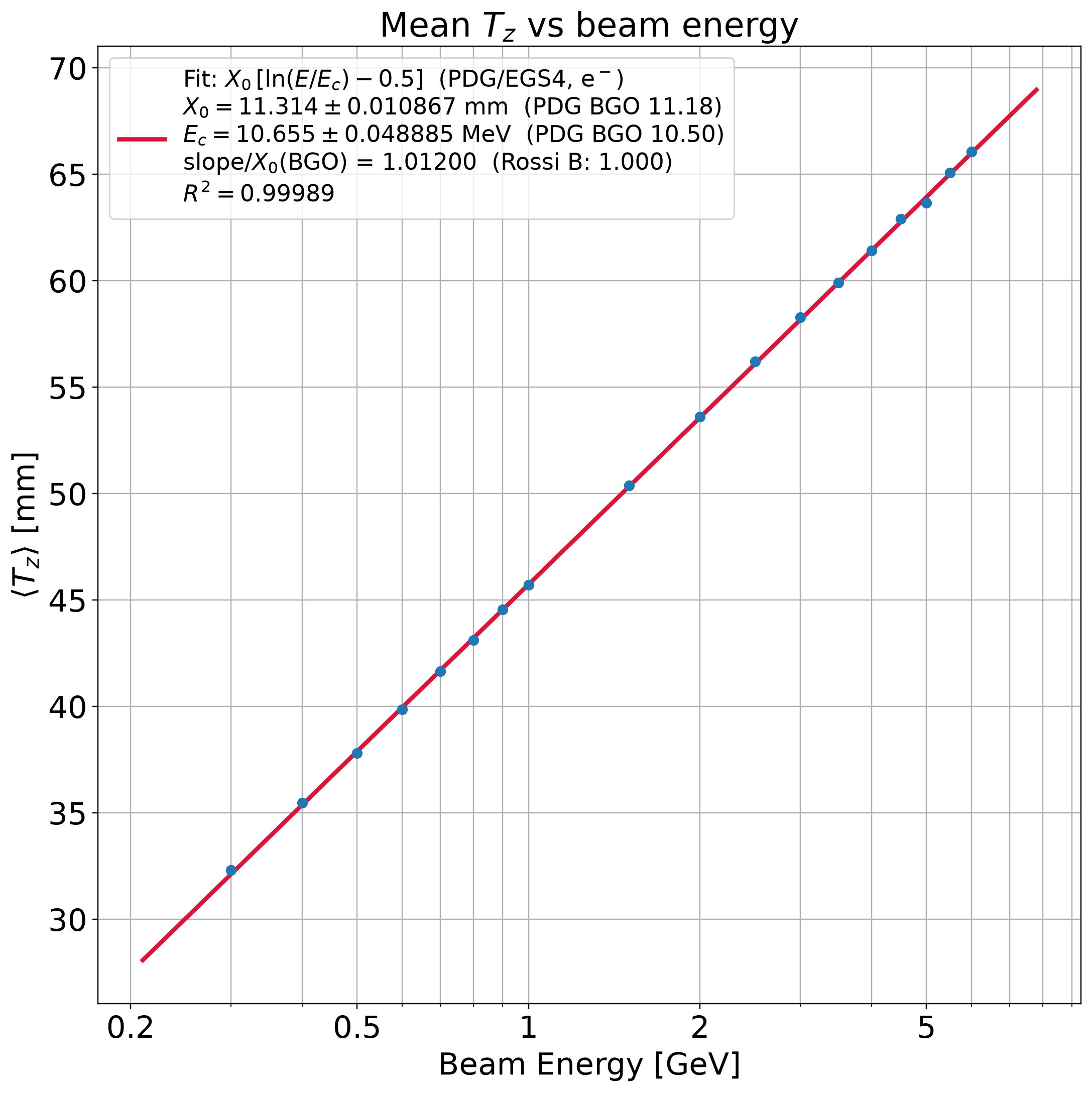}
    \caption{$\langle T_{z} \rangle$ as a function of energy. The blue dots indicate the mean, and the red line indicates a fit to the empirical model in Eq.~\ref{eq:mean_T_z_model}.}
    \label{fig:fig_T_z_mean}
    \end{subfigure}
    \hfill
    \begin{subfigure}{0.45\linewidth}
    \includegraphics[width=\linewidth]{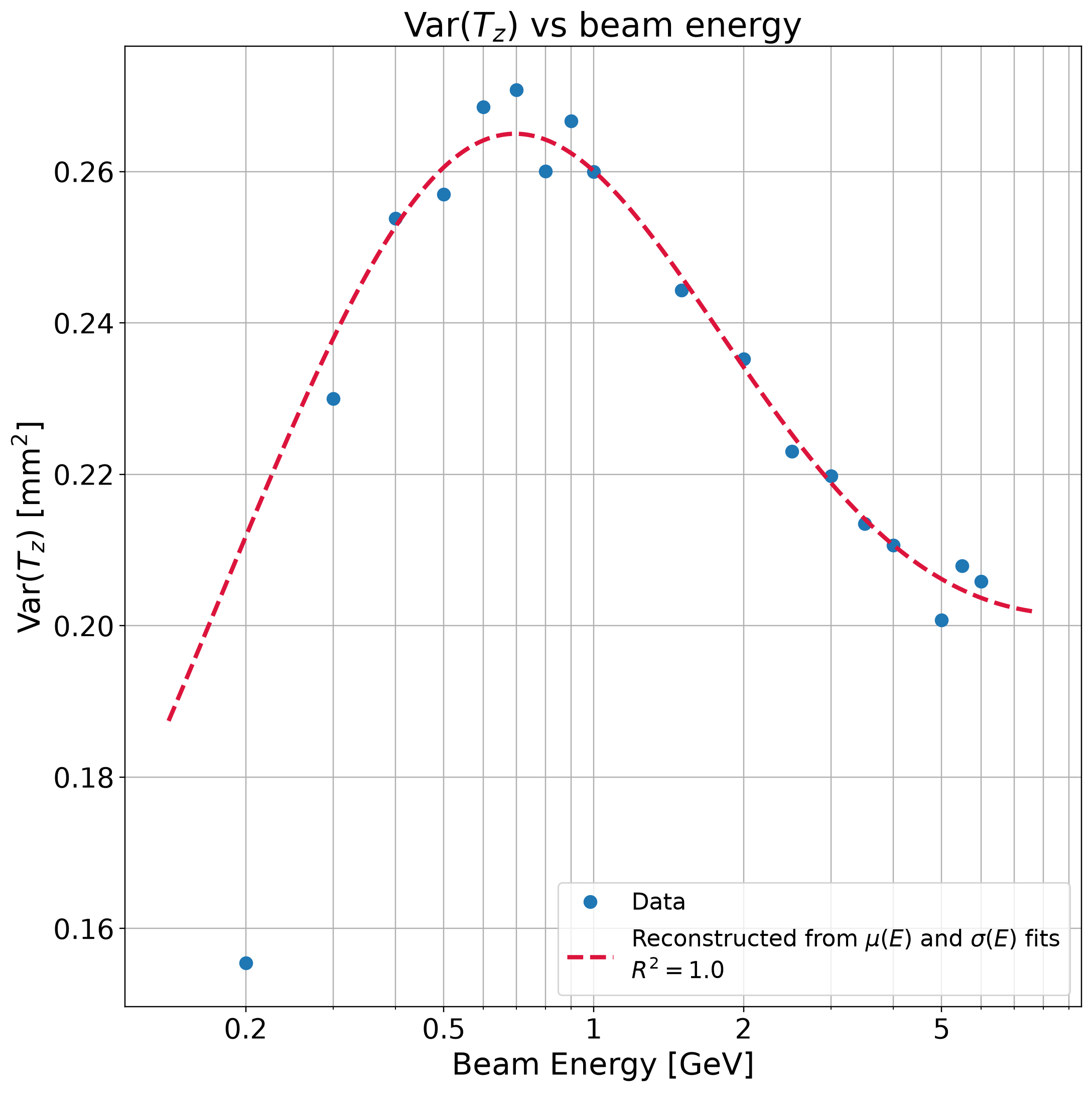}
    \caption{$\operatorname{Var}(T_{z})$ as a function of energy. The blue dots indicate the variance. The red line indicates the value reconstructed from $\langle \log{T_{z}} \rangle$ and  $\operatorname{Var}( \log{T_{z}})$, shown in Fig.~\ref{fig:fig_log_T_z_mean} and Fig.~\ref{fig:fig_log_T_z_variance}}
    \label{fig:fig_T_z_variance}
    \end{subfigure}
    \caption{Distribution of the logarithm of $T_{z}$ and derived mean and variance of $T_{z}$.}
    \label{fig:T_z_results}
\end{figure}

The distributions of the Crystal Ball fits to the logarithms are shown in Fig.~\ref{fig:fig_log_gamma_alpha_z_dist} and Fig.~\ref{fig:fig_log_T_z_dist}. Again, log-normality is generally observed for more than \qty{80}{\percent}  of the data in all cases, with the exception of $T_{z}$ at \qty{0.2}{\giga \electronvolt}, with only \qty{66}{\percent} of the data following a normal distribution. This can be understood as a result of the bias introduced by enforcing $\alpha > 1$, which implies $T_{z} > 0$, leading to tension where the energy distribution starts close to the first layer and is approximately exponential. 

The $\alpha_{z}$ and $T_{z}$ parameters were transformed according to the expectation for log-normally distributed variables, as with $\gamma_{xy}$. The means, $\langle T_{z} \rangle$ and $\langle \alpha_{z} \rangle$ in both cases were fitted with a log-linear form in both cases, shown in Eq.~\ref{eq:log_linear}:

\begin{equation}
    f(E) = A\cdot \left( \log{E} + \log{C} \right)
    \label{eq:log_linear}
\end{equation}

A special case is $\langle T_{z} \rangle$, which was fitted using the known model for the shower maximum of electromagnetic showers of the same form as Eq.~\ref{eq:mean_T_z_model} \cite{groom_passage_2000}:

\begin{equation}
    \langle T_{z} \rangle = X_{0} \cdot \left( \log{\frac{E}{E_{c}}} + C_{e} \right) 
    \label{eq:mean_T_z_model}
\end{equation}

 where $X_{0}$ is the radiation length, defined as the distance required for an electron to lose around \qty{63.2}{\percent} of its energy in matter, $E_{c}$ is the energy at which the average energy losses from radiation processes equal those from ionization \cite{livan_calorimetry_2019}. The constant $C_{e}$ is a correction constant of -0.5 for electron-induced cascades. The PDG values for $X_{0}$ and $E_{c}$ are \qty{11.18}{\milli \meter} and \qty{10.5}{\mega \electronvolt}, respectively \cite{pdg_review_2026, pdg_review_2026_url}. 

Fig.~\ref{fig:fig_alpha_z_mean} shows a log-linear increase. This is expected and implies that showers become less exponential and more spread out longitudinally as a function of energy. Additionally, the variance decreases with energy, indicating that higher-energy showers fluctuate less.

Fig.~\ref{fig:fig_T_z_mean} shows that the radiation length and critical energy of BGO were fitted to be \qty{11.314(0.011)}{\milli \meter} and \qty{10.655(0.049)}{\mega \electronvolt}, respectively. These values agree with the PDG values within \qty{1.1}{\percent} and \qty{1.5}{\percent}, respectively.

The variance of $T_{z}$ cannot be directly fitted, but follows from the monotonic general rational fits performed to $\langle \log{T_{z}} \rangle$ and $\operatorname{Var} \left(\log{T_{z}} \right)$. A dashed red line shows that this behavior can be modeled as the product of two competing processes: the growth of the mean and the shrinking of the variance.

\subsubsection{Correlations}\label{sec:correlations}

The correlations were studied using the Gaussian copula method. The advantage of this method is that it enables the calculation of correlations even in the presence of heavily skewed data. 

A smooth estimate of the ECDF was fitted to the cumulative normalized counts using a monotonic logistic-linked generative additive spline model in \texttt{PyGAM} \cite{daniel_serven_dswahpygam_2025}. Here, the spline used a binomial error distribution and a logit link. The optimal smoothing parameter was found via a grid search over a log-uniform range, with performance minimized using generalized cross-validation. The construction guarantees a smooth, well-bounded, and monotonically increasing ECDF. Both the ECDF and its inverse, the quantile function or percent-point function (PPF), were obtained during this study.

The ECDFs and inverse ECDFs are shown in Appendix Fig.\ref{fig:cdf_ppf_evolution}. The Wasserstein-1 distance and KS-statistic show excellent agreement with the ECDF on average for all variables.

\begin{figure}
    \centering
    \includegraphics[width=\linewidth]{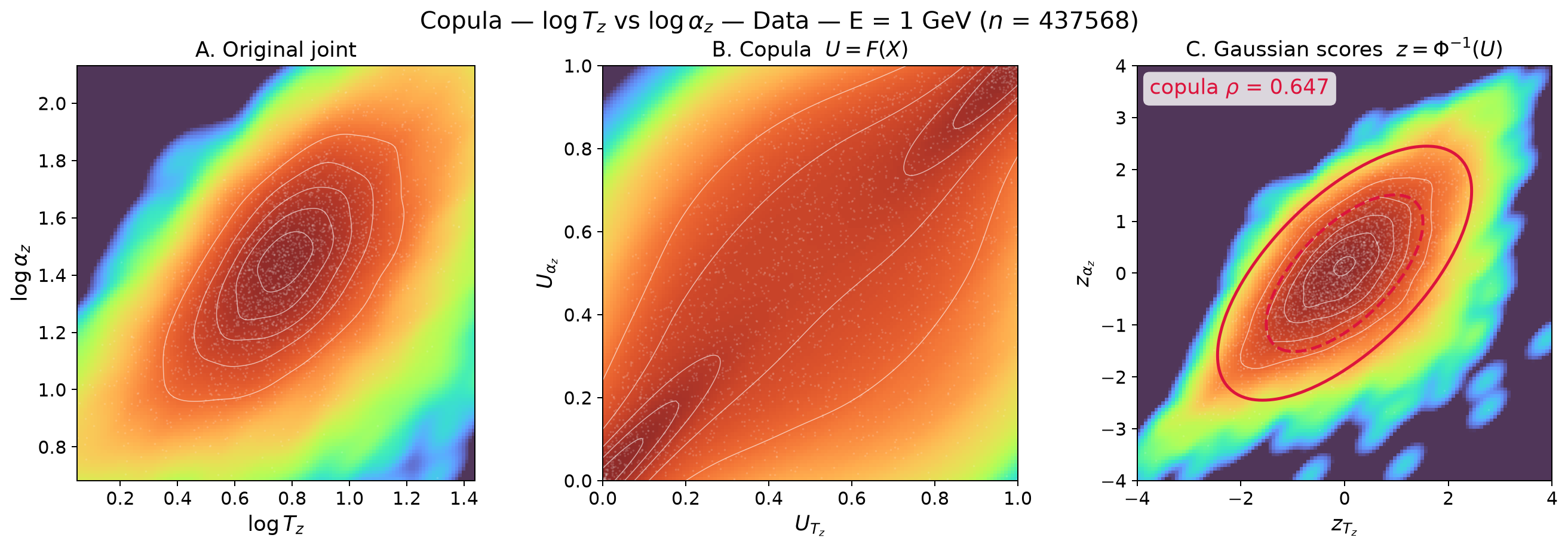}
    \caption{Three stages of the Gaussian copula transformation for the \qty{1}{\giga \electronvolt} data. Blue-to-red indicates low-to-high probability density obtained by kernel density estimation. On the left is the original joint and correlated distribution between $\log{\alpha_{z}}$ and $\log{T_{z}}$. The copula distribution is shown in the middle plot, illustrating the PIT-transformed distributions, $U = F(X)$. The copula transformation to the Gaussian is shown on the right, $Z = \Phi^{-1}(U)$. Contours are shown in white. The dashed and solid red lines mark the 1$\sigma$ and 3$\sigma$ confidence intervals, respectively. The correlation of the normal distribution given by the copula, $\rho$, is shown in the legend. }
    \label{fig:fig_copula_3stage_T_z_alpha_z_data}
\end{figure}

An example 2D copula is shown in Fig.~\ref{fig:fig_copula_3stage_T_z_alpha_z_data} to illustrate the procedure for $\log{T_{z}}$ and $\log{\alpha_{z}}$ only. The central plot illustrates the PIT transform. Both distributions on the $x$ and $y$-axes are uniformly distributed when viewed independently. However, the joint distribution preserves the correlations observed in this plot. The Gaussian-scored distribution is shown on the right. The advantage of this definition of correlation is that it is decoupled from the particular form of the marginal distributions.

\begin{figure}
    \centering
    \includegraphics[width=0.5\linewidth]{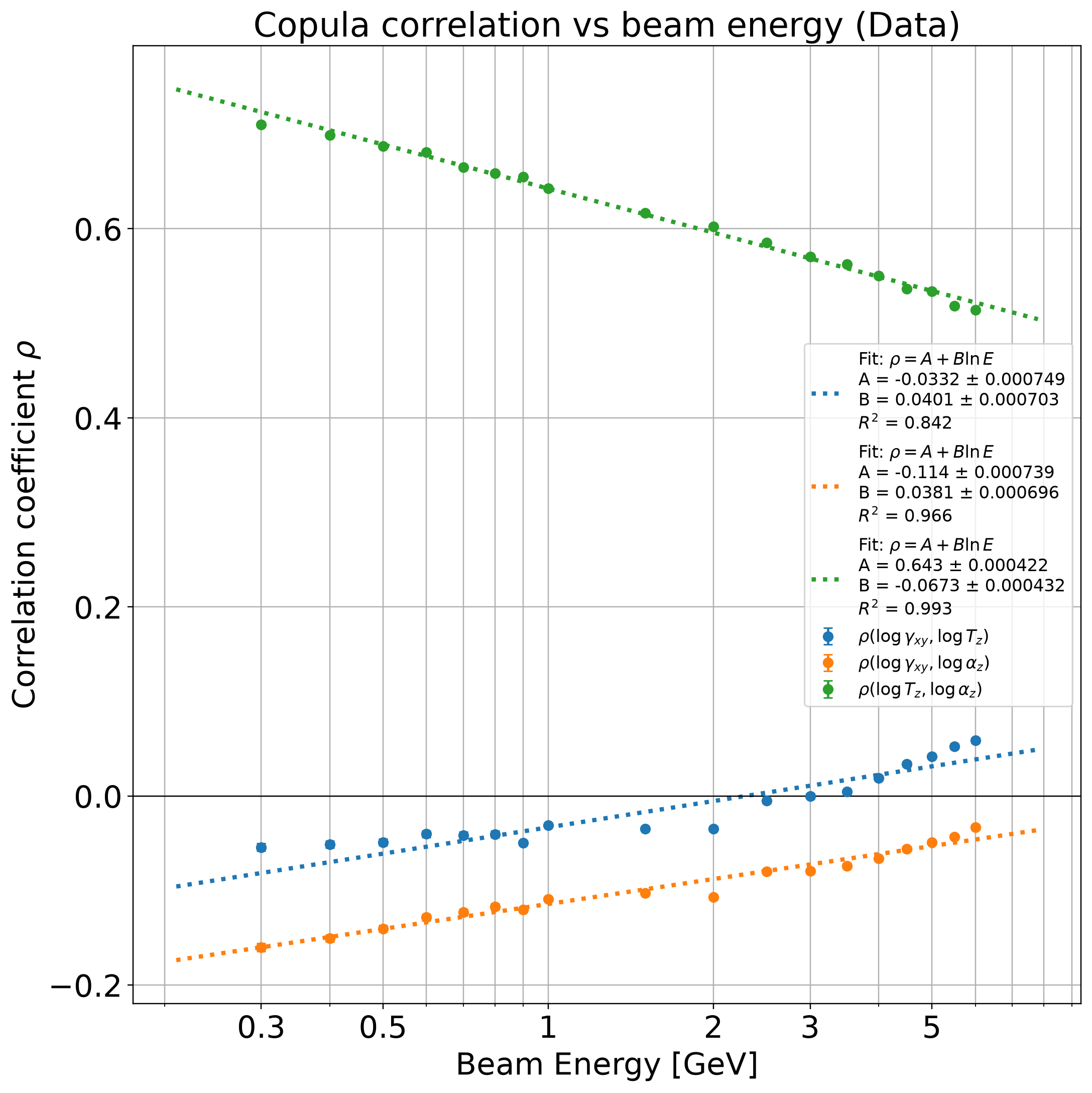}
    \caption{Correlation coefficients $\rho$ in the Gaussian-scored domain. The blue, green, and orange dashed lines indicate fits to the correlation coefficients, $\rho$, of the $\log{\gamma_{xy}}$-$\log{T_{z}}$, $\log{\gamma_{xy}}$-$\log{\alpha_{z}}$ and $\log{T_{z}}$-$\log{\alpha_{z}}$ pairs. The dashed lines in each case indicate a linear fit with particle energy. }
    \label{fig:fig_correlation_vs_energy}
\end{figure}

The correlations of each possible pair between the three variables are shown in Fig.~\ref{fig:fig_correlation_vs_energy}. The strongest is a positive correlation between $T_{z}$ and $\alpha_{z}$ that decreases with increasing energy. A small but non-negligible negative correlation between the shower lateral scale and the longitudinal development was also observed.

\begin{figure}
    \centering
    \includegraphics[width=\linewidth]{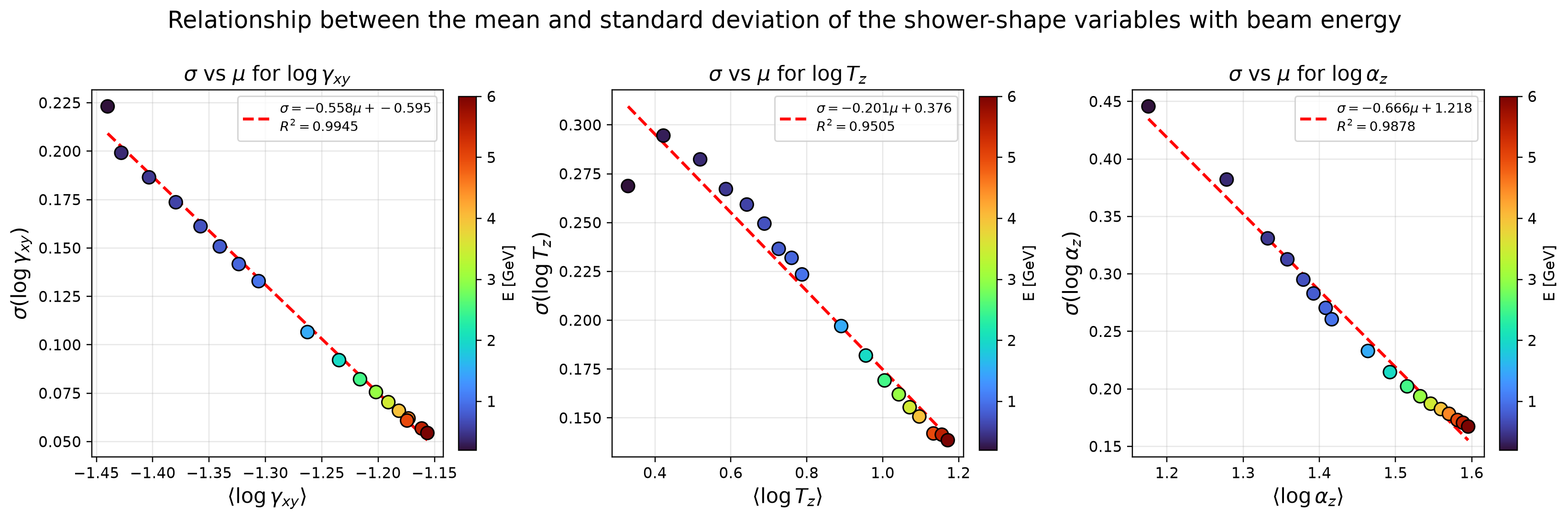}
    \caption{Relationship of the mean and standard deviations of $\log{\gamma_{xy}}$ (left), $\log{T_{z}}$ (middle) and $\log{\alpha_{z}}$ (right). The colored dots, from blue to red, indicate the particle energy. The dashed red line indicates a linear fit to the data. }
    \label{fig:fig_log_sigma_vs_mu}
\end{figure}

The mean and standard deviation of the logarithmic parameters were then compared. In all cases, a negative linear trend was observed. As the mean of all variables increases monotonically with energy, their variance increases linearly with it. In other words, in the energy range studied, showers are shallower and less spread out at low energies and become deeper with increased energy. Accordingly, the shower variance is large at lower energy and decreases with energy.

\subsubsection{Energy Fluctuations of Individual Bars}\label{sec:event_to_event_fluctuations}

A shower is an avalanche process. Fitting the shower shape parameters helps constrain the number of particle generations produced in the shower (e.g., $T_{z}$). Therefore, the remaining event-to-event fluctuations in the energy deposited per unit active volume were studied by conditioning on the shower shape vector, $\theta$, and fitting the distributions in energy deposited in each crystal bar, as well as the relative fractions of energy between individual bars. 

In the following studies, the full set of \qty{0.2}{\giga \electronvolt}, \qty{1}{\giga \electronvolt} and \qty{5}{\giga \electronvolt} $e^{-}$ events are used. \qty{1}{\giga \electronvolt} events are shown in this section and the remainder in the appendix. A shower conditioning vector $\theta$ is also defined using the shower shape information:

\begin{equation}
\theta = \left[c_{x}, c_{y}, \log{\gamma_{xy}}, \log{\alpha_{z}}, \log{T_{z}} \right]^{T}
\end{equation}

Additionally, both SiPMs at each end of the crystal bar were required to simultaneously measure energy above \qty{0.1}{\mathrm{MIP}}. A minimum energy per bar of \qty{0.25}{\text{MIP}} was therefore selected for the study, with a small margin of \qty{0.05}{\text{MIP}} added above the lower limit of \qty{0.2}{\mathrm{MIP}}. The reason is that there is a case where one SiPM on one side of the crystal bar has sufficient energy to meet the threshold, but the other does not. This results in an artificial discontinuity in the energy distribution for a bar and is excluded in the study. The effect of the cut is small compared to the total energy. In the worst case, \qty{0.2}{\giga \electronvolt} showers lose \qty{1.38(1.04)}{\percent} of their energy on average due to this cut.  The fraction is \qty{0.4(0.38)}{\percent} on average across all the data. Consequently, the bias is small, and the fit remains justified.


As an ansatz, the energy distribution of individual bars was fitted with a shower shape-conditioned shifted Gamma distribution:

\begin{equation}
    E_{i} \sim \operatorname{Gamma}(\alpha_i (\theta_{k}), \beta_i (\theta_{k}), \Delta_i(\theta_{k}))
\end{equation}

where $\alpha_i$, $\beta_i$, and $\Delta_i$ are the shape, scale, and location parameters of the Gamma distribution for the $i^{\text{th}}$ bar, and $\theta_{k}$ is the $\theta$ value of the $k^{\text{th}}$ event.

An ansatz was made that the remaining fluctuations would arise from additive, mostly uncorrelated, exponential processes. This is valid because the cell-level variance is dominated by fluctuations within bars. For example, the relative uncertainty on the energy in a single cell was found to be \qtyrange{50}{80}{\percent}, compared to the \qty{2}{\percent \per \sqrt{ E (\giga \electronvolt)}} expected for the total energy according to calorimeter resolution. Event-by-event fluctuations of the energy sum are therefore small or negligible compared to the intrinsic fluctuations within the active volume. 

An MLP was used to perform a conditional fit of $\alpha_i$, $\beta_i$, and $\Delta_i$ as functions of $\theta_{k}$ for each bar with at least 50 data points above the threshold, using unbinned log-likelihood with \texttt{PyTorch} \cite{paszke_pytorch_2019} The model consisted of a 2-hidden-layer MLP with 64 hidden units to map $\theta$ to the $\alpha_{i}$, $\beta_{i}$ and $\Delta_{i}$, with SiLU activation. The models used softplus activations for $\alpha_{i}$ and $\beta_{i}$ and scaled sigmoid activation such that $\Delta_{i} $ was constrained below the minimum energy deposit \qtyrange{0}{2.5}{\mathrm{MIP}}. Adam (learning rate=\num{2e-3}, 4000 steps, batch size 8192) on the negative log-likelihood of the Gamma distribution with an 80/20 train/test split. 

To evaluate the performance of the fit, the raw data counts were compared with the estimate obtained by summing the conditional probabilities, $\hat{E}_{i} = \sum_{k} P(\alpha_{i}(\theta_{k}), \beta_{i}(\theta_{k}), \Delta_{i}(\theta_{k})) $, where the sum runs with index $k$ over all the events, $\hat{E}_{i}$ is the predicted energy per bin and $P$ is the probability assigned to the bin by the conditional distribution of event $k$. This gives an effective count. From the estimate, the reduced-$\chi^{2}$ was calculated using bins chosen by the Freedman–Diaconis rule. The KS statistic was computed unbinned.

\begin{figure}
    \centering
    \includegraphics[width=0.8\linewidth]{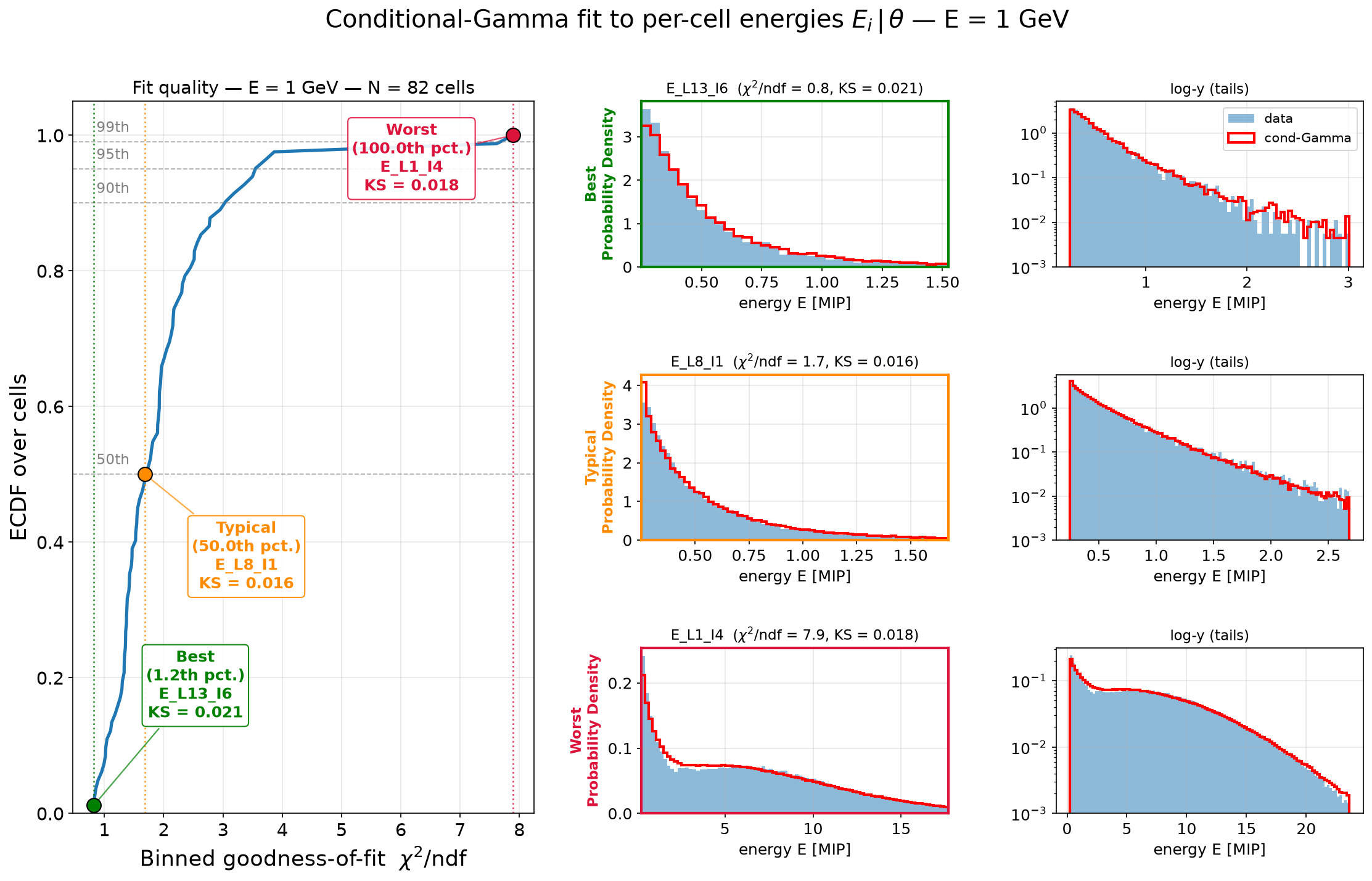}
    \caption{Conditional Gamma distribution fits to the energy deposited per crystal bar. The right-hand plot shows the CDF of the reduced-$\chi^{2}$ distribution over the full set of 82 of 84 cells. The grid on the right shows the best, typical, and worst examples, coloured green, orange, red, and purple, respectively. The blue histogram shows the data, and the red line shows the fitted distribution. The median fit parameters, reduced $\chi^{2}$, and PIT-KS statistic are shown for each figure, using only the test sample of events. In the far-right column, the same distribution is shown on a log scale.}
    \label{fig:fig_cond_gamma_cdf_fits_1GeV}
\end{figure}

The results are shown in Fig.~\ref{fig:fig_cond_gamma_cdf_fits_1GeV} for \qty{1}{\giga \electronvolt}, and in Appendix Figs.~\ref{fig:fig_cond_gamma_cdf_fits_0p2GeV} and \ref{fig:fig_cond_gamma_cdf_fits_5GeV} for \qty{0.2}{\giga \electronvolt} and \qty{5}{\giga \electronvolt}, respectively. The conditional fits indicate excellent agreement across nearly all cells studied, with a median reduced-$\chi^{2}$ of approximately 2 using only the test data.  The worst fit is one of the two bars that initiate the shower: L1\_I4, indicating layer 1 and bar index 4. 

The distributions of log-fractions of energy within an event were then studied. The log fractions have a convenient property that, within the same event, the contributions of fluctuations from the sum are canceled:

\begin{equation}
    \ell_{ij} = \log{\frac{\phi_{i}}{\phi_{j}}} = \log{E_{i}}- \log{E_{j}}
\end{equation}

where $i$ and $j$ are bar indices,  $\phi_{i} = E_{i}/\sum_{k \in \mathrm{event}} E_{k}$ is the energy fraction of energy of bar $i$, where $i$ and $k$ are indices. 

Under the hypothesis that $E_i$ is Gamma-distributed, $\log{E_i}$ is log-Gamma-distributed. The difference of independent Log-Gamma variables follows Fisher's $z$-distribution, as shown in Eq.~\ref{eq:fisher_z}, and can capture correlations between bars by introducing a scaling factor $\bar{s}$, as shown in Eq.~\ref{eq:s_bar_variance_relation}. The log-ratio is studied only by comparing cells within a single event, which further reduces the influence of fluctuations between different showers after accounting for the shape. 

As for the Gamma distribution, an identical MLP with a identical training method to the Gamma case, this time outputting $d_{1}$, $d_{2}$, $\bar{s}$ and $\Delta$ with softplus activation was used to perform a conditional fit of Fisher's $z$-distribution to each bar combination according to Eq.~\ref{eq:fisher_z_fit_1} and Eq.~\ref{eq:fisher_z_fit_2}:

\begin{subequations}
  \begin{equation}
 \ell_{ij} \sim \operatorname{FisherZ}(d_{1,ij}(\theta_{k}), d_{2,ij}(\theta_{k}), \Delta_{ij}(\theta_{k}), \bar{s}_{ij}(\theta_{k}))
 \label{eq:fisher_z_fit_1}
\end{equation}

\begin{equation}
 \ell_{ij} \sim \Delta_{ij}(\theta_{k}) + \bar{s}_{ij}(\theta_{k}) \cdot \log{F(d_{1,ij}(\theta_{k}), d_{2,ij}(\theta_{k}))}
  \label{eq:fisher_z_fit_2}
\end{equation}

\end{subequations}

where $\Delta_{ij}$, $\bar{s}_{ij}$, $d_{1,ij}$ and $d_{2,ij}$ are the parameters for the combination of the $i^{\mathrm{th}}$ and $j^{\mathrm{th}}$ bar and are functions of the $\theta$ for the $k^{\mathrm{th}}$ event. The same conditions, criteria, and metrics were applied to this fit, using an MLP to fit the unbinned log-likelihood, as for the Gamma distribution. 

The estimated distribution of the log-ratios was again obtained using $\hat{\ell}_{ij} = \sum_{k \in \mathrm{event}} P(\Delta_{ij}(\theta_{k}),\bar{s}_{ij}(\theta_{k}), d_{1,ij}(\theta_{k}), d_{2,ij}(\theta_{k})))$, where the sum runs with index $k$ over all the events, $\hat{\ell}_{ij}$ is the predicted log-energy ratio distribution and $P$ is the probability assigned to the bin by the conditional distribution of event $k$. Again, the reduced-$\chi^{2}$ is calculated using the binned statistic, whereas the PIT-KS is applied using the unbinned data.

\begin{figure}
    \centering
    \includegraphics[width=0.8\linewidth]{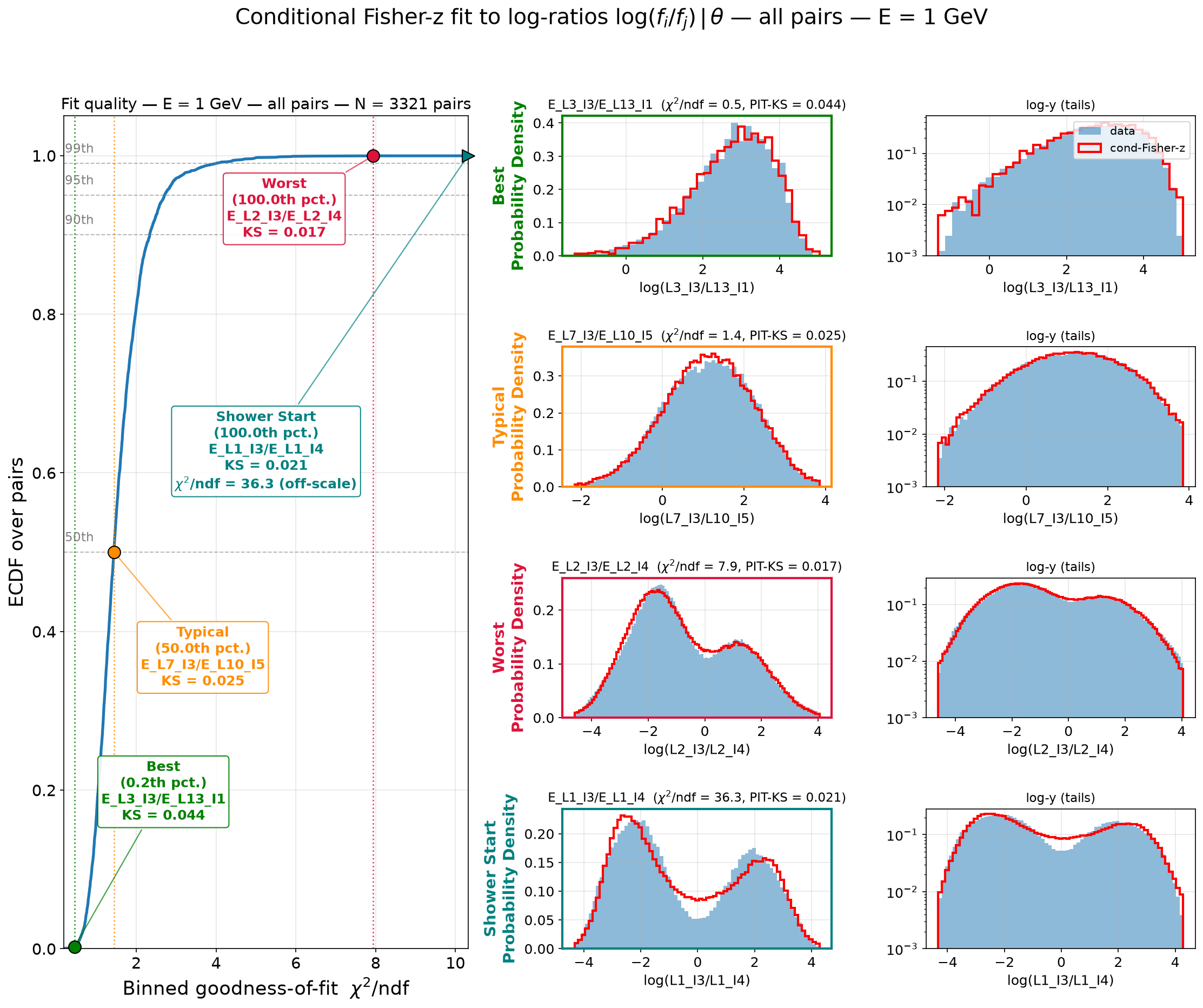}
    \caption{Conditional Fisher-$z$ distribution fits to the log ratio of all possible energy fractions deposited per layer for \qty{1}{\giga \electronvolt} electrons. The shower starting indices/layers are highlighted in light blue. Else, as in Fig.~\ref{fig:fig_cond_gamma_cdf_fits_1GeV}.}
    \label{fig:fig_cond_fisherz_pairs_cdf_fits_1GeV}
\end{figure}

The results are shown for \qty{1}{\giga \electronvolt} in Fig.~\ref{fig:fig_cond_fisherz_pairs_cdf_fits_1GeV} and for \qty{0.2}{\giga \electronvolt} and \qty{5}{\giga \electronvolt} in Appendix Fig.~\ref{fig:fig_cond_fisherz_pairs_cdf_fits_0.2GeV} and Fig.~\ref{fig:fig_cond_fisherz_pairs_cdf_fits_5eV}. The results show that the median reduced-$\chi^{2}$ is close to 1.5, indicating excellent agreement with the model, and that it extends up to around 8 in the worst case. A special case was observed in the bars where the showers start, L1\_I3 and L1\_I4, which show the worst agreement by a considerable margin compared to any other example. This result is most likely attributable to the specific behavior of the shower start. Notably, the next-worst case is the corresponding pair of bars in the next layer, L2\_I3 and L2\_I4.  However, this is a single extreme case, and for nearly any other combination, an excellent goodness-of-fit is observed.

This study shows that the individual bar energies can be modeled as Gamma-distributed, conditioned on the shower shape, with excellent agreement. Additionally, the particular form of the log-ratios has implications for generative modeling of particle showers and is discussed in Sec.~\ref{sec:full_event_modelling}.

\subsection{Generative Modelling}\label{sec:generative_modelling}

Methods for simulating both shower shape parameters and full events are provided using the previously introduced measurements, concepts, and techniques.

\subsubsection{Shower Shape Modelling}\label{sec:shower_shape_modelling}

Two methods for generative modeling of shower shape variables from data parameterizations are discussed. The \texttt{Geant4} fast EM shower generation code for homogeneous calorimeters follows the framework of Ref.~ \cite{grindhammer_parameterized_2000} and is implemented in Ref.~\cite{barberio_gflashhomoshowerparameterisationcc_2004}. While the models share the same longitudinal shower shape model, the lateral model depends on shower depth. As previously mentioned, a simpler exponential model was applied in this study owing to the limited granularity, the structure of the sensitive volume, and the energy sensitivity required for radial fits per layer for \unit{\mega \electronvolt}-scale particle showers. For this reason, a direct comparison with the specific \texttt{Geant4} parameterization is not undertaken in this study. Instead, the general method of generating energy density profiles was evaluated using the shower shape parameterizations extracted in Sec.~\ref{sec:lateral_distributions} and Sec.~\ref{sec:longitudinal_distributions} and compared with a non-parametric copula transformation as a generative method.
 
Let $\theta_{\mathrm{shape}} = \left[\log\gamma_{xy},\; \log\alpha_{z},\; \log T_{z}\right]^{T}$ 
denote the shape subvector of $\theta$, with

\begin{subequations}
\begin{equation}
     \mu_{\mathrm{shape}} = \left[ \langle\log\gamma_{xy}\rangle, \langle\log\alpha_{z}\rangle, \langle\log T_{z}\rangle \right]^{T}
\end{equation}

    \begin{equation}
         S = \operatorname{diag}\!\left[
        \sqrt{\operatorname{Var}(\log\gamma_{xy})},\;
        \sqrt{\operatorname{Var}(\log\alpha_{z})},\;
        \sqrt{\operatorname{Var}(\log T_{z})}
    \right].
    \end{equation}
\end{subequations}

The covariance matrix is constructed as
\begin{equation}
    \Sigma_{\mathrm{shape}} = S\,R\,S,
\end{equation}
and the compatible parameters are sampled via
\begin{equation}
    \theta_{\mathrm{shape}} \sim \mathcal{N}_{3}\!\left(\mu_{\mathrm{shape}},\; \Sigma_{\mathrm{shape}}\right).
\end{equation}

In this formulation, the elements of $\mu$, $D$, and $\rho$ are drawn from the characterizations of energy.

The \texttt{Geant4} fast EM shower generation method (multi-variate normal, MVN), for $\alpha_{z}$ and $T_{z}$ in Ref.~\cite{barberio_gflashhomoshowerparameterisationcc_2004} and in the case studied here including $\gamma_{xy}$, obtains the 
\begin{equation}
\left[\gamma_{xy},\; \alpha_{z},\; T_{z}\right] = \exp\left(\theta_{\mathrm{shape}}\right)    
\end{equation}
This method operates under the assumption that the parameters are log-normally distributed. 

The inverse ECDFs obtained from Sec.~\ref{sec:correlations} and shown in Appendix Fig.~\ref{fig:cdf_ppf_evolution} instead use a mapping based on the copula method (copula). This is achieved using the inverse operation of the forward mapping of the uniform marginals to the physical parameter space:

\begin{equation}
   \left[\gamma_{xy},\; \alpha_{z},\; T_{z}\right] = F^{-1}(\Phi(\theta_{\mathrm{shape}}) | E)     
\end{equation}

where $F^{-1}$ is the inverse-ECDF measured from data, and $\Phi$ is the probit function and $E$ is the energy bin. In the study, it is assumed that the inverse CDFs can be interpolated to estimate the shapes in between the energy bins. 

The performance was measured by sampling \num{2e5} events per particle energy using the $W_{1}$ statistic compared to the original data. The real and reconstructed covariances are also reconstructed from the data using the Frobenius norm of the difference between the generated and data correlation matrices extracted directly from the data according to Eq.~\ref{eq:R_diff}:

\begin{equation}
  \left|R_{\mathrm{data}}-R_{\mathrm{gen}}\right|_{F}=\sqrt{\sum_{i=1}^{n}\sum_{j=1}^{n}(R_{\mathrm{data}, ij}-R_{\mathrm{gen}, ij})^{2}}
  \label{eq:R_diff}
\end{equation}

\begin{figure}
    \centering
    \includegraphics[width=\linewidth]{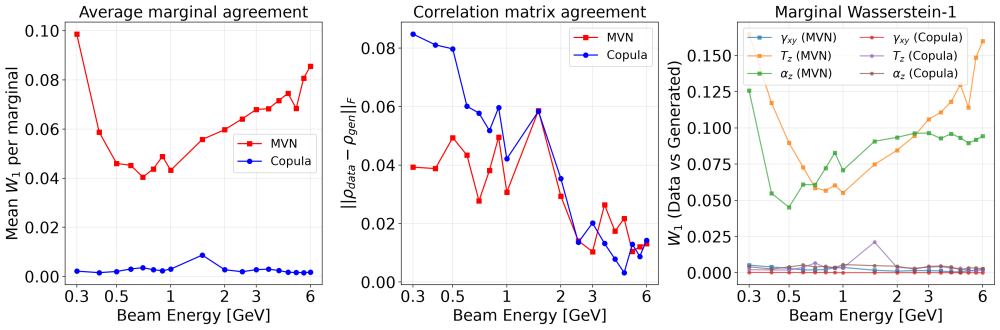}
    \caption{Performance of each shower-shape generation method. The right plot shows the mean $W_{1}$ value for each method, and the middle plot shows the difference in the correlation matrices calculated from data. The red squares and blue circles show the MVN and copula methods, respectively. The right plot shows the marginal W1 split by the generated variable. Blue, orange and green (MVN) vs. red, purple, and brown (copula) show the $W_1$ variable for the $\log{\gamma_{xy}}$, $\log{\alpha_{z}}$, and $\log{T_{z}}$ variables, respectively.}
    \label{fig:fig_shower_shape_generation_W1_covariance}
\end{figure}

The results are shown in Fig.~\ref{fig:fig_shower_shape_generation_W1_covariance}. The comparison of the blue and red lines indicates a greatly reduced $W_{1}$ distance and similar covariance agreement compared to the pure exponential sampling method. The reason for the slightly worse covariance agreement is that the covariance in real data is not the same as for the copula transform. The reason is illustrated in the right plot. Comparing the green and orange lines to the purple and brown lines shows that the MVN method poorly describes the $\log{T_{z}}$ and $\log{\alpha_{z}}$ distributions compared to the copula method.

\begin{figure}
\centering
\begin{subfigure}{0.6\linewidth}
\centering
\includegraphics[width=\linewidth]{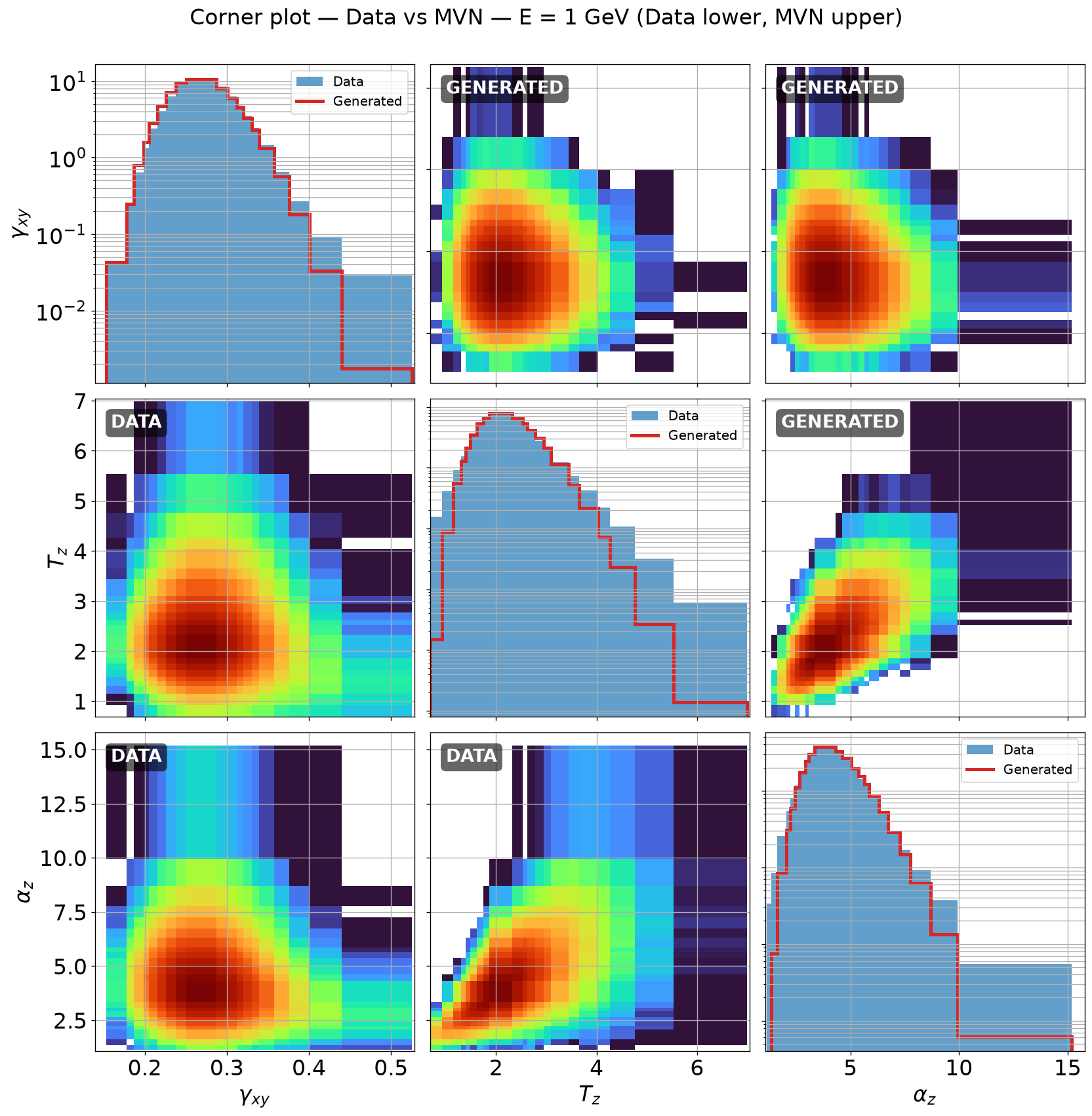}
\caption{}
\label{fig:fig_showershape_gen_corner_1p0GeV_mvn}
\end{subfigure}
\vfill
\begin{subfigure}{0.6\linewidth}
\centering
\includegraphics[width=\linewidth]{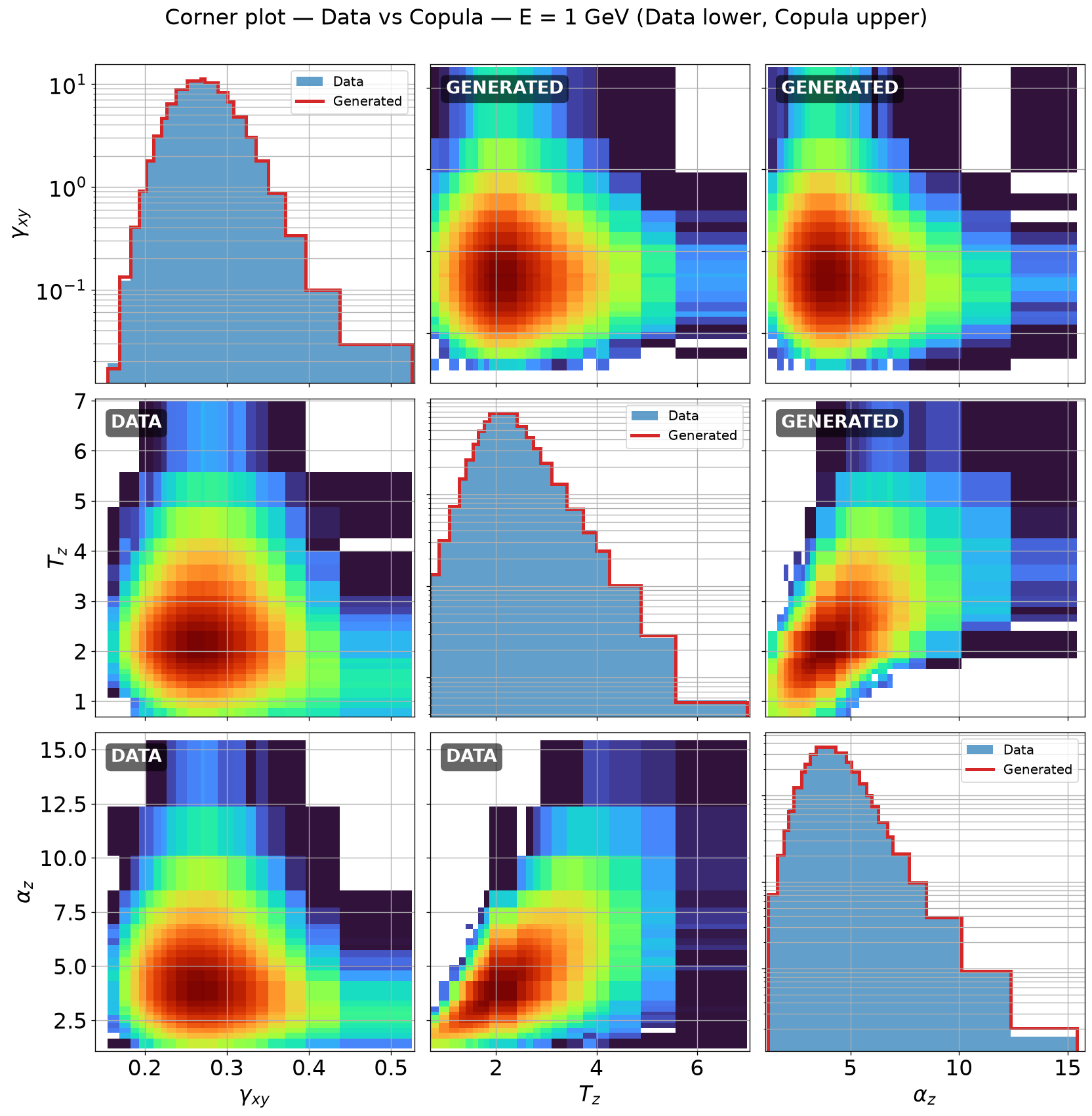}
\caption{}
\label{fig:fig_showershape_gen_corner_1p0GeV_copula}
\end{subfigure}

\caption{1D/2D comparison histograms from the reconstruction of the MVN and copula method, shown in Fig.~\ref{fig:fig_showershape_gen_corner_1p0GeV_mvn} and Fig.~\ref{fig:fig_showershape_gen_corner_1p0GeV_copula} for the \qty{1}{\giga \electronvolt} sample. respectively. The diagonal of the matrix shows the data and generated distributions in solid blue and red, respectively. The upper and lower triangles show the joint distributions. Blue to red colors show the probability density. The upper and lower triangles show the generated and original data distributions, respectively. }
\label{fig:fig_showershape_gen} 
\end{figure}

The \qty{1}{\giga \electronvolt} comparison histogram matrices are shown in Fig.~\ref{fig:fig_showershape_gen_corner_1p0GeV_mvn} and Fig.~\ref{fig:fig_showershape_gen_corner_1p0GeV_copula} for the MVN and copula methods, respectively. The diagonal elements in Fig.~\ref{fig:fig_showershape_gen_corner_1p0GeV_mvn} and the comparison between the joint distributions shown in the upper and lower triangular parts of the matrix reveal that this method underweights the tails of the distribution. This is not observed for the copula method in Fig.~\ref{fig:fig_showershape_gen_corner_1p0GeV_copula}. By contrast, the diagonal shows almost perfect agreement. In addition, the rare dark-indigo regions are better modeled by the copula method. Therefore, it may be concluded that copula transformation is a valid alternative to a naive MVN-based sampling method and can be used to more accurately describe the heavy tails of the distributions. 

The performance of the sampler was measured using a boosted decision tree classifier to distinguish real from generated $\theta_{\mathrm{shape}}$. The quality of the generated events can be measured if a trained classifier between the data and generated events has a receiver operating characteristic (ROC) curve with an area under the curve (AUC) close to 0.5. A boosted decision tree from the \texttt{xGBoost} library was optimized \cite{chen_xgboost_2016}. The model was chosen to have a maximum depth of 6 and up to 500 boosting rounds with early stopping. The main validation sample was split into subsets: a 60/20/20 split defined as the classifier training, classifier validation, and classifier testing sets. For each particle energy, two classifiers are trained using the classifier training set. Then, the best model state is selected using the classifier validation set; finally, the AUC is evaluated on the classifier testing set. The first classifier compares data to data: the data sample is split into two random disjoint halves, and the classifier is trained to distinguish them. This step defines the baseline performance for a perfect generator. The second classifier compares generated to data events: the classifier is trained on data vs. generated events from the fitted model. The random splittings are repeated ten times, and the model is trained and evaluated for each split to measure the uncertainty in the result. This method is also applied in Sec.~\ref{sec:full_event_modelling} for classifier-based metrics.

\begin{figure}
    \centering
    \includegraphics[width=0.5\linewidth]{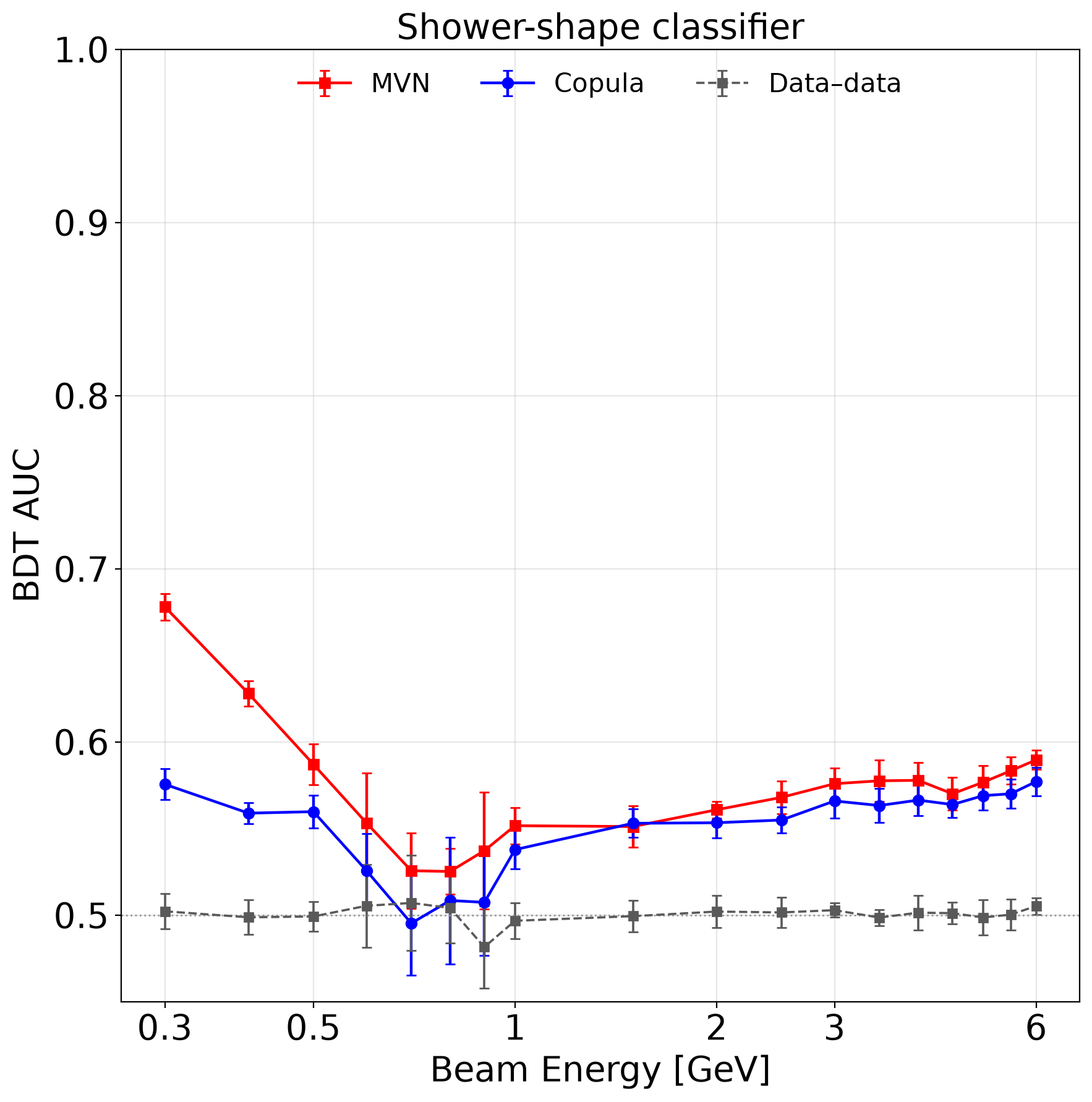}
    \caption{ROC AUC performance for the classifier of $\theta_{\mathrm{shape}}$. The red and blue lines indicate the MVN and copula methods' generated values compared to the data, and the grey line indicates the data compared to the data. Error bars indicate the fluctuation of the result depending on ten different initializations.}
    \label{fig:fig_showershape_gen_classifier_auc_vs_energy}
\end{figure}

The results are shown in Fig.~\ref{fig:fig_showershape_gen_classifier_auc_vs_energy}. The results show that the generated vs. data classifier performance of MVN and Copula methods produces an average AUC of \num{0.573(0.0125)} and \num{0.55(0.0133)} compared to a data vs. data baseline of \num{0.5(0.012)}. Performance is worst at low energy at around \qty{0.3}{\giga \electronvolt}. The copula considerably outperforms the MVN method, owing to its better modeling of the tails of the distribution. The best-modeled energies are at \qtyrange{0.7}{0.9}{\giga \electronvolt}, where the generated shower variables are consistent with the baseline. Both methods have approximately the same AUC of around \numrange{0.57}{0.59} in the range of energy \qtyrange{1}{6}{\giga \electronvolt}. At most energies, the copula outperforms the MVN, though generally the agreement is close at energies greater than \qty{1}{\giga \electronvolt}. 

In summary, both the MVN and copula method can be used as a generative model for shower shape variables, with a minor improvement obtained by using the copula method.

\subsubsection{Full Event Modelling} \label{sec:full_event_modelling}

Consider a compositional vector containing $D$ parts, i.e. $\phi = \left[ \phi_{1}, \phi_{2}, \phi_{3} \dots \phi_{D} \right]^{T}$, where all $\phi_{i} > 0$ and $\sum^{D}_{i=1} \phi_{i} = 1$. The additive log ratio (ALR) transform is defined as:

\begin{subequations}

\begin{equation}
    \operatorname{alr}(\phi) = \left[\ell_1, \ell_2, \dots, \ell_{D-1} \right]^{T}
\end{equation}

\begin{equation}
    \ell_i = \log{\frac{\phi_{i}}{\phi_{D}}}
\end{equation}
    
\end{subequations}

where the transformed vector has $D-1$ dimensions and $\phi_{D}$ could be any fraction in the vector. 

The inverse operation of the ALR is called the Softmax operator:

\begin{equation}
    \operatorname{Softmax}(\ell_{i}) = \frac{e^{\ell_i}}{\sum^{D-1}_{k=1} e^{\ell_{k}}+1}.
    \label{eq:softmax}
\end{equation}

The Softmax is also invariant under common additive shifts. Let $g$ be a vector of inverse-Softmax scores, commonly called logistic units or 'logits'. It is noted that the ALR can be composed with respect to any element $g_{D}$: $\operatorname{Softmax}(g) = \operatorname{Softmax}(g - g_{D})$. In other words, the $D$-dimensional vector $g$ and the $(D-1)$-dimensional ALR vector with $\ell_i = g_i - g_D$ define the same composition: $\operatorname{Softmax}(g) = \operatorname{alr}^{-1}(\ell)$, where the reference fraction enters through the constant term in the denominator. It is noted that $g$ is not necessarily equivalent to $\log{E}$, where $E$ is the vector of energies in the event, due to this property. This reasoning is similar in spirit to the argument that a composition is logistic-normally distributed if the elements of the ALR transform follow a multivariate normal distribution \cite{aitchison_statistical_1982, aitchison_logistic_normal_1980}. 

At this stage, an additional conditioning vector $\eta$ is defined:

\begin{equation}
\eta = \left[c_{x}, c_{y}, \log{\gamma_{xy}}, \log{\alpha_{z}}, \log{T_{z}}, \log{E_{\mathrm{particle}}} \right]^{T}
\end{equation}

where $\eta$ is the condition $\theta$ concatenated with the log of the particle energy; $E_{\mathrm{particle}}$. This is necessary only because an absolute energy scale is required to describe the distribution of censored (dark) active bars in the event below threshold.    

Sec.~\ref{sec:event_to_event_fluctuations} demonstrated that the ALR transform of the energy fractions for any fixed combination of crystal bars, the log ratio of the energy between each bar, followed a scaled Fisher's $ z$-distribution conditioned on the shower-shape parameters, $\theta$, with excellent agreement. Therefore, a joint distribution over the fractions with this property can be obtained by defining $g$ to be correlated Log-Gamma variates, $g_i \sim \operatorname{LogGamma}(\Delta_{i}(\eta) \alpha_i(\eta), \beta_i(\eta))$. This choice is advantageous as no reference cell is required, and all physical content is carried by pairwise differences $g_i - g_j = \log(\phi_i/\phi_j)$, which are the log-ratios studied in Sec.~\ref{sec:event_to_event_fluctuations}. 

Four caveats are highlighted. Firstly, special cases, such as the shower-start layer, exhibit additional fluctuations that must be modeled differently. can be resolved by including a $\eta$-conditional spline to weight $\zeta$ after it is generated. Rational-Quadratic Splines (RQS) are an attractive method for this purpose as they are invertible, continuous, monotonic, and have well-defined gradients, and were introduced for the specified purpose of re-weighting probability densities in Ref.\cite{durkan_neural_2019}

Secondly, $\ell \sim \operatorname{FisherZ} | \eta$ with scaling $\bar s$ and offset $\Delta$ , implied by the study of $\ell \sim \operatorname{FisherZ} | \theta$ from Sec.~\ref{sec:event_to_event_fluctuations},  requires correlations between bars to be expressed. This can be resolved by using a Gaussian copula to model the correlations between the bars. These correlations can be modeled with an MVN and transformed to the marginal LogGamma distribution via a copula. 

Thirdly, calorimeter energy deposits are drawn from a censored distribution (i.e., above an energy threshold). Consequently, the probability distribution for energies below the threshold is ill-defined. This can be resolved by defining a censored likelihood that uses the expected limit of the fraction given $\theta$ to account for the missing probability mass due to the cut, as discussed in Appendix Sec.~\ref{sec:censored_fits}. The latent variables of the factor analysis generative model discussed in Appendix Sec~\ref{sec:factor_analysis} are used to make the calculation of the log-likelihood tractable for a particular latent.

Fourthly, the marginal model must be able to describe both the dark and active bars in the event, which enters through a censored probability term that implicitly includes the marginal as part of the copula transform. The dark probability mass must be describable by the CDF of the marginal distribution in the censored Tobit loss term. Without a probability mass term in the marginal, the same $\alpha$, $\beta$, and $\Delta$ must alone describe both active and dark bars. To resolve this pathology, a zero-inflated Log Gamma distribution is defined: 

\begin{equation}
    \operatorname{ZILG}(x; \alpha, \beta, \Delta, \pi) = \pi \cdot \delta(x) +  (1 - \pi)\cdot\operatorname{LogGamma}(\alpha, \beta, \Delta) 
\end{equation}

where $\delta(x)$ is the Dirac delta function, $\pi$ is the per-cell 'dark' probability. Zero inflation is necessary as the Log-Gamma distribution alone cannot describe the dark cells. The point mass therefore absorbs the inactive fraction, which cannot be described by the Log-Gamma distribution and would cause bias in the marginal parameter fits of $\alpha$, $\beta$ and $\Delta$ if not included.

Alg.~\ref{alg:sample} presents a generative sampling method for showers. In the following, $i$ represents the index of a single bar in the calorimeter. The model accepts the $E_{\mathrm{sum}}$, $\eta$ and an activity mask, $a$, as inputs. $a$ is used only during training. During generation, a generated mask defined as $\tilde a$ is used instead. It is derived by thresholding the generated log fractions at $\log(\frac{E_{\mathrm{cut}}}{E_{\mathrm{sum}}})$.

Firstly, latent vectors are drawn that encode the event-wise fluctuations:  $w_{\nu} \in \mathbb{R}^{4}$ for the Gamma mean parameter, $\nu$; $w_{\pi} \in \mathbb{R}^{4}$ for the 'active' fraction of sensors, $\pi$; and $w_{c} \in \mathbb{R}^{8}$, which describes the conditional mean of the copula variable, $m$. MLPs are used to predict conditional variables. The conditioning vector $\eta$ for an event is passed into a 2-hidden-layer encoder MLP with 128 hidden units and SiLU activation, followed by individual output layers for all the conditional distribution parameters: the shower shape parameter, $\bar \alpha = \operatorname{MLP}_{\alpha}(\eta)$; the shower mean, $\bar \nu = \operatorname{MLP}_{\nu}(\eta)$; a logit for the scaling of the location parameter $\Delta$, $\bar \Delta = \operatorname{MLP}_{\Delta}(\eta)$; the per-cell active probability, $\bar \pi = \operatorname{MLP}_{\pi}(\eta)$; the private noise, $\bar \Psi = \operatorname{MLP}_{\Psi}(\eta)$ and factor loadings $U = \operatorname{MLP}_{U}(\eta)$. Two additional single-hidden-layer MLPs with 128 hidden units and SiLU activation are used to model perturbations in $\nu$, $\delta_{\nu} = \operatorname{MLP}_{\delta_{\nu}}(\eta, w_{\nu})$; and $\pi$, $\delta_{\pi} = \operatorname{MLP}_{\delta_{\pi}}(\eta, w_{\pi})$. $\bar \Delta_{\max} \in \mathbb{R}^{D}$ is a learnable cap for the location parameter, $\Delta$, with an element for each bar in the calorimeter.

At this stage, the distribution parameters are calculated: $\alpha = \alpha_{\min} + \operatorname{Softplus}(\bar \alpha)$, where $\alpha_{\min} > 0$ is an empirical limit for minimum $\alpha$, $\nu = \operatorname{Softplus}(\bar \nu + \delta_{\nu}) + \nu_{\min}$, where $\nu_{\min}$ is an empirical limit for the minimum $\nu$; the shower rate parameter $\beta = \alpha \nu^{-1}$; $\pi = \operatorname{Sigmoid}(\bar\pi + \delta_{\pi})$; $\Delta = \Delta_{\max} \odot \operatorname{Sigmoid}(\bar \Delta)/E_{\mathrm{sum}}$, where $\odot$ is the Hadamard product, $\Delta_{\max} = \min(\operatorname{Softplus}(\bar \Delta_{\max}), E_{\mathrm{cut}})$ and $E_{\mathrm{sum}}$ is used to convert the cut to fraction units, and $\Psi = \epsilon + \operatorname{Softplus}(\bar \Psi)$.  $\mathcal{M}(\eta)$ is also defined as an elementwise quadratic spline with 8 bins, bounded to the range $\left[ -6, 6\right]$. The factor loadings are capped relative to each bar's total variance, $\frac{|U_{i}|^{2}}{\sigma^{2}_{i}} < \rho_{\max}$, enforced by scaling $U_{i}$ by a factor of $\min\!\left(1,\; \frac{c_0\,\sqrt{\Psi_i}}{\lVert U_i \rVert}\right)$, where $c_0 = \sqrt{\frac{\rho_{\max}}{1 - \rho_{\max}}}$. This choice prevents $s_{i}$ from collapsing to zero, since $s^{2}_{i} = \frac{\Psi_{i}}{\sigma^{2}_{i}} = 1 - \frac{|U_{i}|^{2}}{\sigma^{2}_{i}} \geq 1 - \rho_{\max}$. 

In the next stage, the conditional moments of the copula variable are calculated according to Eq.\ref{eq:cond_moments}, $m_{i} = \frac{U_{i} \cdot w_{c}}{\sigma_{i}}$ and $s_{i} = \frac{\sqrt{\Psi_{i}}}{\sigma_{i}}$. The generative model for the copula variable conditioned on $\eta$ and $w_{c}$ is $z = m + s \odot \varepsilon$, where $ \varepsilon$ is a random draw from a standard normal distribution with dimension $D$, and $z$ is the Gaussian copula score conditioned on $w_{c}$ and $\odot$ is the Hadamard product. Then, $z$ is passed through a conditional inverse spline mapping, where each element $z_{i}$ is mapped by the spline conditioned on shape parameters, $\zeta_{i} = \mathcal{M}_{i}^{-1}(z_{i}; \eta)$. This choice allows the marginals to be warped independently of the copula, acting as a flow from a known marginal to a more complex distribution, which provides sufficient flexibility for the model to describe richer distributions, e.g., the multimodality at the shower-start layers discussed in Sec.~\ref{sec:event_to_event_fluctuations}. Then, the zero-inflated Log-Gamma inverse CDF is applied to the warped score,  $g_{i} = F^{-1}_{\operatorname{ZILG}(\alpha_{i}, \beta_{i}, \Delta_{i}, \pi_{i})}(\Phi(\zeta_{i}))$. An activity mask $\tilde a$ is defined according to the censoring threshold, $\tilde a_i = \mathbb{1}\!\left[g_i > \log\phi_{\mathrm{cut}}\right]$ and $\log\phi_{\mathrm{cut}} = \log E_{\mathrm{cut}} - \log E_{\mathrm{sum}}$. A common re-scaling of all cell energies shifts every $g_{i}$ by the same constant, to which the Softmax is invariant (exact for $\Delta = 0)$. The physical scale therefore enters through $E_{\mathrm{sum}}$ via the censoring threshold and $\Delta$. The event is then normalized using a masked log-Softmax operation, $\log\phi_i = g_i - \log\!\sum_{j:\tilde a_j=1}\exp(g_j)$. Finally, the dark cells where $\tilde a = 0$ are masked with $-\infty$, so that $\exp(\log{\phi}) = 0$. The result is the sampled log energy fractions, $\log{\phi}$. 

The full generative method is described in Alg.~\ref{alg:full_sample}. The energy sum, $E_{\mathrm{sum}}$ and shower conditioning vector $\theta$ are sampled from their respective energy-dependent distributions, $E_{\mathrm{sum}} \sim f_{\mathrm{calo}}(E_{\mathrm{particle}})$ and  $\theta \sim f_{\mathrm{\theta}}(E_{\mathrm{particle}})$, where $E_{\mathrm{particle}}$ is the particle energy. $f_{\mathrm{calo}}$ here is the calorimeter resolution distribution (i.e., in the perfect case, $f_{calo}(E_{\mathrm{particle}}) = \mathcal{N}\!\big(E_{\mathrm{particle}},\, a^{2} E_{\mathrm{particle}}\big)$) It is assumed $(E_{\mathrm{sum}} \perp \theta) \mid E_{\mathrm{particle}}$ in this configuration, based on the argument that the fluctuations in the shower avalanche development are independent of the fluctuations of the number of signal quanta, which is also assumed in Ref.~\cite{grindhammer_parameterized_2000}. The logarithm of the particle energy is appended to $\theta$ to obtain an extended conditioner, $\eta$. Then, $\log{\phi}$ is drawn from the conditioned sampling method shown in Appendix Alg.~\ref{alg:sample}, $\log{\phi} \sim f_{\mathrm{sample}}(\eta, E_{\mathrm{sum}}, E_{\mathrm{cut}})$, where $E_{\mathrm{sum}}$ and $E_{\mathrm{cut}}$ are used only for defining an activity mask, and not in the sampler. Finally, the per-bar energies are obtained as $E = E_{\mathrm{sum}} \cdot \phi$. 

The model can be fitted using QMC marginal log-likelihood, as described in Appendix Alg.~\ref{alg:nll}. The same steps are taken as in Appendix Alg.~\ref{alg:sample} up until after the calculation of the conditional moments. At this point, the inverse of the sampling chain is performed to map input log fractions $\log{\phi}$ to the standardized normal scores, $z_{i} = \mathcal{M}\!\Bigl(\Phi^{-1}\!\bigl(F_{\mathrm{ZILG}(\alpha_i, \beta_i, \Delta_{i}, \pi_{i})}(\log\phi_{i})\bigr)\Bigr)$. Additionally, the energy cut as a log-ratio, $\log{\phi_{\mathrm{cut}}} = \log{E_{\mathrm{cut}}} - \log{E_{\mathrm{sum}}}$, is passed through the same transform to obtain the equivalent cut in $z^{\mathrm{cut}}_{i}$ for a particular event. 

The probability for a particular $x$ is obtained by the chain rule:

\begin{align*}
    f(x_{i} \mid w) &= f(z_{i} \mid w) \cdot \left|\diff{z}{\zeta}\right| \cdot   \left|\diff{\zeta}{x}\right|\\
     &= \frac{1}{s_{i}}\varphi \left(\frac{z_{i} - m_{i}}{s_{i}}\right) \cdot \mathcal{M}^{\prime}(\zeta) \cdot \frac{f_{\operatorname{ZILG}(\alpha_{i}, \beta_{i}, \Delta_{i}, \pi_{i})}(x_{i})}{\varphi(\zeta_{i})}
\end{align*}

Where $w$ represents the particular latent draw, $w \equiv \left( w_{\nu},\, w_{\pi},\, w_{c} \right)$ and $\varphi$ is the standard normal density. Therefore, the log-likelihood is defined as:

\begin{align*}
        \log f(x_i \mid w) &=  \Bigl[ \underbrace{\log f_{\mathrm{ZILG}}(x_i;\, \alpha_i, \beta_i, \Delta_i, \pi_i)}_{\text{marginal}} + \underbrace{\log \mathcal{M}_{i}^{\prime}(\zeta_i)}_{\text{flow}} \\ &+ \underbrace{\log \varphi\!\left(\frac{z_i - m_i}{s_i}\right) - \log s_i}_{\text{copula}} - \underbrace{\log \varphi(\zeta_i)}_{\text{PIT}} \Bigr]
\end{align*}

The censored probability mass of the event can be inferred from the total log probability that the copula score $z_{i}$ fell below the threshold $z^{\mathrm{cut}}_{i}$ for a particular event and latent draw $w$: 

\begin{equation}
    \setlength{\jot}{2pt}
    \log f_{i} =
        \begin{cases}
          \log{f(x_{i} | w)} & a_i = 1, \\[2pt]
          \log \Phi\!\left(\dfrac{z_i^{\rm cut} - m_i}{s_i}\right), & a_i = 0.
        \end{cases}     
\end{equation}

In practice, QMC as described in Sec~\ref{sec:censored_fits} is used to estimate the loss over several draws of $w$. $K$ samples of $w$ are drawn, and the expectation value of the log-likelihood is determined as an average over $K$. The log-likelihood weight, $\omega$: 
\begin{equation}
\log \tilde \omega^{(k)} = \sum_{i=1}^{D} \log f_i^{(k)}, \qquad k = 1,\dots,K
\end{equation}

\begin{align*}
   \mathcal{L}(x) &=  \log{\mathbb{E}(\tilde \omega}) \\ 
   &= \log \sum_{k=1}^{K}\!\bigl(\exp \bigl( \log \tilde \omega^{(k)}\bigr)\bigr) \;-\; \log K. 
\end{align*}
 
The conditional MLPs were trained using an NVIDIA RTX 5060 Laptop (GPU) with the Adam optimizer using a learning rate of \num{1e-3}, with cosine annealing, for \num{2e4} steps, using $K=8$ particles for the QMC and a batch size of 1024 events. The training was performed using a random 80/20 split of the data for training and validation of the model. After training, generated events are sampled using Appendix Alg.~\ref{alg:sample}, but using the same $E_{\mathrm{sum}}$ and $\theta$ values as the data validation dataset instead of sampling them. This decouples the classifier performance of the shower shapes and full showers, such that it can be studied in solation. 

The model's performance was evaluated according to the conventions of Ref.~\cite{krause_calochallenge_2025}, which evaluates a variety of generative models for particle showers. Manifold-based metrics are used to measure the realism and diversity of the generated showers compared to data. Four measurements are introduced: Precision, Recall, Density, and Coverage (PRDC) \cite{sajjadi_assessing_2018, naeem_reliable_2020}.

A manifold is first defined using the nearest neighbors of data $x_{1}\dots x_{n}$:

\begin{equation}
\mathrm{manifold}(x_1, \ldots, x_n) = \bigcup_{i=1}^{n} \operatorname{B}\!\left(x_i, \mathrm{NND}_k(x_i)\right), 
\end{equation}

where $\operatorname{NND}$ is the $k^{\mathrm{th}}$ nearest neighbor distance of one sample to another, $\operatorname{B}(x, r)$ is a sphere around $x$ of radius $r$, and $\bigcup$ is the union operator. In this study, $k=5$.

Precision counts the fraction of generated events that are contained within the union of spheres around the data events. Precision measures the plausibility of the generated events with respect to the data:

\begin{equation}
\mathrm{precision} = \frac{1}{m} \sum_{j=1}^{m} \mathbb{1}_{\,y_j \in \mathrm{manifold}(x_1, \ldots, x_n)},
\end{equation}

Recall measures the complementary case to precision: how many data events are contained within the union of spheres around the generated events:

\begin{equation}
\mathrm{recall} = \frac{1}{n} \sum_{i=1}^{n} \mathbb{1}_{\,x_i \in \mathrm{manifold}(y_1, \ldots, y_m)},
\end{equation}

Instead of the binary membership in the union used for precision, the Density the average number of per-event nearest-neighbor spheres that contain each generated event:

\begin{equation}
\mathrm{density} = \frac{1}{k\,m} \sum_{j=1}^{m} \sum_{i=1}^{n} \mathbb{1}_{\,y_j \in \mathrm{B}(x_i, \mathrm{NND}_k(x_i))}, 
\end{equation}

High-quality generated events will be found where spheres densely overlap, because they are close to many data events. This choice helps to account for outliers, which may spuriously overextend the manifold.

Finally, Coverage measures the proportion of data events with nearest-neighbor spheres containing at least one generated event:

\begin{equation}
\mathrm{coverage} = \frac{1}{n} \sum_{i=1}^{n} \mathbb{1}_{\,\exists\, j \ \mathrm{s.t.}\ y_j \in \mathrm{B}(x_i, \mathrm{NND}_k(x_i))}.  
\end{equation}

Realism is measured by precision and density, while diversity is measured by recall and coverage. 

The PRDC metrics are measured using the \texttt{prdc} Python library \cite{clovaai_prdc_2026}. The full vector of log energy fractions $\log{\phi} \in \mathbb{R}^{D}$ is computed and then used to calculate PRDC metrics on a validation sample of data events compared to the corresponding generated events, using two random disjoint halves of each set of data. If all statistics are close to the baseline, then the model can be determined to be both realistic and diverse. Dark cells are given the per-event fraction threshold, $\log{\phi_{\mathrm{cut}}}$. Errors are measured by repeating the split ten times and using the mean and standard deviation over the repeated splittings and measurements using different random seeds.     

A second metric is a classifier output with the same method as defined for the classifier method in Sec.~\ref{sec:shower_shape_modelling}, which follows the conventions of Ref.~\cite{krause_calochallenge_2025}. A dense neural network (DNN) is defined that takes the energy fraction $E/E_{\mathrm{particle}} \in \mathbb{R}^{D}$, where $E$ is the vector of hit energies and $\log_{10}{E_{\mathrm{particle}}}$ is the input. The model was implemented in \texttt{PyTorch} \cite{paszke_pytorch_2019}. The DNN is trained to classify data from generated events. The DNN consists of two hidden layers with 2048 units and leaky ReLU activation, and a final layer that outputs a sigmoid-activated probability, which describes the probability that an event was generated. The model is trained for 50 epochs using the Adam optimizer, a learning rate of \num{2e-4}, and a batch size of 1000, using the main validation sample.

A third metric is again a classifier output using the global aggregate statistics of the shower event using the same DNN model as for the hit-level classifier and the method defined in Sec.~\ref{sec:shower_shape_modelling}. The total shower energy, $E_{\mathrm{sum}}$, the number of hits above threshold, $N_{\mathrm{hit}}$, the the center of gravity, or the energy-weighted mean average position of the shower ($\mathrm{CoG}_{X}$, $\mathrm{CoG}_{Y}$, $\mathrm{CoG}_{Z}$) the average distance to the center-of-gravity in $x$ and $y$, ($\langle|X-\mathrm{CoG}_{X}|\rangle$, $\langle|Y-\mathrm{CoG}_{Y}|\rangle$) and the fraction of energy deposited in each longitudinal layer of the calorimeter ($f_{\text{layer},\,l-1}$). A table showing the equations for these variables is available in Appendix.~\ref{tab:bdt_params}. The classifier is otherwise trained and evaluated in the same way as for the full event vector.

\begin{figure}
    \centering
    \includegraphics[width=0.8\linewidth]{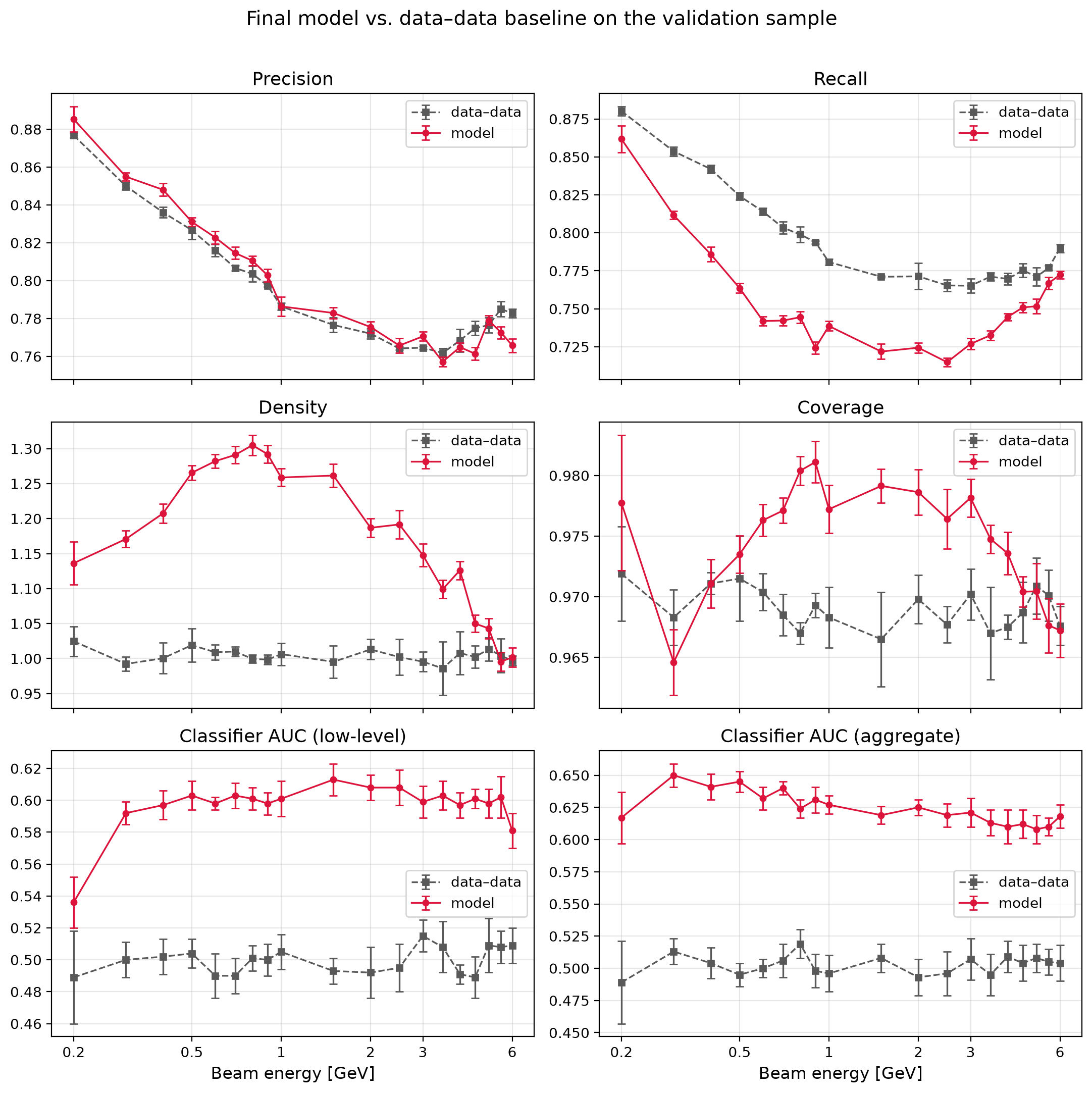}
    \caption{PRDC metrics (top two rows) and classifier metrics (bottom row) for the copula shower model. The red line indicates the generated values compared to the data, and the grey line indicates the data compared to the data. Error bars indicate the fluctuation of the result depending on ten different initializations. }
    \label{fig:fig_metrics_vs_energy}
\end{figure}

The results are shown in Fig.~\ref{fig:fig_metrics_vs_energy}. The precision is at or above the baseline given by the data-data baseline, indicating that generated events are around as plausible as data events in the manifold. The recall is around \qtyrange{3}{8}{\percent} lower than the baseline, while the density is greater than one and the coverage is at the baseline. This result implies a small over-concentration of events in the high-density regions of the manifold. Generated events are therefore of high quality at the cost of mildly under-representing rare events. The classifier for the event fractions obtains an AUC averaged over all beam energies of \num{0.597(0.016)}, while the aggregate feature classifier obtains a similarly averaged AUC of \num{0.624(0.012)}. For reference, a comparison is made to the top-5 submissions to CaloChallenge for electrons, using the \texttt{Geant4} \texttt{Par04} calorimeter model \cite{par04}. This is a cylindrical Si-W calorimeter design with 45 layers, 9 radial bins, and 16 angular bins (6480 voxels), covering a depth of 36 $X_{0}$. The top five entries for electron showers (CaloDREAM \cite{favaro_calodream_2025}, CaloDiffusion \cite{amram_denoising_2023}, CaloScore \cite{mikuni_score_based_2022, mikuni_caloscore_2024}, SuperCalo \cite{pang_calorimeter_2024}, Convolutional L2L-Flows \cite{buss_convolutional_2024}, results taken from Table C15, Ref.~\cite{krause_calochallenge_2025}), obtained AUC scores for the full event classifier  AUC of \num{0.531(0.003)} \num{0.577(0.004)}, \num{0.595(0.003)}, \num{0.694(0.006)} and \num{0.708(0.004)}; and for the aggregate classifier AUC obtained \num{0.521(0.002)}, \num{0.591(0.009)}, \num{0.666(0.002)}, \num{0.757(0.004)} and \num{0.737(0.002)}, respectively. The presented statistical model achieves similar performance, which suggests it is a viable alternative to these generative methods. As a caveat, the \texttt{Par04} simulated calorimeter is considerably more granular than the one under test in this study and is not homogeneous, the results are not directly comparable. 

\begin{figure}
    \centering
    \includegraphics[width=\linewidth]{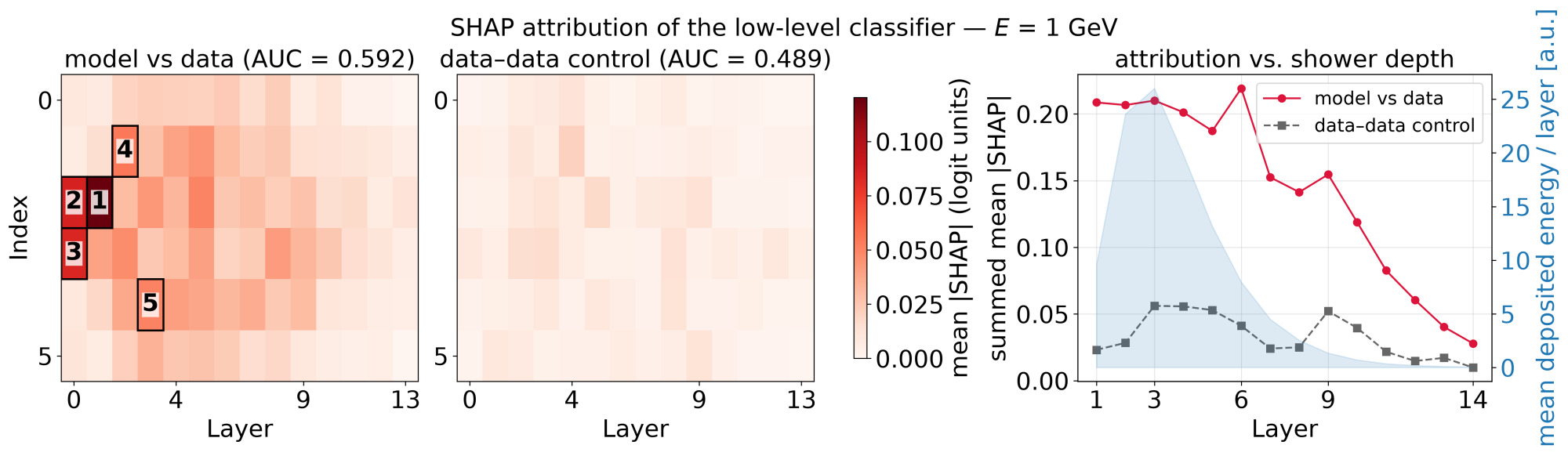}
    \caption{Left and center: mean absolute SHAP values as a function of the bar index and layer for the model vs. data and data vs. data, respectively. The x and y axes indicate the bar index and layer index, respectively. White-to-red indicates the increasing relevance of the bar to any decision made by the classifier. The top five contributing cells are highlighted for the generated SHAP values  Right: the mean SHAP value per layer vs. the layer index. Red solid and black dashed lines indicate the model vs. data and data vs. data, respectively. The average energy deposited per layer is shown by the light blue shaded area.}
    \label{fig:fig_shap_classifier_1p0GeV}
\end{figure}

In order to understand how the classifier selects between the data and the generated event, SHapley Additive exPlanations (SHAP) \cite{lundberg_local_2020} is used. This is a model-agnostic technique for interpreting machine learning models, rooted in game-theoretic Shapley values \cite{shapley_notes_1951}. This method quantifies the significance of each input variable to the model's decision-making process by measuring the average marginal effect of including that variable across all subsets of the other features. This can be achieved in practice by replacing the variable with random values from a background distribution, which approximates 'removing' that feature from the model and is similar in spirit to the 'leave-one-out' method of jackknife re-sampling. and studying the deviation of the output of the model compared to its average output. However, specialized algorithms like the \texttt{GradientExplainer} in the eponymous \texttt{SHAP} software library \cite{noauthor_shapshap_2026} can compute exact approximate SHAP values for a particular test sample. The mean absolute SHAP value for each bar of the validation sample indicates the total importance of a particular bar to the classifier's decision, which helps to identify how the generative model can be improved. Fig.~\ref{fig:fig_shap_classifier_1p0GeV} shows the mean SHAP values for the model vs. data and data vs. data cases. From this plot, the key driver for the classifier is the energy deposited in the first two layers and in the center of the calorimeter. This result is expected based on Fig.~\ref{fig:fig_cond_gamma_cdf_fits_1GeV} and Fig.~\ref{fig:fig_cond_fisherz_pairs_cdf_fits_1GeV}, which show that the marginal agreement of the proposed model fails considerably at the initiation of the shower. While a spline flow was used to warp the marginals at the normal stage, the model may improve by modifying the structure of the model in future iterations.

\begin{figure}
    \centering
    \includegraphics[width=0.8\linewidth]{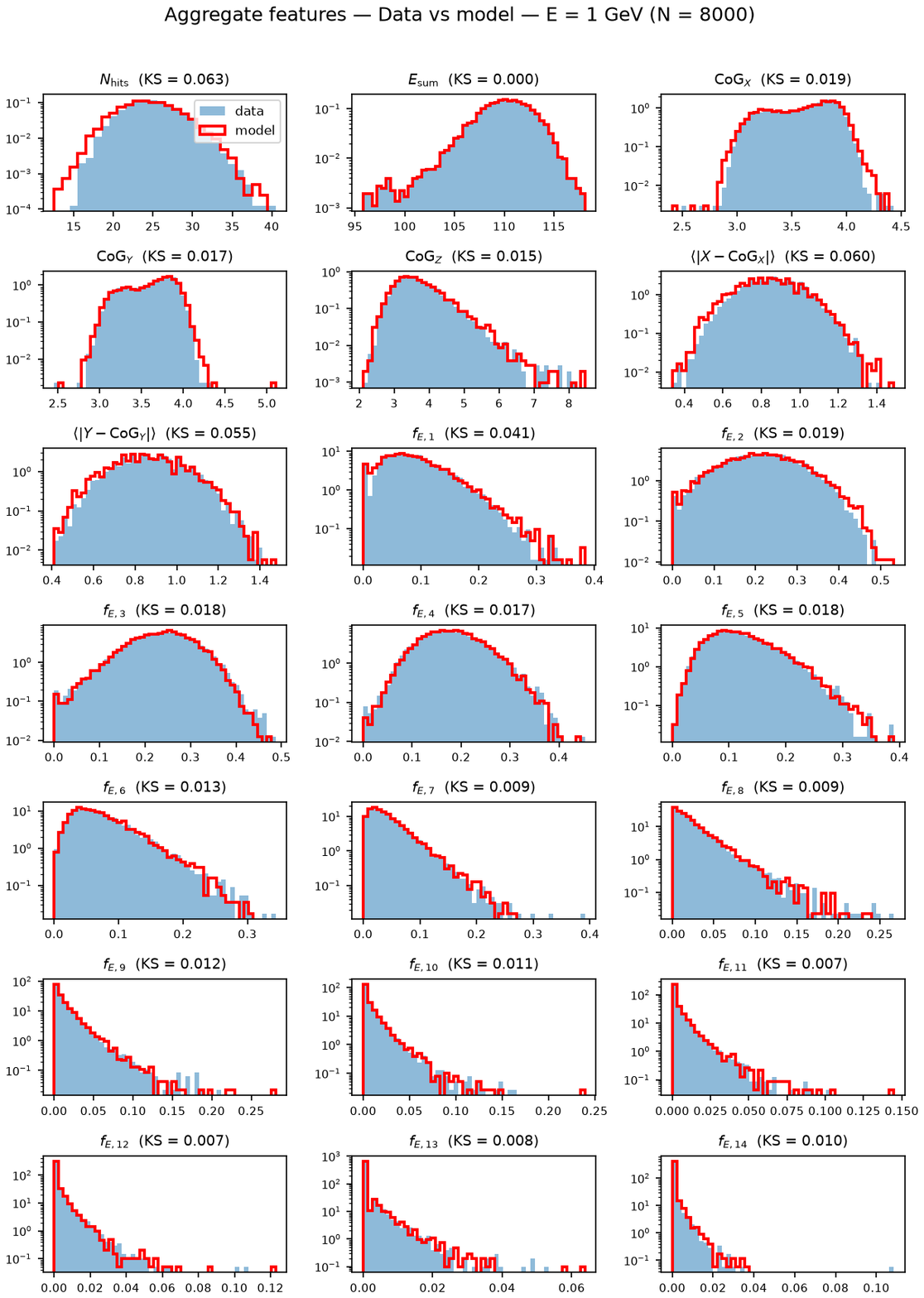}
    \caption{Comparison of the distribution of aggregate variables for the \qty{1}{\giga \electronvolt} sample. The aggregate variables of Table~\ref{tab:bdt_params} are shown. The blue indicates the data validation sample and the red indicates the generated events. $E_{\mathrm{sum}}$ is shown for reference only, since Alg.~\ref{alg:full_sample} utilizes the same $E_{\mathrm{sum}}$ and $\theta$ as the data event at generation time. The $E_{\mathrm{sum}}$ distribution is perfect as it is already resampled from the true distribution, and is shown for reference.}
    \label{fig:fig_features_final_model_E1}
\end{figure}

Fig.~\ref{fig:fig_features_final_model_E1} shows the distributions of each aggregate parameter studied for \qty{1}{\giga \electronvolt} 8000 data and generated events, and shows that all distributions are in good agreement with the data. It is noted that $N_{\mathrm{hits}}$ is slightly overdispersed. The \qty{0.2}{\giga \electronvolt} and \qty{5}{\giga \electronvolt} samples are shown in Appendix Fig. \ref{fig:fig_features_final_model_E0.2} and Appendix Fig. ~\ref{fig:fig_features_final_model_E5}, respectively, which show that the overdispersion affecting the number of active cells is most prominent at low energy and decreases to negligible influence at higher energy.

Example event displays showers are shown for \qty{0.2}{\giga \electronvolt}, \qty{1}{\giga \electronvolt}, and \qty{5}{\giga \electronvolt}, that illustrate the performance of the model visually.

\begin{figure}
    \centering
    \includegraphics[width=0.9\linewidth]{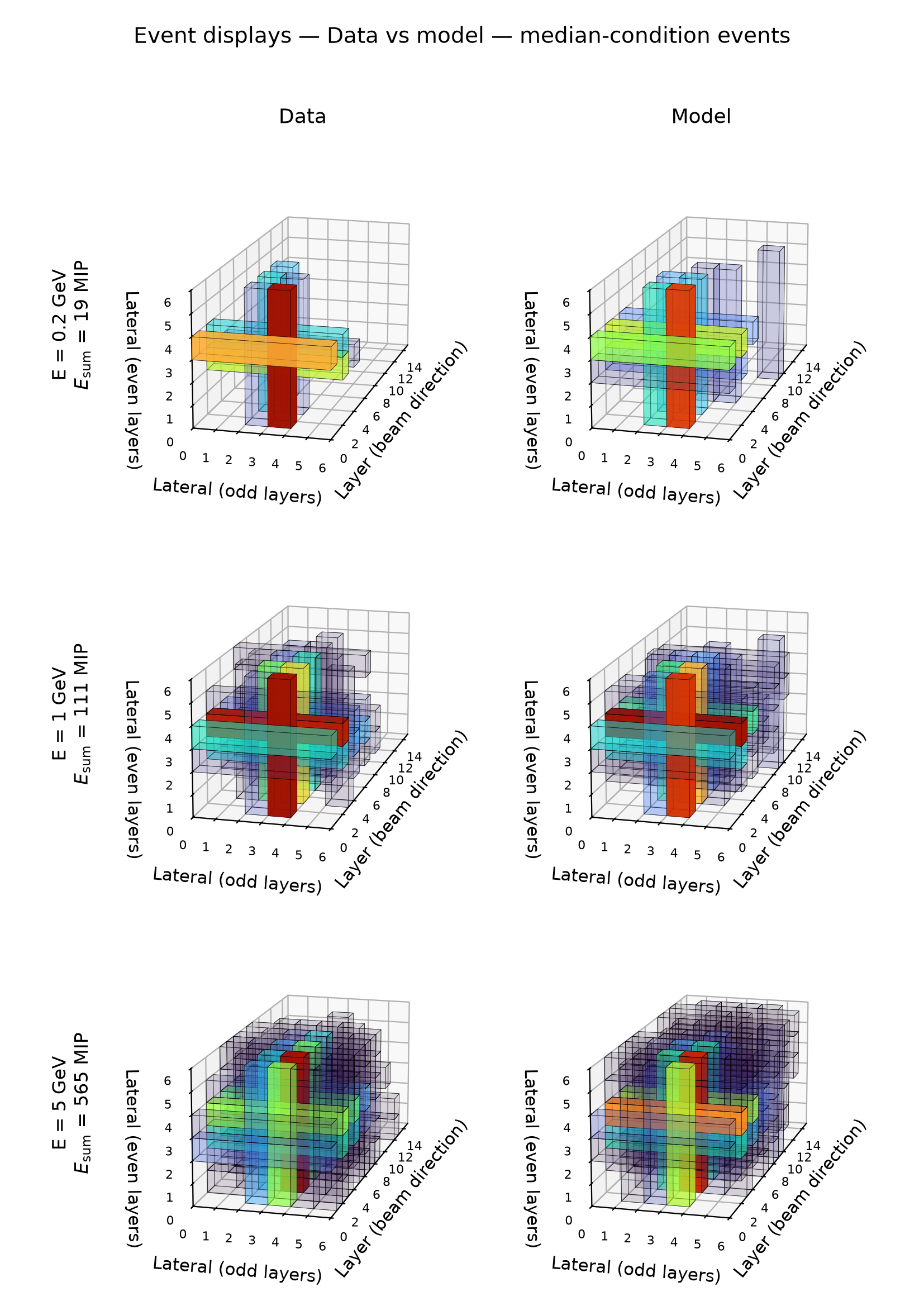}
    \caption{Example event displays showing data and generated examples on the left and right, respectively, and \qty{0.2}{\giga \electronvolt}, \qty{1}{\giga \electronvolt}, and \qty{5}{\giga \electronvolt} showers on the top, middle, and bottom rows. Transparent blue to solid red indicates energy relative to the maximum energy deposit. The energy sum is shown on the left of each plot in MIP units.}
    \label{fig:fig_event_displays_A4}
\end{figure}

\begin{figure}
    \centering
    \includegraphics[width=0.6\linewidth]{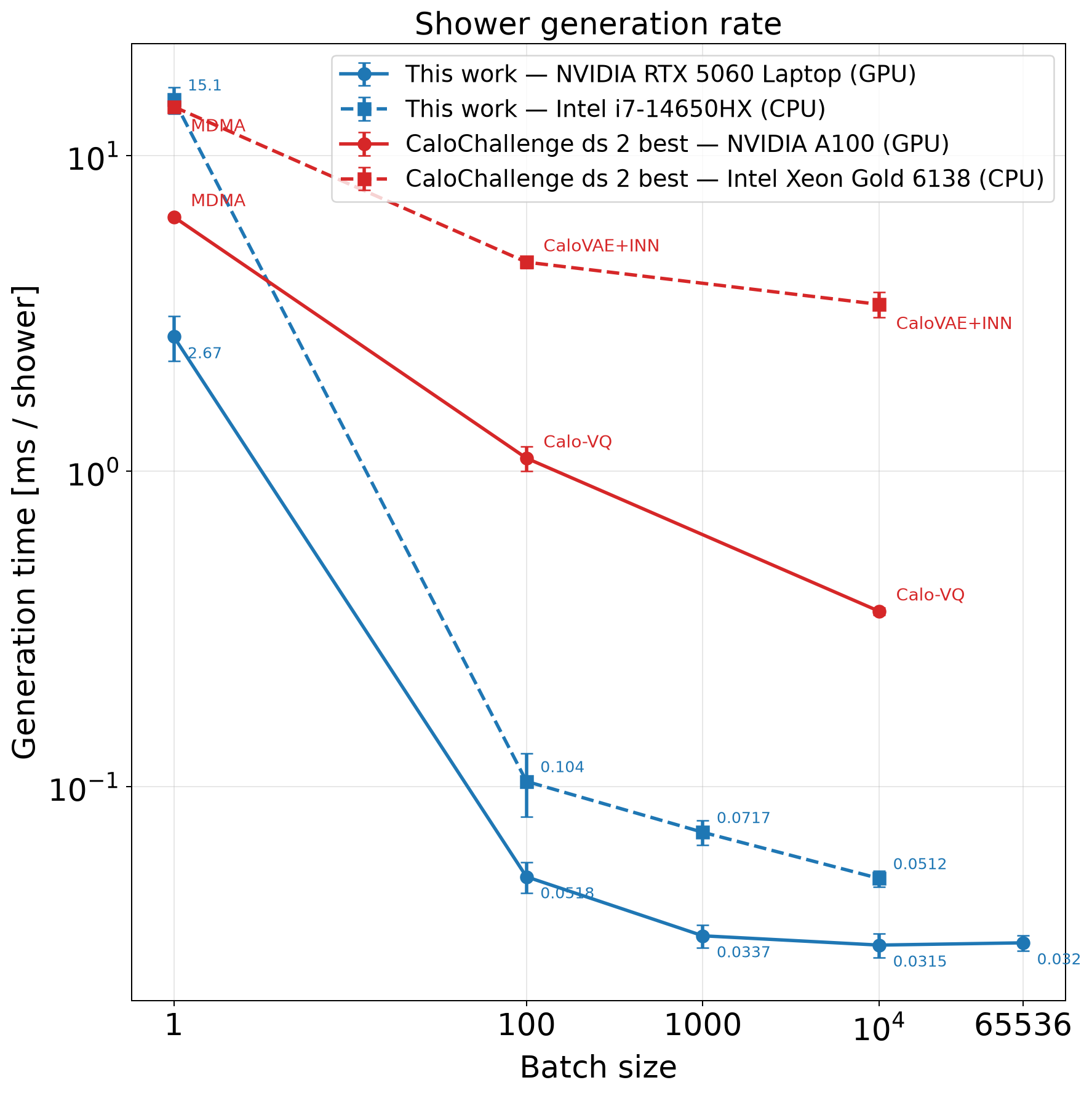}
    \caption{Shower generation time per event as a function of batch size. The blue lines indicate the shower copula model and the red lines the fastest models applied to the electron dataset 2 of CaloChallenge at each batch size. The dashed lines indicate CPU inference, and the solid lines indicate GPU inference.  Points for this work are the mean of five repeated measurements, with the standard deviation shown.}
    \label{fig:fig_speed_comparison}
\end{figure}

The event output rate was measured for batch sizes from 1-\num{1e4} events per batch using 
an Intel i7-14650HX CPU and an NVIDIA RTX 5060 Laptop GPU. This was compared to the fastest shower models of CaloChallenge measured on an NVIDIA A100 GPU and Intel Xeon Gold 6138 CPU (Tables~C21 and~C22 of Ref.~\cite{krause_calochallenge_2025}). The results are shown in Fig.~\ref{fig:fig_speed_comparison}

The results show that the performance for multi-batch generation saturates at around \qty{0.032(0.003)}{\milli \second \per \mathrm{shower}} on GPU and \qty{0.051(0.003)}{\milli \second \per \mathrm{shower}} on CPU. For batches of 1, 100, and 10000, the copula model is $2.4\times$, $21 \times$, and $11 \times$ faster than the best-performing model on GPU, and $0.95 \times$, $44\times$, and $66\times$ faster on CPU. However, this result comes with the caveat that the granularity between the calorimeters is different.

In conclusion, the idea that the energy deposits from electron showers in the HGCCAL calorimeter crystal bars are Gamma distributed conditional on the shower shapes and particle energy is strongly indicated by the agreement of the statistical model developed in this section.

\section{Conclusion}

Electron data taken with a highly granular crystal ECAL during a 2025 testbeam campaign at the PS-T9 beamline at CERN were fitted with standard shower shape models. The models were found to be good descriptions, on average, of the shower shape, with the median fit having an average $W1$ value over all three spatial dimensions of around 0.1 bar widths (\qty{2}{\milli \meter}), which was well within the spatial resolution of the calorimeter and indicates good agreement with the fitted models.


The distributions of the shower maximum and longitudinal and lateral shape parameters were found to be approximately log-normal. From the mean and variance of these distributions, the radiation length, critical energy, and  radius were measured: $X_{0} = \qty{11.314(0.011)}{\milli \meter}$ and $E_{c} = \qty{10.655(0.049)}{\mega \electronvolt}$, and $R_{M} = \qty{25.992(0.007)}{\milli \meter}$. The radiation length and critical energy agreed with the PDG values within \qty{1.1}{\percent} and \qty{1.5}{\percent}, while the Molière radius was around \qty{3.4}{\milli \meter} larger compared to PDG, which is attributed to the crystal bar structure. As a caveat, the particular definition used in this study may not match the definition in PDG. The latent Gaussian correlations between variables were measured via copula transform and were observed to be strongest between the shower maximum and the shower shape parameter, with $\rho = \num{0.65} - {0.07} \log {E_{\mathrm{particle}}}$. The standard deviation of the distribution of the logarithm of the shower shape variables was found to decrease linearly with the mean in all cases. 

The remaining fluctuations in the energy were found to follow a Gamma distribution conditioned on the shower shape parameters. The model agreed using two tests: the first directly to the distributions of energy in a single bar, fitted by a Gamma distribution, and then to the logarithm of the fractions of energy deposited between any two combinations of bars, fitted with a Fisher's-$z$ distribution, and was found to have a median reduced $\chi^{2}$ of around 2 and 1.5 in each case. 

Generative methods were provided for both shower shapes using an existing fast simulation method from Ref.~\cite{grindhammer_parameterized_2000}, and copula transforms were explored. It was found that, while a log-normal distribution could describe the bulk of the fitted distribution, superior agreement could be obtained using Gaussian copula sampling. A full-event sampling method was then developed using a copula transform by mapping a correlated MVN to Log-Gamma distributions to produce correlated log-fraction compositions by exploiting the observation that calorimeter log energy fractions were Fisher's-$z$ distributed. This model was found to describe electron shower data at a similar level to existing shower generation models, with the advantage of being a minimum of $\times 11$ faster at generation on GPU than the fastest model of Ref.~\cite{krause_calochallenge_2025}. This result should be taken with the caveat that the study in Ref.~\cite{krause_calochallenge_2025} used a simulated calorimeter of higher granularity than the one in this study. 

Future studies could attempt to extend the models developed in this paper to sampling calorimeters and hadron or photon showers, which would then provide a complete picture of the statistical models that can be used to describe particle showers.

\section*{Acknowledgements}
We gratefully acknowledge the CERN management for its support and hospitality and the accelerator staff for reliable and efficient operation of the test beam. 
This research was supported by the National Key Research and Development Program of China (Grant Number 2023YFA1606300 and by  MOST National Key Programme for S\&T Research and Development (Grant Number: 2023YFA1606303) .

\clearpage
 
\clearpage

\begin{appendices}

\section{Statistical Methods}\label{sec:statistical_methods}

A variety of uncommon probability density functions and statistical concepts are used in this study. They are discussed and introduced in this section.

\subsection{Empirical Cumulative Distribution Functions}\label{sec:ecdf}
The cumulative distribution function (CDF) describes the probability that a sample falls below a certain threshold $t$,  $P(X < t) = F(x) = \int^{x}_{-\infty} f(t).\mathrm{d}t$ 

The empirical cumulative distribution function (ECDF) is the distribution function that describes an empirical set of data, and can be obtained trivially by counting the fraction of samples falling below the threshold: 

\begin{equation}
\widehat {F}_{n}(t)={\frac {{\text{cumulative count of elements in the sample}}\leq t}{n}}={\frac {1}{n}}\sum _{i=1}^{n}\mathbf {1} _{X_{i}\leq t}
\end{equation}

\subsection{Gaussian Copulas}\label{sec:gaussian_copula}

In the following, a joint distribution refers to a distribution of two or more possibly correlated parameters, i.e., $f(x_{1},x_{2})$. A marginal distribution refers to the distribution of the single variable projected in one dimension,  e.g., $f(x_{1}) = \int^{\infty}_{-\infty} f(x_{1},x_{2}).\mathrm{d}x_{2}$. The CDF of the marginal distribution is $F_{1}(x) = \int^{x}_{-\infty} f(t)\,\mathrm{d}t$.

By Sklar's theorem, there exists a 'copula', $C$, which is a mathematical function that couples several individual and possibly correlated marginal distributions, $F_{i}$, relating them by a single joint distribution, $H$ \cite{sklar_fonctions_1959}:
\begin{equation}
    H(x_{1},x_{2}\dots)= C(F_{1}(x_{1}), F_{2}(x_{2})\dots),
\end{equation}

A Gaussian copula assumes that after mapping variables $x_{i}$ individually to a standard normal distribution, $z_{i}$, the joint distribution of the $z_{i}$ can be described by a multivariate normal distribution (MVN) with correlation matrix $R$. All correlation is then carried by $R$, while the distribution shapes are described by the per-variable mapping. 

Two stages are used for the mapping. Firstly, the 'Probability Integral Transform' (PIT), $u_{i} = F_{i}(x_{i})$, is used. This results in a variable, $u_{i}$, which is a percentile value of the original distribution $u_{i}\in \left[0, 1 \right)$, and is uniformly distributed. Then, the data can be mapped to the normal distribution using the inverse CDF or probit function, $z_{i} = \Phi^{-1}(u_{i})$, where $\Phi^{-1}$ is the normal probit function. Written together, $z_{i} = \Phi^{-1}(F_{i}(x_{i}))$.



This method is used in two ways in this study. Firstly, DSCB distributions do not admit tractable multidimensional fits. Due to heavy tails, the correlation can be conflated with the fluctuations within its own marginal distribution. A Gaussian copula can be used to reduce this effect, as it maps the distribution to percentiles before calculating the correlation coefficient. This step decouples the shapes of specific marginal distributions (e.g., tails) from their dependencies. The two-stage method of estimating the marginals and the dependence separately is a standard method called Inference Functions for Margins (IFM) \cite{joe_estimation_1996}. Secondly, marginal distributions may be correlated, which is challenging to model without complex underlying multivariate distributions. Fortunately, for the reasons stated in the previous point, correlated marginal distributions can be obtained by sampling from an MVN with a particular correlation matrix and applying the inverse mapping, $\hat{x}_{i} = F^{-1}_{i}(\Phi(z_{i}))$. This modeling technique has considerable advantages over 'black box' methods, as a particular marginal can be enforced while keeping correlations between variables.

\subsection{Factor Analysis}\label{sec:factor_analysis}

Factor analysis is a method for describing the variability among correlated variables using fewer dimensions than the full covariance matrix \cite{bartholomew_latent_2011}. This has the advantage that it enables mapping from a smaller, uncorrelated vector space, called a 'latent space', which carries the underlying structure of the data.

A covariance matrix of this type arises from the generative model:

\begin{subequations}

\begin{equation}
      y = Uw + \eta,  
\end{equation}
    
\begin{equation}
     w \sim \mathcal{N}(0, I_{m}),
    \qquad \eta \sim \mathcal{N}(0, \Psi),
    \label{eq:factor_model}
\end{equation}
\end{subequations}

for which, since $w$ and $\eta$ are independent,
\begin{equation}
    \Sigma \;=\; \mathrm{Cov}(y) \;=\; UU^{T} + \Psi.
    \label{eq:factor_analysis}
\end{equation}

Here, $n$ is the dimension of the variables and $m$ is a smaller 'latent dimension'. $\Sigma$ is a covariance matrix. $U$ is an $n\times m$ matrix called a 'factor loading', whose rows $U_i$ give the latent weights for observed variable $i$. Latent weights refer to the 'hidden structure' of the correlations.  $w$ is a 'latent vector' that describes the underlying fluctuations in the event. $\eta$ is a 'private noise' variable drawn from $\Psi$, which is a diagonal matrix of noise variances, $\Psi = \text{diag}(\psi_1, \psi_2, \dots, \psi_n)$ and $\psi_{i}>0$. 

The term $U U^{T}$ is symmetric and positive semi-definite. Since $\Psi$ is always positive definite, $\Sigma$ is forced to be symmetric and positive definite, which is a requirement for a covariance matrix.

$y_{i}$ is defined as the latent Gaussian variable at index $i$ for a given draw of the latent vector $w$: 

\begin{equation}
    y_i = U_i \cdot w + \eta_i
\end{equation}

The marginal per-row standard deviation of $y_{i}$ is defined as:

 \begin{equation}
    \sigma_{i} = \sqrt{|U_{i}|^{2} + \psi_i}
\end{equation}

The normalized conditional moments of the standardised variable, $z_{i} = y_{i}/\sigma_{i}$ are defined:

\begin{equation}
    m_i = \frac{U_i \cdot w}{\sigma_i},
    \qquad
    s_i = \frac{\sqrt{\psi_i}}{\sigma_i}
    \label{eq:cond_moments}
\end{equation}

Conditional on a draw of $w$ and defining $S = \operatorname{diag}{\left[\sigma_{1}, \sigma_{2} \dots \sigma_{n} \right]}$, the standardized score of each variable is normal with moments defined as:

\begin{equation}
z = S^{-1} y,  \qquad  z_{i} \,|\, w \sim \mathcal{N}(m_i, s_i^{2})
\label{eq:cond_normal}
\end{equation}

A correlation matrix, $R$, can be defined: 
\begin{equation}
    R = S^{-1} \Sigma S^{-1}
\end{equation}

$R$ is a fixed matrix. Any dependence on a particular draw of $w$ enters only through the conditional moments of Eq.~\ref{eq:cond_moments}.

Marginally over $w$, each $z_{i}$ is a standard normal variable. These are the copula variables $z_{i} = \Phi^{-1}(F_{i}(x_{i}))$.

Any two correlated normal variables generated in this way are correlated only through their shared latent weights: $\operatorname{Cov}(y_{i},y_{j})=U_{i} \cdot U_{j}$ for $i \neq j$. This follows from the fact that $\operatorname{Cov}(w) = I_{m}$ and $\operatorname{Cov}(\eta) = \Psi$, the latter of which is diagonal. $U_{i}$ and $U_{j}$ are fixed parameters and do not vary across draws of $w$. Therefore, given a draw of $w$, the shared component $U_{i}\cdot w$ is constant. The only other source of variation is the private noise $\eta_{i}$,  which is uncorrelated with the private noises of the other variables. Therefore, given a drawn latent vector $w$, the correlations between any $y_{i}$  and $y_{j}$  where $i \neq j$ are zero. Therefore, each $z_{i}$ given $w$ is described by an independent one-dimensional normal distribution with the moments of Eq.~\ref{eq:cond_normal}. This is relevant because it allows the conditional CDF of each cell to be evaluated in closed form, which is otherwise intractable due to the correlations. This fact is used to determine the probability of falling below a certain threshold, which is relevant for the discussion of Sec.~\ref{sec:censored_fits}.

\subsection{Censored Fits}\label{sec:censored_fits}

Calorimeter shower data is typically censored. Censoring in this context refers to the energy threshold required to cut away noise. This is referred to as 'left-censored,' referring to the low (left-tailed) part of the energy distribution. This means that some probability density cannot be described by data below the threshold. 

Since the density cannot be estimated from data below $t$, a censored cell contributes not a density but a probability. The log probability can be estimated by using a piecewise definition that utilizes the density above the threshold and the probability mass that sits below it: 

\begin{equation}
\log{f(x; t)} = 
\begin{cases} 
      \log{f(x)}  & x \ge t \\
       \log{F(t)} & x < t 
\end{cases}
\label{eq:censored_log_likelihood}
\end{equation}

where $f(x)$ is the probability density of $x$, $t$ is a threshold, and $F(t)$ is the CDF evaluated at $t$. 
This is called a 'Tobit model' \cite{tobin_estimation_1958, amemiya_tobit_1984}

 The marginal distributions of the energy (or its fraction) deposited in individual active volumes of a highly granular calorimeter are known or can be assumed as an ansatz. A shower is an avalanche process; therefore, the energy deposited in different volumes of active material during a shower is correlated. As per the discussion of Sec.~\ref{sec:gaussian_copula}, a copula can be used to describe the correlations between variables while retaining the marginals.

In this paper, a cell below the threshold is defined as 'dark', and above the threshold is defined as 'active'. The correlation structure complicates the probability mass that a particular cell is 'dark ', as it depends on whether other cells were 'dark' too. The log probability mass term of Eq.~\ref{eq:censored_log_likelihood} therefore requires integration of the joint density over a multidimensional rectangle (orthant), for which no closed form exists. 

The factor analysis formulation provides a convenient sidestep for this issue. In Sec.~\ref{sec:factor_analysis}, defining $z_{i}$ as the standardized latent Gaussian variable corresponding to either energy or its fraction in an event, each $z_{i}$ given $w$ is conditionally independent. This allows the integration to be factorized per $w$ draw, and each censored energy deposit/fraction contributes the one-dimensional tail $\Phi( (z^{t}_{i} - m_{i})/s_{i})$, where $z^{t}_{i}$ is the threshold of measured energy mapped to the Gaussian latent space via $z^{t}_{i} = \Phi^{-1}( F_{i}(t_{i}) )$. The correlation is restored by marginalizing the product of the conditional distributions over $w$, $f(z_{1}, \dots, z_{n}) = \int^{\infty}_{-\infty} \prod_{i} f(z_{i} | w) \cdot f(w) .\mathrm{d}w$.

In practice, integration over $w \sim N(0, I_{m})$ is achieved via quasi-Monte Carlo (QMC) integration. Numerical integration on grids or similar regular sampling methods, such as Gaussian–Hermite quadrature, fail in high dimensions because the number of sampling points grows exponentially. Quasi-random sample points from a Sobol sequence  of length K generate quasi-random, approximately uniform points on the unit hypercube \cite{sobol_distribution_1967}. Gaussian samples can be generated using the PIT transform. This choice ensures the latent space is sampled evenly, so no region of the latent distribution is systematically under- or over-represented for a given number of samples, K.

\subsection{Similarity Measures}\label{sec:similarity_measures}

Several similarity measures beyond the standard reduced $\chi^{2}$ are used in this study. 

The Kolmogorov-Smirnov (KS) statistic measures the maximum fractional disagreement between two empirical cumulative distribution functions (CDFs):

\begin{equation}
    D_n = \sup_{x} \left|  F(x) - F_n(x) \right|
\end{equation}

where $\sup$ represents the supremum, or the largest absolute distance. This represents a 'worst case' diagnostic for the agreement between two distributions.

When a single CDF is not available, such as when the CDF may be conditional, a similar statistic is the PIT-KS statistic. This quantity measures the calibration quality of a predictive model. It is defined as the maximum absolute fractional disagreement between the theoretical CDF provided by the model and the ECDF of the data in question:

\begin{equation}
    D^{\mathrm{PIT}}_{n} = \sup_{u \in [0,1]} \left| \hat{F}_{n}(u) - u \right|
\end{equation}

where $u_{i} = F(x_{i}; \theta_{i})$ and $\theta_{i}$ are model parameters corresponding to datum $i$. The fundamental idea is that overall, the CDF should produce a uniform percentile distribution if it correctly describes the modeled distribution.

The Wasserstein-1 statistic, or 'Earth-Mover's Distance ', describes the average 'distance' required to transport the probability mass from one distribution to the other. It is defined as the integral over the absolute distance in probability between two CDFs:

\begin{equation}
    W_{1} = \int^{\infty}_{-\infty} \left| F(x) - G(x) \right| 
\end{equation}

where $F$ and $G$ are two CDFs.

\subsection{Non-Standard Probability Density Functions}\label{sec:non_standard_distributions}

The double-sided Crystal Ball function (DSCB) is a piecewise distribution described by Eq.~\ref{eq:DSCB}:

\begin{subequations}
    \begin{equation}
    \operatorname{DSCB}(x;\mu,\sigma,\alpha_{L},n_{L},\alpha_{R},n_{R})=\begin{cases}A_{L}\cdot(B_{L}-\frac{x-\mu}{\sigma})^{-n_{L}},&  \frac{x-\mu}{\sigma}<-\alpha_{L}\\\exp(-\frac{(x-\mu)^{2}}{2\sigma^{2}}),&-\alpha_{L}\leqslant  \frac{x-\mu}{\sigma}\leqslant\alpha_{R}\\A_{R}\cdot(B_{R}-\frac{x-\mu}{\sigma})^{-n_{R}},& \frac{x-\mu}{\sigma}>\alpha_{R}\end{cases}.
    \label{eq:DSCB}
    \end{equation}

    \begin{equation}
    A_{L / R}=\left( \frac{n_{L / R}} {\left| \alpha_{L / R} \right|} \right)_{L / R}^{n_{L/ R}} \cdot\operatorname{e x p} \left(-\frac{\left| \alpha_{L / R} \right|^{2}} {2} \right)
    \end{equation}

    \begin{equation}
        B_{L / R}=\frac{n_{L / R}} {\left| \alpha_{L / R} \right|}-\left| \alpha_{L / R} \right|
    \end{equation}
\end{subequations}

where $\mu$ and $\sigma$ represent a normally-distributed core region bounded by $\alpha_{L/R}$ and with power law tails defined by $n_{L/R}$. This distribution is used to fit unknown distributions with a core fraction that is normally distributed with heavy or asymmetrical tails. It is used in this study to fit heavy-tailed distributions with a predominantly Gaussian core without requiring selection cuts, and to estimate the valid normal regions of those distributions. 

Fisher's $z$ distribution describes the distribution of the difference between two Log-Gamma distributed variables \cite{fisher_distribution_1924, johnson_continuous_1995, weisstein_wolfram_nodate}:

\begin{equation}
    \operatorname{FisherZ}\left(z; d_{1}, d_{2}\right)=\frac{2d_{1}^{d_{1}/2}d_{2}^{d_{2}/2}}{\mathrm{B}\left(\frac{d_{1}}{2},\frac{d_{2}}{2}\right)}\frac{e^{d_{1}z}}{\left(d_{1}e^{2z}+d_{2}\right)^{\left(d_{1}+d_{2}\right)/2}}
    \label{eq:fisher_z}
\end{equation}

where $d_{1/2}$ are shape variables and $\mathrm{B}$ is the beta function,  $B(d_{1},d_{2})=\int_{0}^{1}t^{d_{1}-1}(1-t)^{d_{2}-1}dt$. Variables can also be written using Eq.~\ref{eq:fisher_z_variate}

\begin{equation}
    z = \Delta + \bar{s}\log{f}
    \label{eq:fisher_z_variate}
\end{equation}

where $\Delta$ and $\bar{s}$ are position and scale parameters and  $f$ is a random variable drawn from the $F$-distribution, 

\begin{equation}
    F(x; d_1, d_2) = \frac{\left(\dfrac{d_1}{d_2}\right)^{d_1/2} x^{\,d_1/2 - 1}\left(1+\dfrac{d_1}{d_2}x\right)^{-(d_1+d_2)/2}}{B\!\left(\dfrac{d_1}{2},\dfrac{d_2}{2}\right)}, \qquad x > 0
    \label{eq:f_distribution_pdf}
\end{equation}

Where $d_{1/2}$ are shape parameters and $B(d_1, d_2) = \int_0^1 t^{d_1-1}(1-t)^{d_2-1}\,dt$

The classical Fisher's-$z$ defines an scaling of factor $\bar{s} = \frac{1}{2}$, $z = \tfrac{1}{2}\log F$.  In this study, it is redefined on $\log F$ so that the independent-Gamma limit has $\bar{s} = 1$ and $\Delta = \log(d_1/d_2) = \log(\beta_1\alpha_1/\beta_2\alpha_2)$. This choice changes only the parameterization, not the family of distributions.

The scaling $\bar{s}$ is related to the covariance. Fisher's-$z$ describes the difference between two variates. Writing $g_{i}=\log{\gamma_{i}}$ for the Log-Gamma variates, where $\gamma_{i} \sim \mathrm{Gamma}(\alpha_{i})$, their moments are:"

\begin{subequations}
\begin{align}
\operatorname{Var}(g_i) &= \psi_1(\alpha_i) \\
\operatorname{Cov}(g_1, g_2) &= c(\rho)\,\sqrt{\psi_1(\alpha_1)\,\psi_1(\alpha_2)} \\
\operatorname{Var}(g_1 - g_2) &= \psi_1(\alpha_1) + \psi_1(\alpha_2)
    - 2\,c(\rho)\sqrt{\psi_1(\alpha_1)\,\psi_1(\alpha_2)}
\end{align}
\end{subequations}
where $\alpha_{1}$ and $\alpha_{2}$ are the Gamma shape parameters, $d_{1/2} = 2\alpha_{1,2}$, $\psi_1$ is the trigamma function, and $c(\rho)$ is the 'correlation transfer' function, describing the correlation as propagated through a PIT.  

The scaling factor $\bar{s}$ in Eq.~\ref{eq:fisher_z_variate} scales the variance of the uncorrelated distribution where $g_1 \perp g_2$ by a factor of $\bar{s}^{2}$. Therefore, defining $\operatorname{Var}(z_1 - z_2) = \bar{s}^{2} \cdot \operatorname{Var}_{\perp}(z_1 - z_2)$, then it follows that:
\begin{equation}
\bar{s} = \sqrt{\frac{\operatorname{Var}(z_1 - z_2)}{\operatorname{Var}_{\perp}(z_1 - z_2)}} = \sqrt{1 - 2\,c(\rho)\,\frac{\sqrt{\psi_1(\alpha_1)\psi_1(\alpha_2)}}{\psi_1(\alpha_1) + \psi_1(\alpha_2)}}
\label{eq:s_bar_variance_relation}
\end{equation}

 The correlation of the two Log-Gamma distributions in a grid range of $\alpha_{1,2} \in [1, 10]$ was studied. Firstly, the generative model $\zeta = LZ$ was used to obtain the correlated Gaussian scores from $Z \sim \mathcal{N}_{2}(0, I)$, where $L$ is the Cholesky decomposition of the correlation matrix $R$ given by the true correlation coefficient, $\rho$. The correlation-transfer function $\rho \mapsto c(\rho)$ was then evaluated using Gauss–Hermite quadrature, a method for numerical integration. The ratio $c(\rho)/\rho$ is calculated over the range of values and evaluated.

 \begin{figure}
    \centering
    \includegraphics[width=\linewidth]{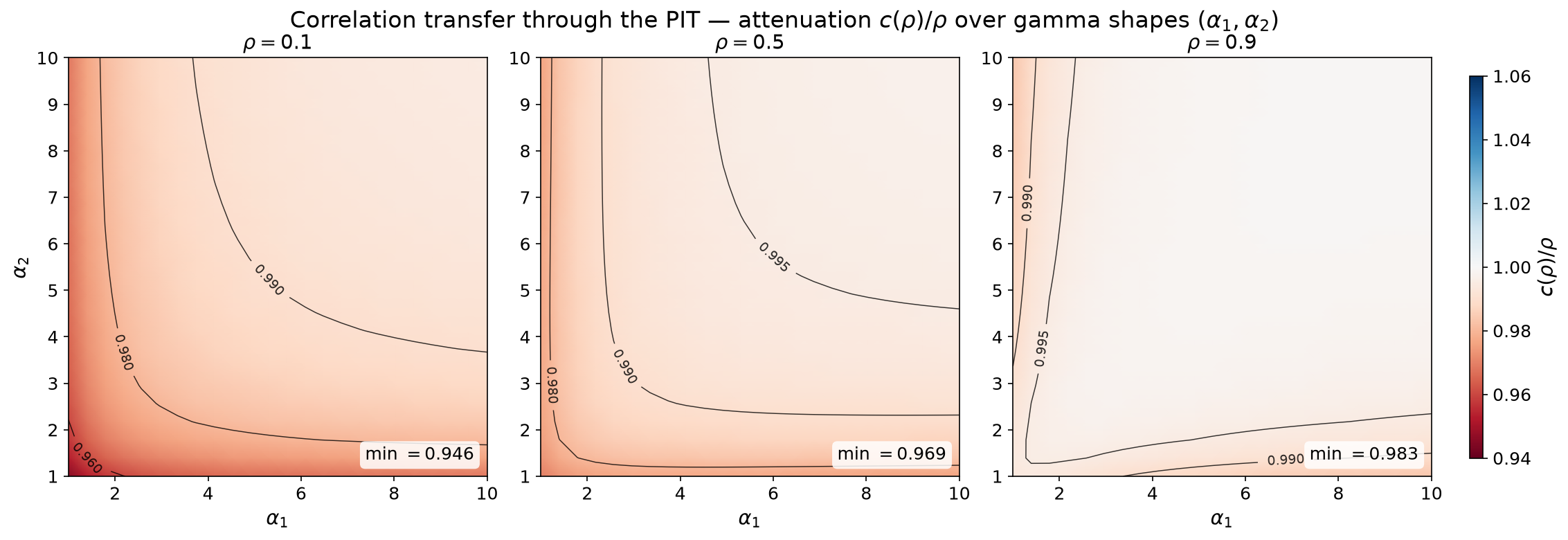}
    \caption{Ratio of the correlation-transfer function $c(\rho)$, describing the correlation between two Log-Gamma variables passed through a PIT transform with Gaussian copula correlation $\rho$. Each axis indicates a scan of the correlation at different values of two shape parameters, $\alpha_{1}$ and $\alpha_{2}$, where the expectation values of the copula transform distribution were evaluated using Gauss-Hermite quadrature. The plots show $\rho = 0.1$ (left), $\rho = 0.5$ (middle), $\rho=0.9$ (right). The red-to-white color indicates ratios lower than to equal to 1. }
    \label{fig:fig_corr_transfer_through_pit}
\end{figure}

 The results are shown in Appendix Fig.~\ref{fig:fig_corr_transfer_through_pit}. The correlation coefficient of the two Log-Gamma variables was found to be within \qty{6}{\percent} of the Gaussian $Z$ correlation coefficient, i.e $c(\rho) \simeq \rho$. Therefore where $\alpha_{1} \simeq \alpha_{2}$, the scaling $\bar{s} \simeq \sqrt{1 - \rho}$. MC integration yields equivalent results. 

\begin{figure}
    \centering
    \includegraphics[width=\linewidth]{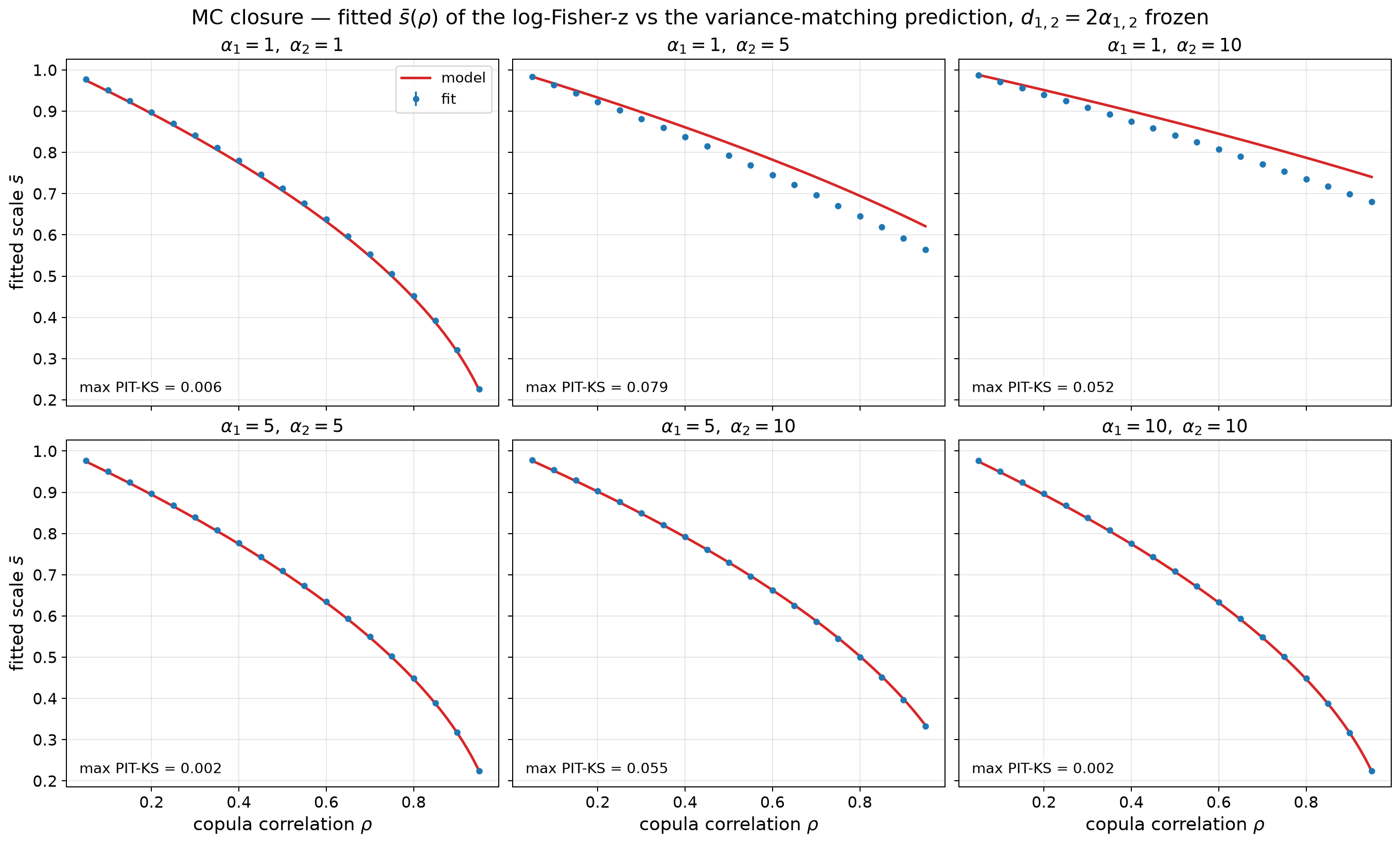}
    \caption{Scans of $\rho$ vs. the fitted values of $\bar{s}$ from fits of the scaled Fisher's-$z$ distribution with fixed parameters to MC samples of the difference between correlated Log-Gamma variables for combinations of $\alpha_{1/2} \in [1,5,10]$. The blue dots indicate the fitted value of $\bar{s}$ as a function of $\rho$. The model of Eq.~\ref{eq:s_bar_variance_relation} is shown as red lines. The PIT-KS score for the fits is shown in the bottom left of each plot.}
    \label{fig:fig_fisherz_scale_vs_rho_scan}
\end{figure}

 Fisher's-$z$ was then fitted to simulations of the log-ratio of two sets of \num{4e5} Gamma variates according to the same method, fitted using unbinned log-likelihood with the \texttt{iminuit} fit library \cite{dembinski_scikit-hepiminuit_2025}. The $d_{1}$, $d_{2}$ and $\Delta$ parameters were fixed at their nominal values. Therefore, the only remaining degree of freedom was the scale $\bar{s}$, which should be approximated by Eq.~\ref{eq:s_bar_variance_relation}. The results are shown in Fig.~\ref{fig:fig_fisherz_scale_vs_rho_scan}.

 The PIT-KS remains at or below the few-per-mille level except at $\alpha \simeq 1$, $\rho \simeq 1$, where it rises to $\sim$0.03–0.07. It can be observed that for $\alpha_{1/2} > 1$, there is excellent agreement between the predicted relationship between the scaling factor $\bar{s}$ and $\rho$, with a small error for the case $\alpha_{1/2}\simeq 1$, $\alpha_{2/1} > 1$, where the approximation $c(\rho) \simeq \rho $ is no longer valid. 

 This result indicates that a scaled Fisher's-$z$ is sufficient to capture correlations between Log-Gamma variates. This observation is relevant for the study of Sec.~\ref{sec:event_to_event_fluctuations}.

\subsection{Multi Layer Perceptrons} \label{sec:mlps}
 Multi-layer perceptrons (MLPs) are feed-forward neural networks which are used in this study as function approximators for conditional fitting and generative modeling. They consist of layers of nodes that correspond to a series of sequentially applied linear transformations, each of which is passed through a nonlinear 'activation' function before the next transformation is applied \cite{cybenko_approximation_1989}. All MLPs in this study use the sigmoid Linear Unit (SiLU), $\operatorname{SiLU}(x) = x \cdot (1 + \exp(-x))^{-1}$, for hidden layers, the sigmoid activation, $\operatorname{Sigmoid}(x) = (1 + \exp(-x))^{-1}$, the Softplus activation $\operatorname{Softplus}(x) = \log(1 + \exp(x))$ or the Softmax operation, which is introduced separately in Sec.~\ref{sec:full_event_modelling}, Eq.~\ref{eq:softmax}.

\newpage

\section{Derivation of Lateral Profile Model}\label{sec:lateral_profile_pdf}

It is noted that the Bessel-K distribution has been derived before (e.g., Ref.~\cite{bonamente_properties_2025}), but it is presented in the context of projected EM shower modeling to illustrate the validity of the study.

\subsection{Probability Density Function} 

A phenomenological and well-validated EM shower shape distribution is defined as: 

\begin{equation}
\diff{E}{S} = E_{T} \cdot \frac{e^{-\frac{r}{\gamma}}}{2\pi \gamma^{2}}
\label{eq:Molière_derivation_dE_dS}
\end{equation}

where $E_{T}$ is the total deposited energy, $\gamma$ describes the scale of the shower fluctuations from multiple scattering, and $r$ is the distance from the shower core position, $r = \left|X - C \right|$, where $X$ is the spatial position of the energy deposit $X = \left[x,y\right]$, and $C$ is the core position of the shower $C = \left[ c_{x}, c_{y} \right]$, and $\mathrm{d}S$ is the surface of an infinitesimal ring around the core axis, $\mathrm{d}S = 2\pi r~\mathrm{d}r$  

It is first noted that this distribution can be described in terms of the Gamma distribution, with a shape parameter $\alpha=2$:
\begin{equation}
\diff{E}{R} = \diff{E}{S} \cdot \left|\diff{S}{R}\right| = E_{T} \cdot \frac{r e^{\frac{-r}{\gamma}}}{\gamma^{2}}
\end{equation}

\begin{equation}
\Gamma(r; \alpha = 2, \beta = \gamma) = \frac{r e^{\frac{-r}{\gamma}}}{\gamma^{2}}
\label{eq:Molière_derivation_gamma_r}
\end{equation}
 
The Molière radius is therefore defined in terms of $\gamma$ according to the \qty{90}{\percent} percentile of the Gamma distribution, which is given by the inverse of the regularized upper incomplete Gamma function, $Q^{-1}(2, t)$, where $Q(2, t) = \int^{\infty}_{t} s\, e^{-s}\,\mathrm{d}s = e^{-t}(t + 1)$. The \qty{90}{\percent} percentile is given by solving $Q(2, t) = 0.1$. In practice, $\rho_{M} \simeq 3.89~\gamma$.      
   
The distribution of $x$ is now derived by integrating out $y$. As multiple scattering is uncorrelated at any azimuth angle around the core, the distribution does not vary for any projection of the shower profile, and thus this derivation holds for any orthogonal direction along the shower axis. 

 Substituting $y = \pm\sqrt{r^{2} - |x|^{2}}$ into Eq.\ref{eq:Molière_derivation_dE_dS} with two equally contributing branches, giving a factor of 2, and letting $\mathrm{d}y = \left| \diff{y}{r} \right| \mathrm{d}r = \frac{r}{\sqrt{r^{2} - |x|^{2}}}~\mathrm{d}r$:

\begin{equation}   
\diff{E}{x} = 2 \cdot E_{T} \cdot \int^{\infty}_{|x|} \frac{r}{\sqrt{r^{2} - |x|^{2}}}\frac{e^{-\frac{r}{\gamma}}}{2 \pi \gamma^{2}}~ \mathrm{d}r 
\end{equation} 

Let $r = |x|\cosh(t)$, as this has a convenient property:

\begin{equation}
   \diff{r}{t} = |x|\sinh{t}
\end{equation}

\begin{equation}
   \sqrt{r^2 - |x|^2} = \sqrt{|x|^2 \cosh^{2}{t} - |x|^{2}} = |x|\sinh{t}
\end{equation}

\begin{equation}                                                                                            
\diff{E}{x} = E_{T} \cdot \frac{|x|}{\pi \gamma^{2}} \cdot \int^{\infty}_{0} \cosh(t) \cdot e^{-\frac{|x|}{\gamma}\cosh(t)}~ \mathrm{d}t
\end{equation}      

The integral is an identity \textbf{10.32.9} from Ref.~\cite{NIST_DLMF} for the Modified Bessel Function of the second kind, $K_\nu(z)$: 

\begin{equation}
K_\nu(z) = \int_0^\infty e^{-z \cosh t} \cosh(\nu t) \, dt, \quad |\operatorname{ph} z| < \frac{1}{2} \pi
\label{eq:Molière_radius_modified_bessel_Knu}
\end{equation}

Where the phase condition on $z$ is irrelevant in this case, because $\gamma$ is real. 

The distribution of energy is therefore given by the so-called 'Bessel-K' distribution: 

\begin{equation}
    \diff{E}{x} = E_{T} \cdot \operatorname{BesselK}(x; \gamma) =  E_{T} \cdot \frac{|x|}{\pi \gamma^{2}} K_{1} \left(\frac{|x|}{\gamma}\right)
    \label{eq:bessel_k_pdf_derivation}
\end{equation}

If displaced, a location parameter $c$ can be introduced:

\begin{equation}
    \diff{E}{x} =  E_{T} \cdot \operatorname{BesselK}(x; c, \gamma) = E_{T} \frac{|x-c|}{\pi \gamma^{2}} K_{1} \left(\frac{|x-c|}{\gamma}\right)
\end{equation}

\subsection{Cumulative Distribution Function}

The CDF is provided as it is helpful for binned likelihood fits. The parameter $E_{T}$ is omitted for brevity. 

Substituting $u = \frac{|x-c|}{\gamma}$, $\mathrm{d}u = \gamma^{-1} \mathrm{d}x $, the PDF per unit $u$ is $f(u) = \frac{u}{\pi} K_{1}(u)$ for $u \geq 0$, and the CDF is:                                                      
\begin{equation}
F(x) = \frac{1}{\pi}\int^{x}_{-\infty} |v| \cdot K_{1} (|v|) ~ \mathrm{d}v                                   \label{eq:Molière_radius_derivation_f_u}                                                                   \end{equation}      

Any symmetric probability distribution has the property:

\begin{equation}
    F(u; \gamma) = \frac{1}{2} + \operatorname{sign}(x - c) \cdot G\left( u; \gamma \right)
\end{equation}

where $F(x) = \frac{1}{2} + \operatorname{sign}(x-c) \cdot G\left( \frac{|x-c|}{\gamma} \right)$ and:       

\begin{equation}
 G(u; \gamma) = \int^{u}_{0} f(v)~\mathrm{d}v = \frac{1}{\pi} \int^{u}_{0} v \, K_{1}(v)~\mathrm{d}v.      
\label{eq:Molière_radius_derivation_G}
\end{equation}

Integrating Eq.~\ref{eq:Molière_radius_derivation_G} by parts, with $\int K_{1}(v)~\mathrm{d}v = -K_{0}(v)$:                                                                                                           
\begin{equation}
G(u; \gamma) = \frac{1}{\pi} \left[ -v K_{0}(v) \right]^{u}_{0} + \frac{1}{\pi} \int^{u}_{0} K_{0}(v)~\mathrm{d}v 
\end{equation}

An identity is introduced: 

\begin{equation}
    \int^{u}_{0} K_{0}(v)~\mathrm{d}v = \frac{\pi u}{2} \left[ K_{0}(u)L_{-1}(u) + K_{1}(u)L_{0}(u) \right] =\frac{\pi u}{2} A(u);
    \label{eq:Molière_radius_derivation_identity}
\end{equation}

where $L_{0}$ and $L_{-1}$ are Modified Struve functions, 


The by-parts identity $\int K_{1}(v)~\mathrm{d}v = -K_{0}(v)$ follows from \textbf{10.29.3} in Ref.~\cite{NIST_DLMF}. Eq.~\ref{eq:Molière_radius_derivation_identity} can be proved by differentiating the ansatz $A(v)$ by parts and using the identities $K^{\prime}_{0}(v) = -K_{1}(v)$ (\textbf{10.29.3}), $K^{\prime}_{1} = -K_{0} - \frac{K_{1}}{v}$ (\textbf{10.29.2}), $L^{\prime}_{0} = L_{-1}$ (\textbf{11.4.29} with $\nu=0$), and $L^{\prime}_{-1} = L_{0} - \frac{L_{-1}}{v} + \frac{2}{\pi v}$ (which follows from \textbf{11.4.25} and \textbf{11.4.26} with $\nu = -1$) in Ref.~\cite{NIST_DLMF}: 

\begin{equation}
\diff{}{v} A(v) = - \frac{1}{v}\cdot A(v) + \frac{2K_{0}(v)}{\pi v}                                    
\end{equation}

Differentiating by parts: 

\begin{align*}
\diff{}{v} \frac{\pi v}{2} A(v) &= \frac{\pi}{2} \left[ A(v) + v A^{\prime}(v) \right] \\ 
&=  \frac{\pi}{2}  \left[ A(v) + v \left( -\frac{1}{v} A(v) + 2 \frac{K_{0}(v)}{\pi v} \right) \right] \\   
&= K_{0}(v) 
\end{align*}    

\begin{align*}
    G(u; \gamma) & =  \frac{1}{\pi} \left( 
    -u K_{0}(u) + \frac{\pi u}{2} A(u) \right) \\
    & = 
    -\frac{u}{\pi} K_{0}(u) + \frac{u}{2} A(u)  \\
\end{align*}

Therefore, the CDF is defined as: 

\begin{equation}
    F(u; \gamma, c) = \frac{1}{2} + \operatorname{sign}(x - c) \left(-\frac{u}{\pi} K_{0}(u) + \frac{u}{2} A(u)  \right) 
    \label{eq:bessel_k_cdf_derivation}
\end{equation}

with $u=\frac{|x - c|}{\gamma}$. 

\subsection{Statistical Moments}

Because the distribution is symmetrical, the odd moments (mean, skew) are 0. 

The even moments of the standardised Bessel-K variable $U=(X-c)/\gamma$ (i.e. $c=0$, $\gamma=1$ ) are therefore:

\begin{align*}
\mathbb{E}[U^K]
&= 2 \int_0^\infty u^K \cdot f(u) \, du \\ 
&= 2 \int_0^\infty u^K \cdot \frac{u}{\pi} K_1(u) \, du \\ 
&= \frac{2}{\pi} \int_0^\infty u^{K+1} K_1(u) \, du.
\end{align*}

Combining with \textbf{10.43.19} ($\mu = K+2$, $\nu = 1$) we obtain:

\begin{equation}
\mathbb{E}[U^{K}] = \frac{2^{K+1}}{\pi} \cdot \Gamma \left( \frac{K+3}{2} \right) \Gamma\!\left( \frac{K+1}{2} \right) 
\end{equation}

Variance and kurtosis can be derived as follows:

\begin{equation}
 \mathbb{E}[U^{2}] = \frac{8}{\pi} \cdot \Gamma \left( \frac{5}{2} \right) \cdot \Gamma\left( \frac{3}{2} \right) = 3
\end{equation}

\begin{equation}
 \mathbb{E}[U^{4}] = \frac{32}{\pi} \cdot \Gamma \left( \frac{7}{2} \right) \cdot \Gamma\left( \frac{5}{2} \right) = 45 
\end{equation}

The variance is therefore:
\begin{equation}
\sigma^{2}_{U} = 3 
\label{eq:bessel_k_variance_derivation}
\end{equation}

The excess kurtosis is therefore
\begin{equation}
\kappa_4 = \frac{45}{3^2} - 3 = 5 - 3 = 2. 
\label{eq:bessel_k_excess_kurtosis_derivation}
\end{equation}

For a general scale $\gamma$, $\mathbb{E}[(X-c)^{K}] = \gamma^{K}\,\mathbb{E}[U^{K}]$, so $\sigma^{2}_{X} = 3\gamma^{2}$.

\subsection{Generating Random Variables}
The generation of random variables follows immediately from the definition given in Eq.~
\ref{eq:Molière_derivation_gamma_r}.
If:

\begin{subequations}
\begin{equation}
   r \sim \operatorname{Gamma}(\alpha=2, \beta=\gamma) 
\end{equation}

\begin{equation}
   \theta \sim \operatorname{Uniform}(-\pi, \pi)
\end{equation}
\end{subequations}

then:

\begin{equation}
r \cdot \cos{\theta} \sim \operatorname{BesselK}(\gamma)
\label{eq:bessel_k_sampling}
\end{equation}

\subsection {Monte Carlo Closure}

Monte Carlo closure was performed using \num{1e6} events generated via the polar decomposition of Eq.~\ref{eq:bessel_k_sampling}. The probability density function, cumulative distribution function, PIT-transform uniformity, and bootstrap convergence of the first four standardized moments are studied. The agreement is measured using reduced $\chi^{2}$ for the PDF, $W_{1}$ for the CDF, and the PIT-KS statistic for the PIT. Definitions are available in Appendix Sec.~\ref{sec:similarity_measures}. For the moment convergence test, for each sample size $N$, $500$ independent polar samples are drawn and the moments are computed for each sample. The mean and $\pm 1\sigma$ percentile interval of these estimates are then computed.

\begin{figure}[htbp]
    \centering
    \includegraphics[width=0.7\textwidth]{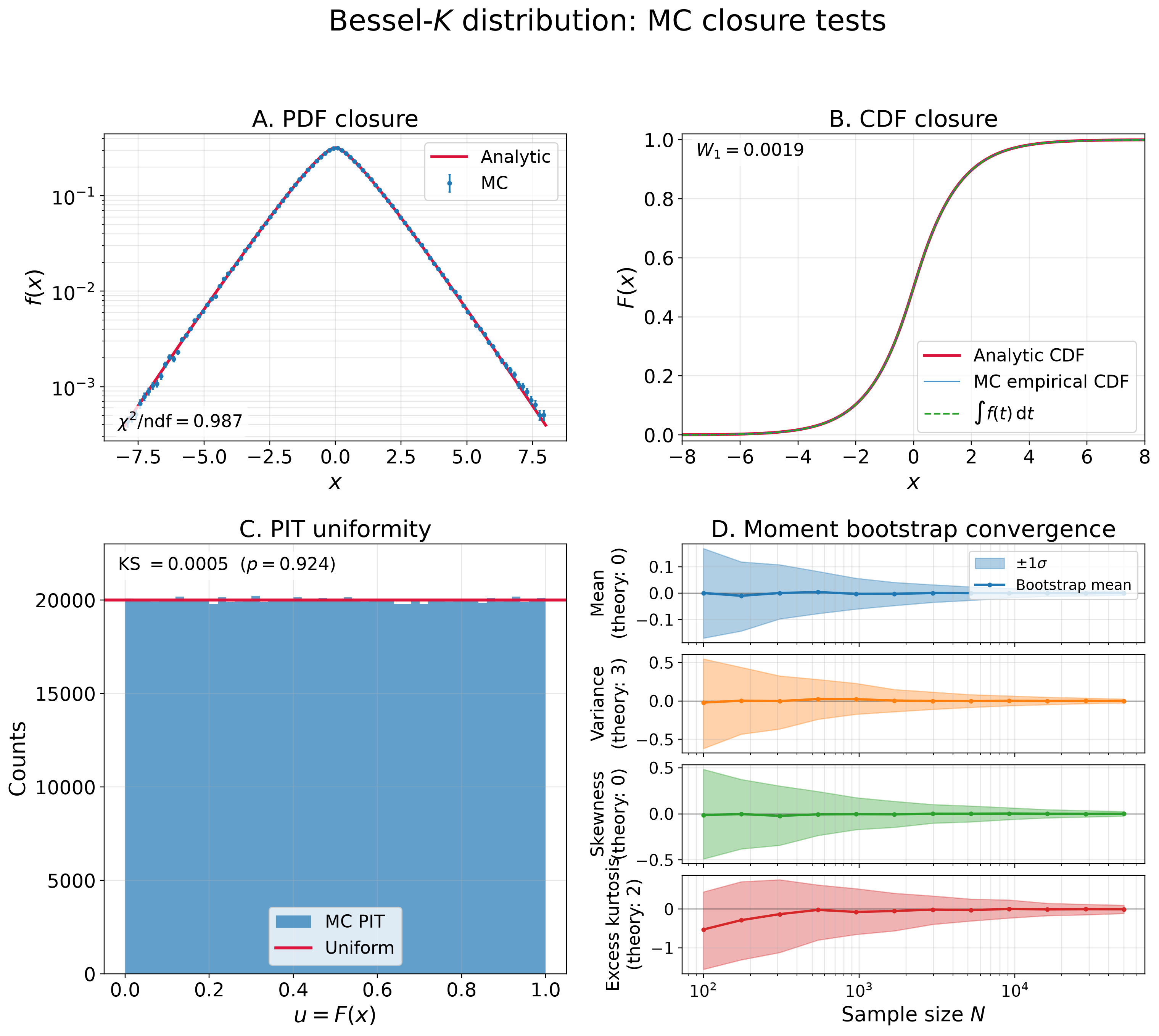}
    \caption{Monte Carlo closure tests of the Bessel-$K$ distribution,
    $\operatorname{BesselK}(x;\gamma=1)$. Top left: binned probability density of \num{1e6} events across four orders of magnitude. The blue dots indicate the binned and normalized MC histogram, and  the red line the PDF of Appendix Eq.~\ref{eq:bessel_k_pdf_derivation}.
    Top right: the solid blue and red lines and dashed green line show the empirical CDF, the analytical CDF of Appendix Eq.~\ref{eq:bessel_k_cdf_derivation}, and the numerical cumulative integral of Eq.~\ref{eq:bessel_k_pdf_derivation}, respectively. Bottom left: the blue histogram shows the PIT $u = F(x)$ applied to the MC,  while the red line shows the uniform distribution scaled to the counts. Bottom right: the first four moments compared to theory. The solid line shows  the deviation of the bootstrap mean from the theoretical value. The shaded  band shows the $\pm 1\sigma$ percentile interval.}

    \label{fig:bessel_k_closure}
\end{figure}

The results are shown in Fig.~\ref{fig:bessel_k_closure}. The binned MC density agrees with the analytic PDF of Appendix Eq.~\ref{eq:bessel_k_pdf_derivation} across four orders of magnitude with reduced $\chi^{2}=0.987$. The empirical CDF agrees with the analytic CDF of 
Eq.~\ref{eq:bessel_k_cdf_derivation} with $W_{1}=0.0019$. The PIT is uniform, with KS statistic $=0.0005$ ($p=0.924$). Finally, the mean, variance, skewness, and excess kurtosis (mean and skewness are 0 by symmetry; variance and excess kurtosis are given by Appendix Eqs.~\ref{eq:bessel_k_variance_derivation} and \ref{eq:bessel_k_excess_kurtosis_derivation}) are compared to the bootstrapped values. The estimators are unbiased, with uncertainties that scale approximately as $N^{-\frac{1}{2}}$. Therefore, the MC closure indicates the distribution is well defined and in agreement with expectations.

\section{Supplementary Figures and Tables}


\begin{table*}[tbp]
\centering
\caption{Number of electron-shower events per beam energy passing the 
data-quality selection. The total sample comprises 2,853,392 events.}
\label{tab:event_counts}
\setlength{\tabcolsep}{6pt}
\begin{tabular}{l rrrrrrr}
\toprule
$E_{\mathrm{beam}}$ [GeV] & 0.2 & 0.3 & 0.4 & 0.5 & 0.6 & 0.7 & 0.8 \\
\midrule
events & 6,944 & 56,015 & 73,809 & 83,325 & 86,856 & 88,190 & 90,269 \\
\addlinespace[0.8em]
$E_{\mathrm{beam}}$ [GeV] & 0.9 & 1 & 1.5 & 2 & 2.5 & 3 & 3.5 \\
\midrule
events & 179,296 & 437,568 & 175,668 & 173,827 & 172,714 & 170,794 & 168,612 \\
\addlinespace[0.8em]
$E_{\mathrm{beam}}$ [GeV] & 4 & 4.5 & 5 & 5.5 & 6 & Total & \\
\midrule
events & 167,286 & 165,191 & 246,302 & 157,865 & 152,861 & 2,853,392 & \\
\bottomrule
\end{tabular}
\end{table*}

\begin{table}[htbp]
    \centering
    \small
    \renewcommand{\arraystretch}{1.5}
    \setlength{\tabcolsep}{4pt}
    \caption{Event-level aggregated parameters used as inputs to the BDT. 
    The table lists the variable name, symbol, and definition for each. 
    Here, $i$ is the bar index (left$\to$right, up$\to$down) running from 1 to 6, 
    $l$ is the layer index from 1 to 14, $E_{\mathrm{sum}}$ is the total deposited energy in the event, 
    $N_{\mathrm{hits}}$ is the number of bars above threshold, and ``even''/``odd'' refer to even/odd bar indices (for transverse) or layers (for longitudinal). $f_{\text{layer},\,l-1}$ is defined such that $f_{\text{layer},\,0}$ corresponds to the first layer.}
    \label{tab:bdt_params}
    \begin{tabular}{@{}>{\raggedright\arraybackslash}p{0.24\linewidth}>{\centering\arraybackslash}p{0.20\linewidth}>{\centering\arraybackslash}p{0.46\linewidth}@{}}
        \toprule
        Variable Name & Symbol & Definition \\
        \midrule
        Number of Hits 
            & $N_{\mathrm{hit}}$
            & $\sum_{i} \mathbb{1}\!\left[E_i > E_{\mathrm{cut}} \right]$\\[8pt]
        Center of Gravity, $x$ 
            & $\mathrm{CoG}_{X}$ 
            & $\frac{\sum_{i=2,\,\text{even}} E_{\mathrm{bar},i}\,i}{E_{\mathrm{sum,\,even}}}$ \\[8pt]
        Center of Gravity, $y$ 
            & $\mathrm{CoG}_{Y}$ 
            & $\frac{\sum_{i=1,\,\text{odd}} E_{\mathrm{bar},i}\,i}{E_{\mathrm{sum,\,odd}}}$ \\[8pt]
        Center of Gravity, $z$ 
            & $\mathrm{CoG}_{Z}$ 
            & $\frac{\sum_{l=1}^{14} E_{\mathrm{bar},l}\,l}{E_{\mathrm{sum}}}$ \\[8pt]
        Mean Distance from CoG, $x$ 
            & $\langle|X-\mathrm{CoG}_{X}|\rangle$ 
            & $\frac{\sum_{i=2,\,\text{even}} |i-\mathrm{CoG}_{X}|}{N_{\mathrm{hit,\,even}}}$ \\[8pt]
        Mean Distance from CoG, $y$ 
            & $\langle|Y-\mathrm{CoG}_{Y}|\rangle$ 
            & $\frac{\sum_{i=1,\,\text{odd}} |i-\mathrm{CoG}_{Y}|}{N_{\mathrm{hit,\,odd}}}$ \\[8pt]
        Fraction of Shower Energy in Layer $l$ 
            & $f_{\text{layer},\,l-1}$ 
            & $\frac{E_{\mathrm{bar},l}}{E_{\mathrm{sum}}}$ \\
        \bottomrule
    \end{tabular}
\end{table}

\begin{figure}
    \centering
    \includegraphics[width=0.9\linewidth]{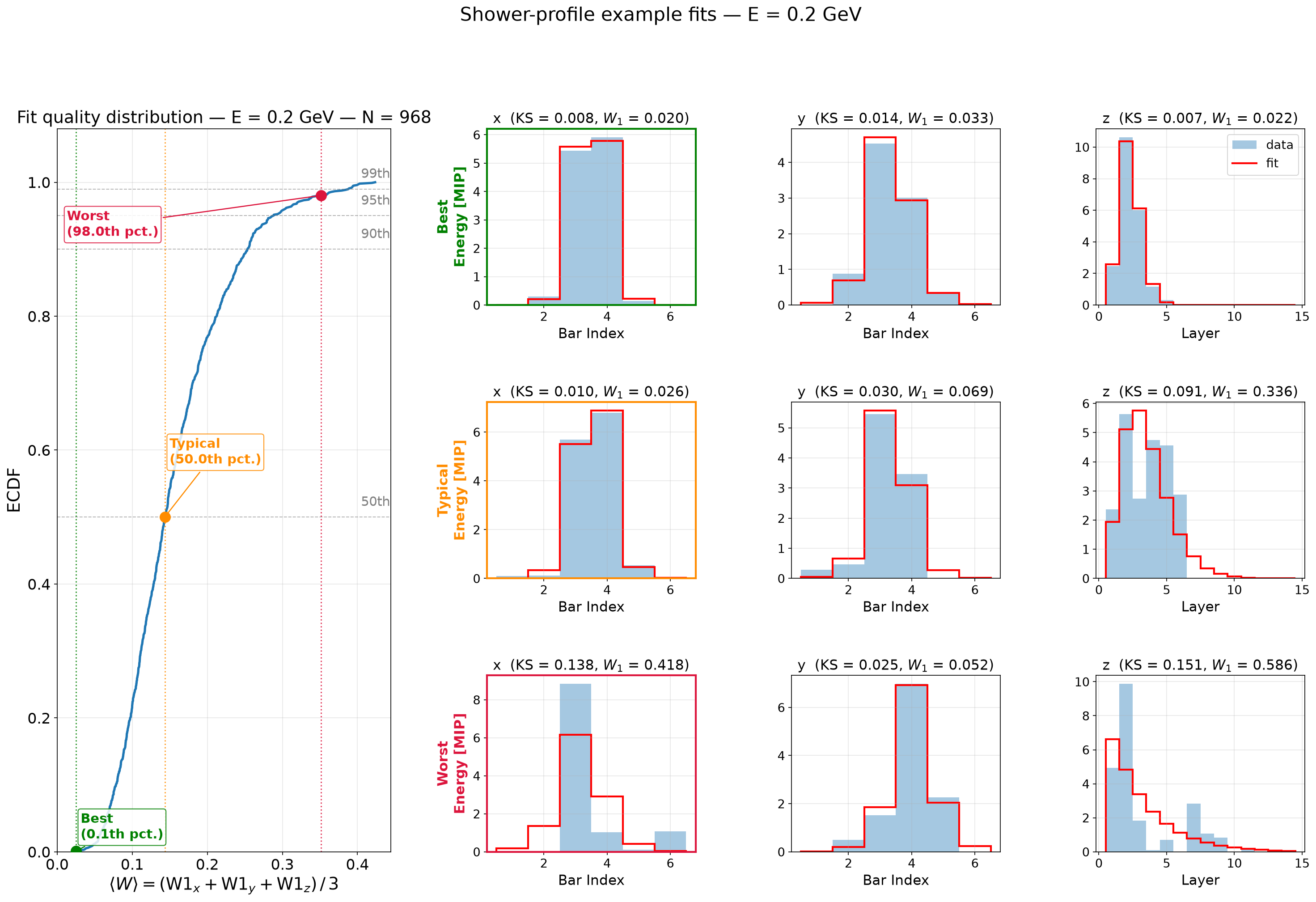}
    \caption{Profile fits for \qty{0.2}{\giga \electronvolt} electron showers. Else, as in Fig.~\ref{fig:fig_example_fits_profile_1GeV}}
    \label{fig:fig_example_fits_profile_0p2GeV}
\end{figure}

\begin{figure}
    \centering
    \includegraphics[width=0.9\linewidth]{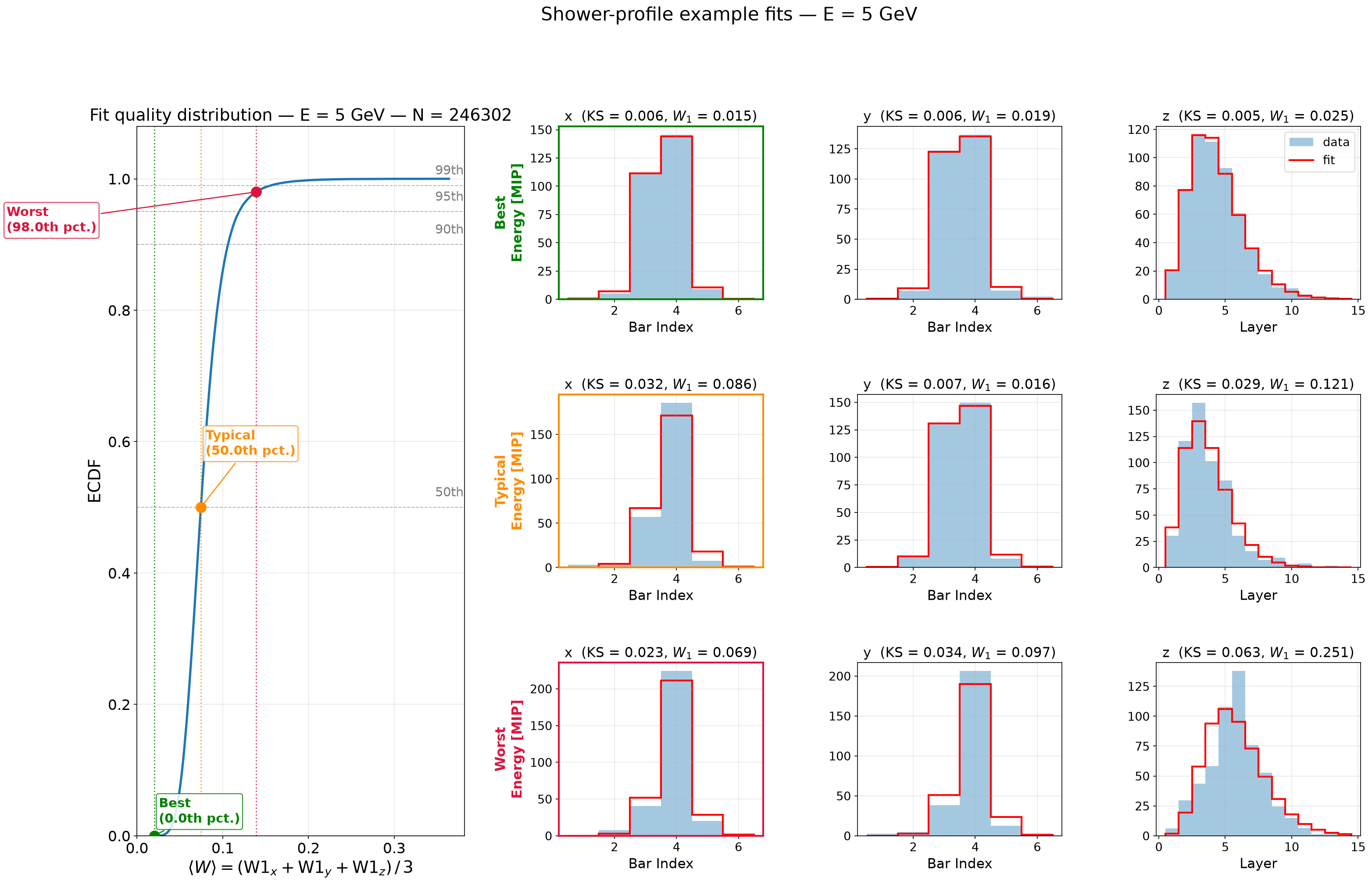}
    \caption{Profile fits for \qty{5}{\giga \electronvolt} showers. Else, as in Fig.~\ref{fig:fig_example_fits_profile_1GeV}}
    \label{fig:fig_example_fits_profile_5GeV}
\end{figure}

\begin{figure}
\centering
    \begin{subfigure}{0.4\linewidth}
    \includegraphics[width=\linewidth]{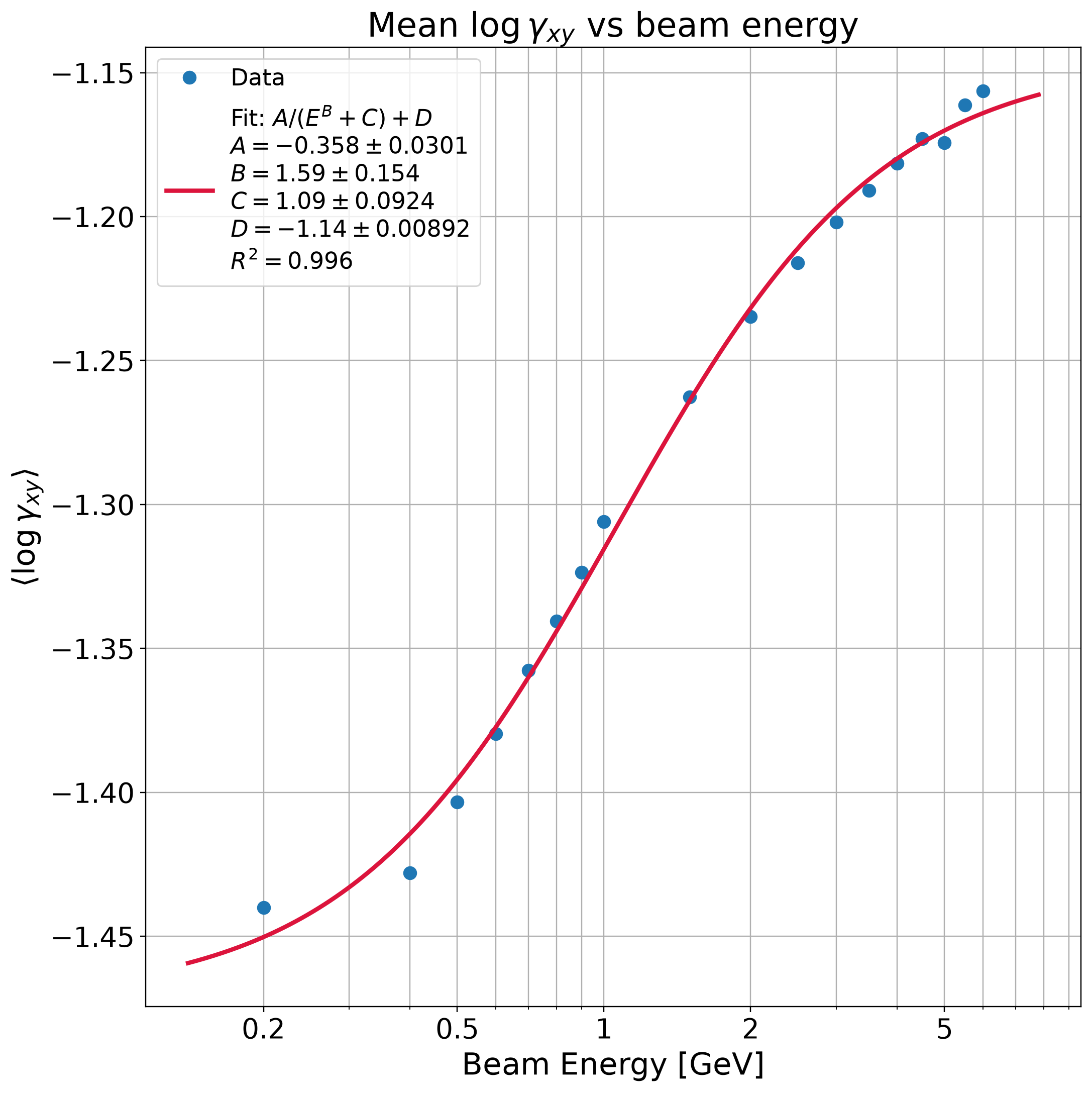}
    \caption{$\langle \log{\gamma_{xy}} \rangle$ as a function of energy. Else as in Fig.~\ref{fig:fig_rho_m_mean}.} 
    \label{fig:fig_log_gamma_xy_mean}
    \end{subfigure}
    \hfill
    \begin{subfigure}{0.4\linewidth}
    \includegraphics[width=\linewidth]{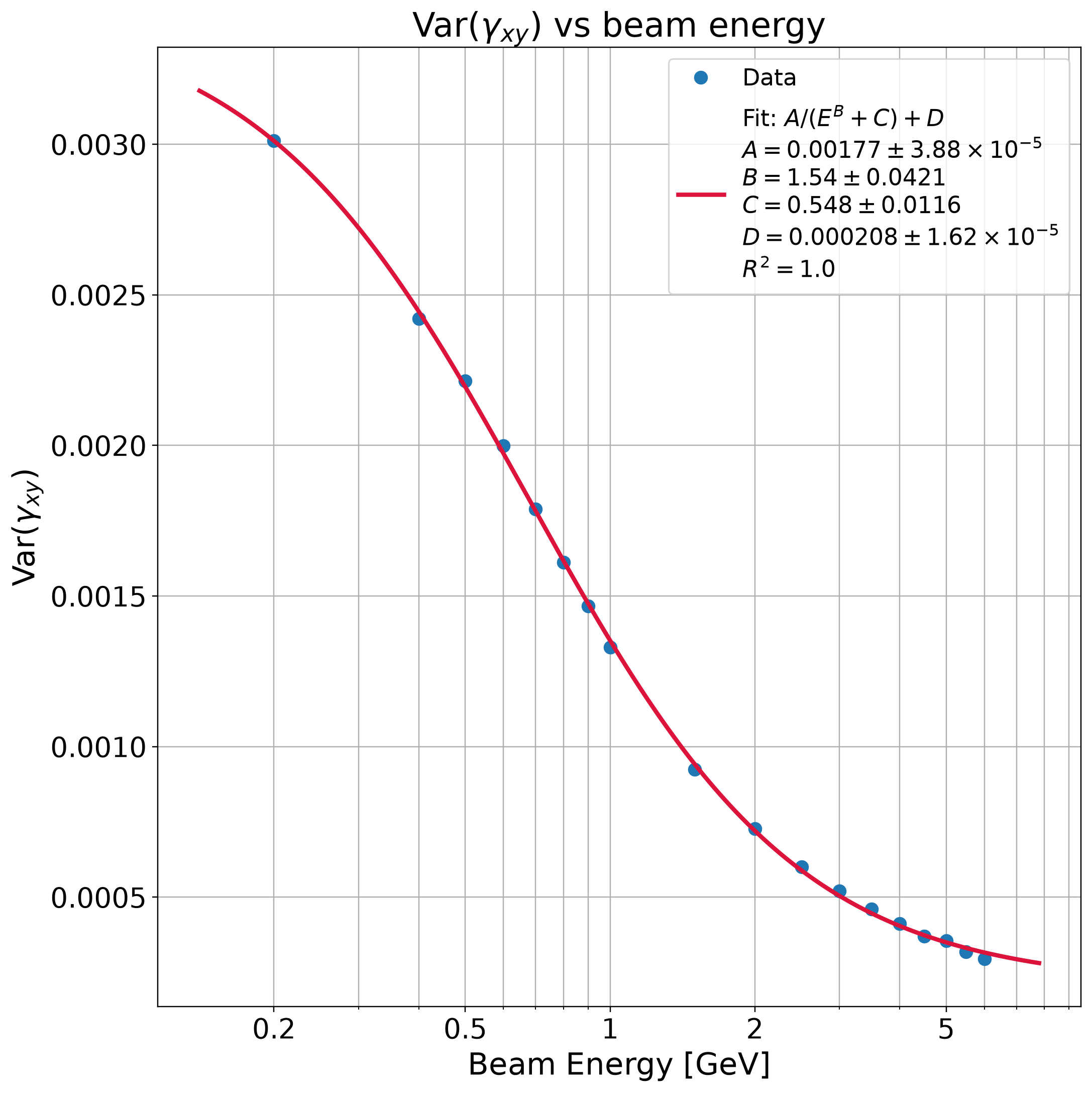}
    \caption{$\operatorname{Var}( \log{\gamma_{xy}})$ as a function of energy. Else as in Fig.~\ref{fig:fig_rho_m_mean}.}
    \label{fig:fig_log_gamma_xy_variance}
    \end{subfigure}
    \vfill
    \begin{subfigure}{0.4\linewidth}
    \includegraphics[width=\linewidth]{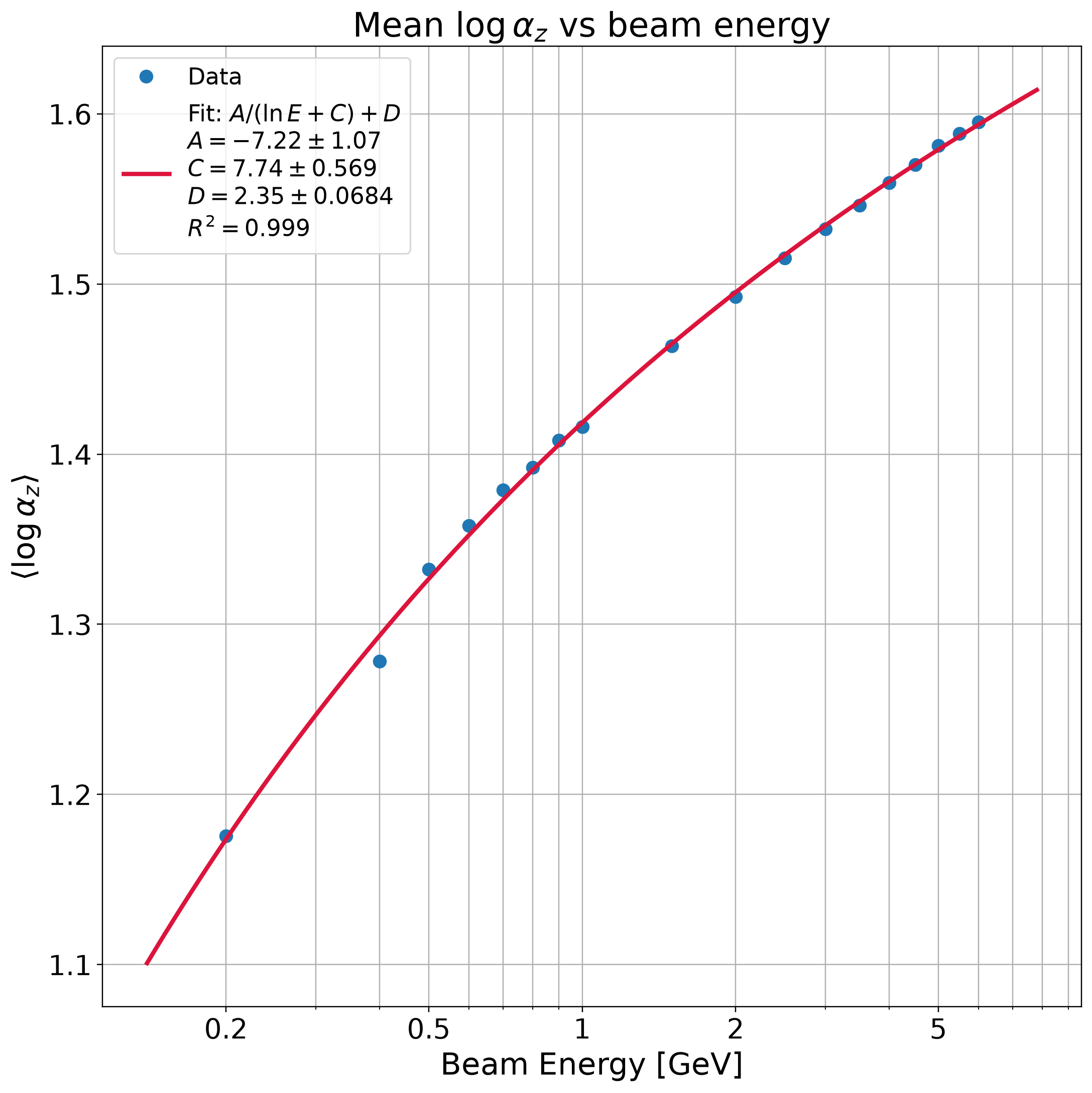}
    \caption{$\langle \log{\alpha_{z}} \rangle$ as a function of energy. Else as in Fig.~\ref{fig:fig_rho_m_mean}.} 
    \label{fig:fig_log_alpha_z_mean}
    \end{subfigure}
    \hfill
    \begin{subfigure}{0.4\linewidth}
    \includegraphics[width=\linewidth]{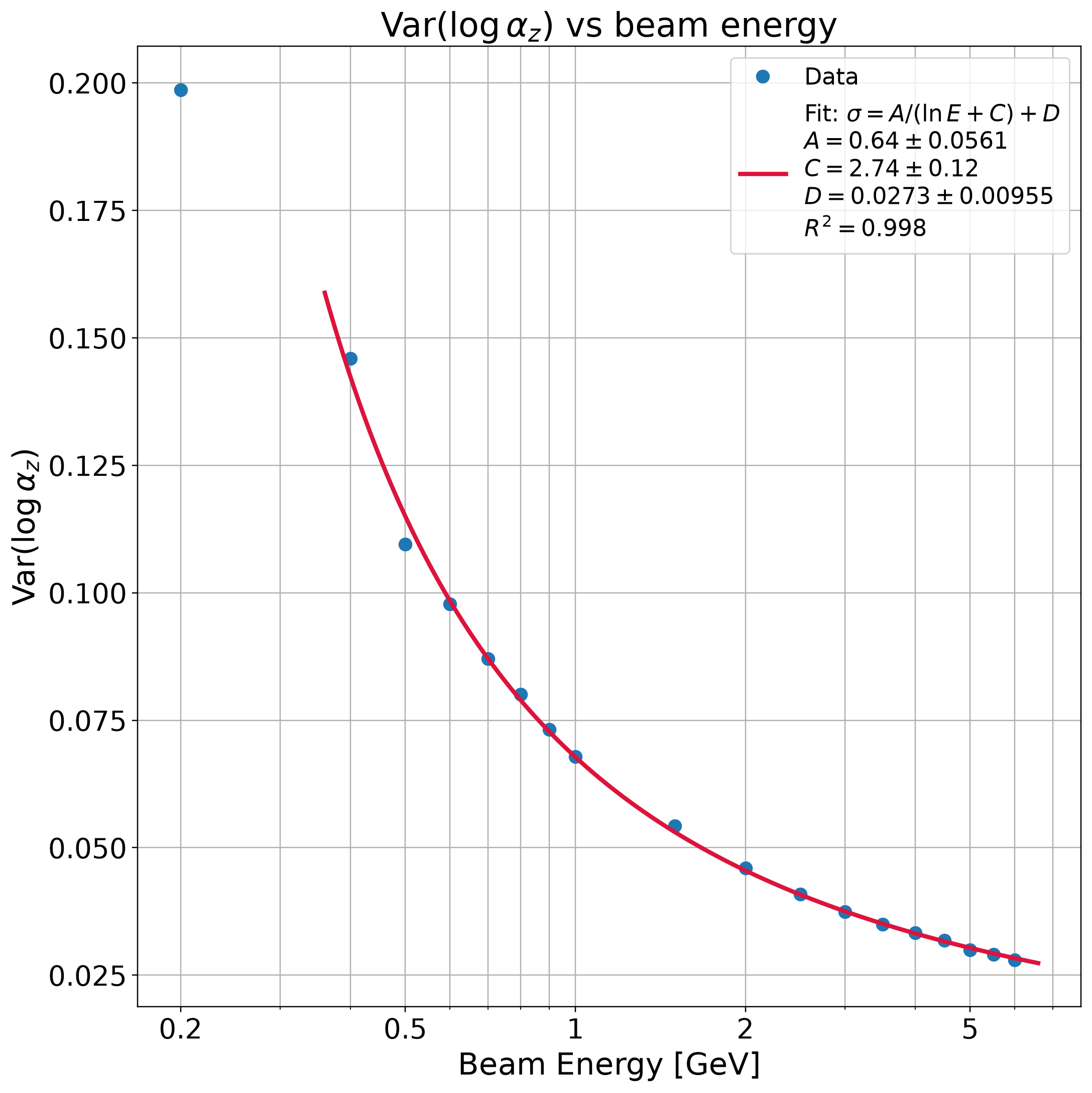}
    \caption{$\operatorname{Var}( \log{\alpha_{z}})$ as a function of energy. Else as in Fig.~\ref{fig:fig_rho_m_mean}.}
    \label{fig:fig_log_alpha_z_variance}
    \end{subfigure}
    \vfill
    \begin{subfigure}{0.4\linewidth}
    \includegraphics[width=\linewidth]{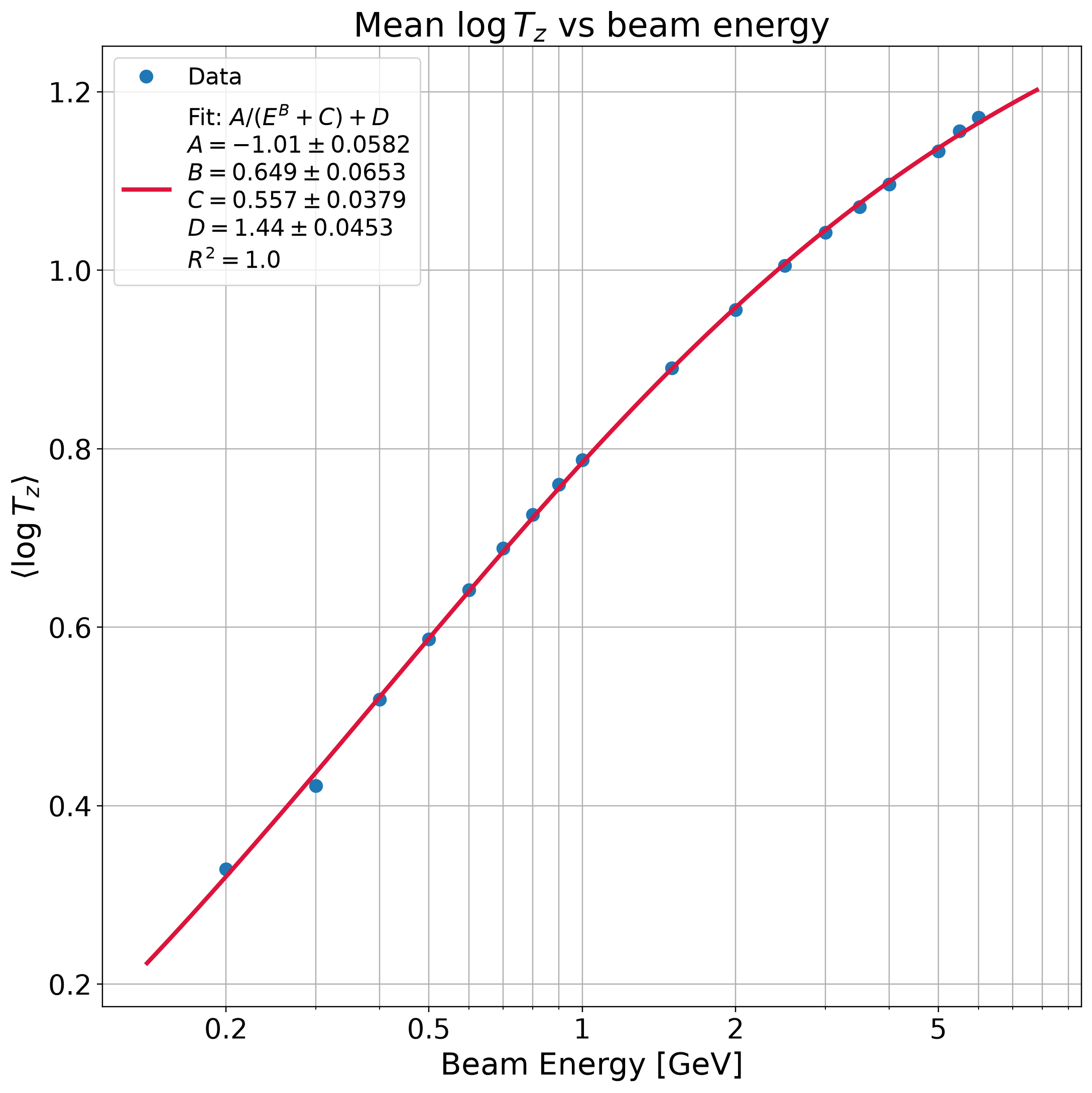}
    \caption{$\langle \log{T_{z}} \rangle$ as a function of energy. Else as in Fig.~\ref{fig:fig_rho_m_mean}.} 
    \label{fig:fig_log_T_z_mean}
    \end{subfigure}
    \hfill
    \begin{subfigure}{0.4\linewidth}
    \includegraphics[width=\linewidth]{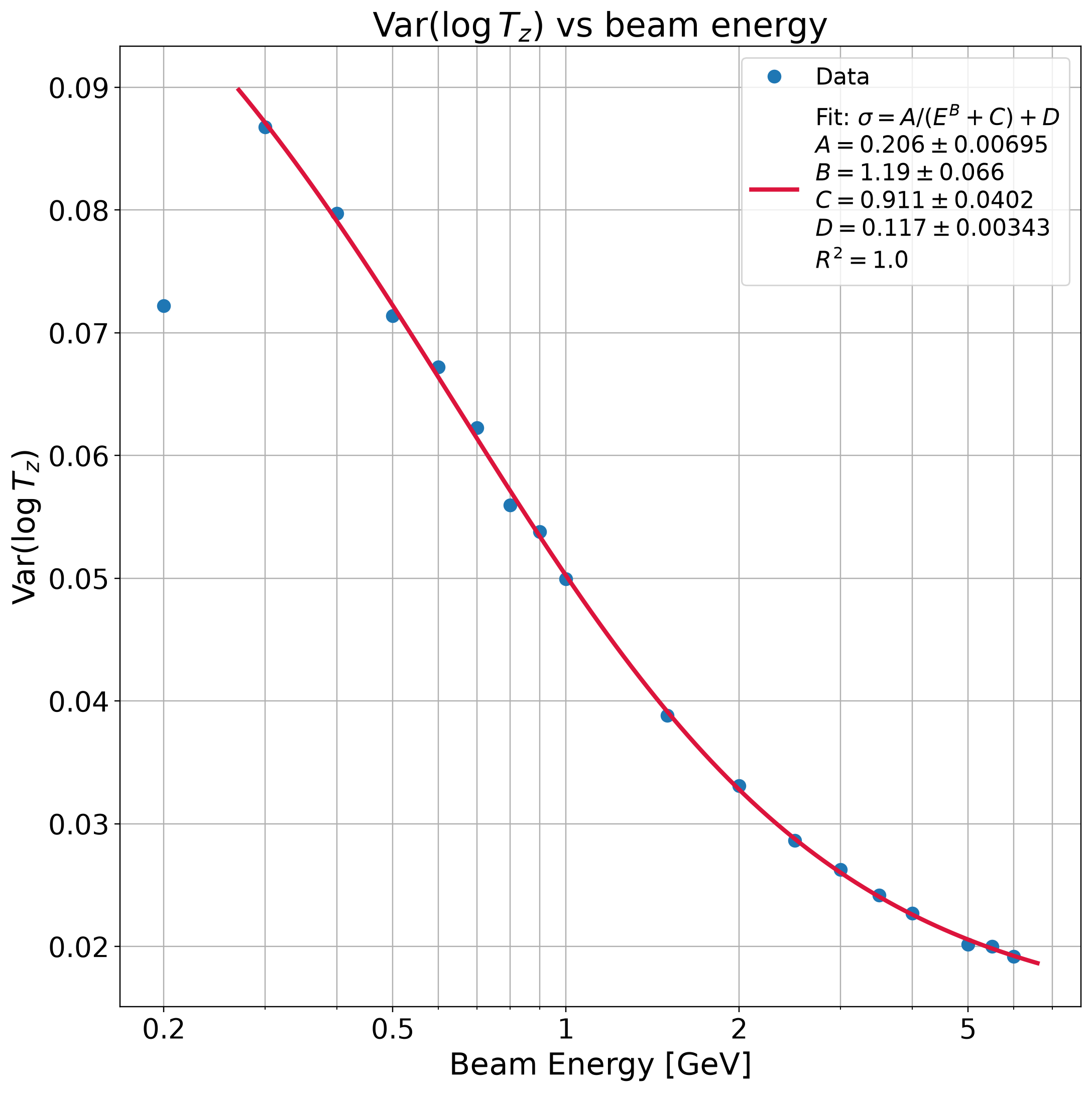}
    \caption{$\operatorname{Var}( \log{T_{z}})$ as a function of energy. The blue dots indicate the variance. Else as in Fig.~\ref{fig:fig_rho_m_mean}.}
    \label{fig:fig_log_T_z_variance}
    \end{subfigure}

    \caption{Distributions of the mean and variance of the logarithms of $\gamma$, $\alpha_{Z}$, and $T_{z}$ shower shape parameters. }

    \label{fig:log_shapes_results}
\end{figure}

\begin{figure}
\centering
    \begin{subfigure}{0.4\linewidth}
    \includegraphics[width=\linewidth]{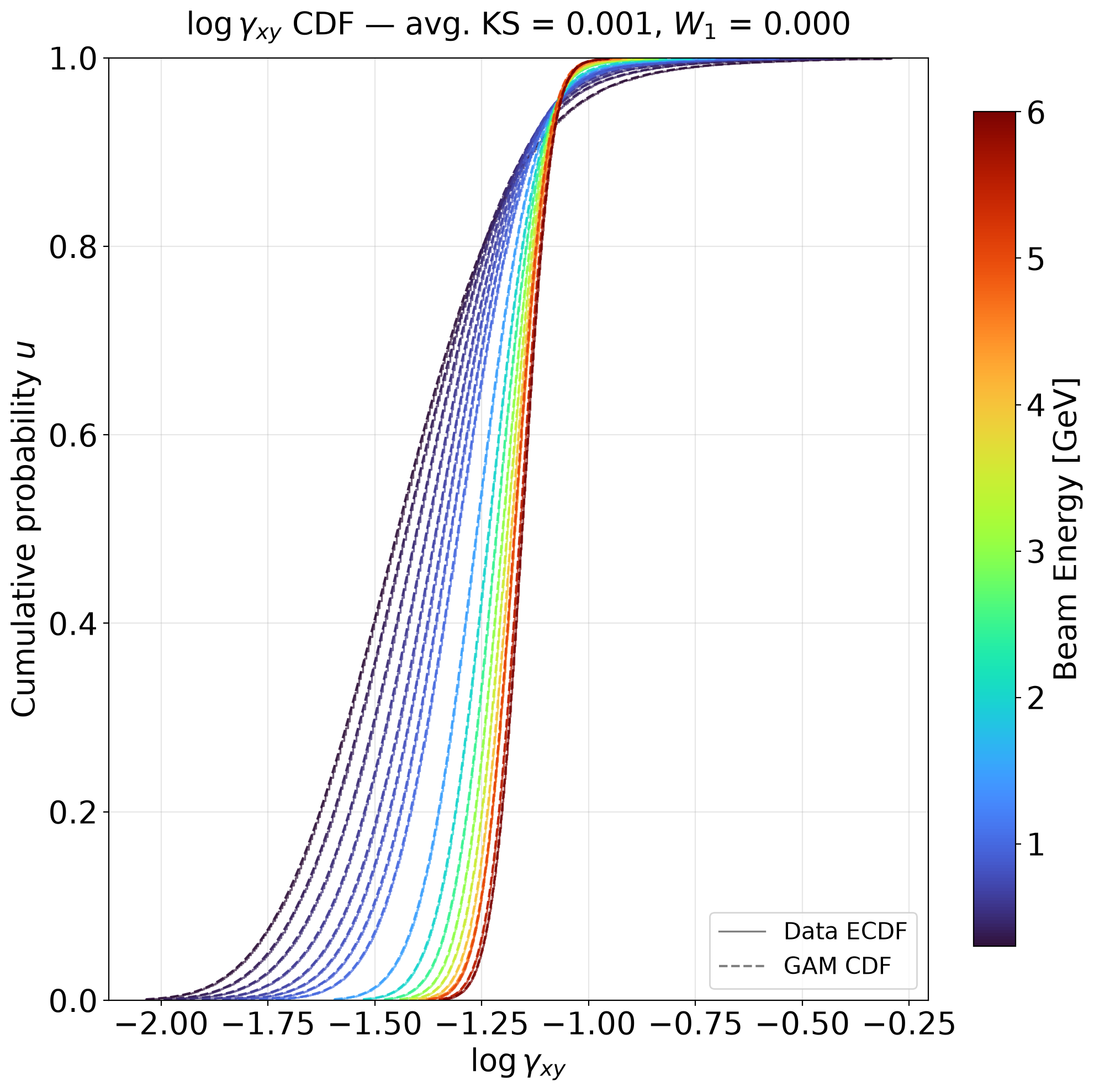}
    \caption{ECDF of the $\log{\gamma_{xy}}$ variable. } 
    \label{fig:fig_cdf_evolution_gamma_xy}
    \end{subfigure}
    \hfill
    \begin{subfigure}{0.4\linewidth}
    \includegraphics[width=\linewidth]{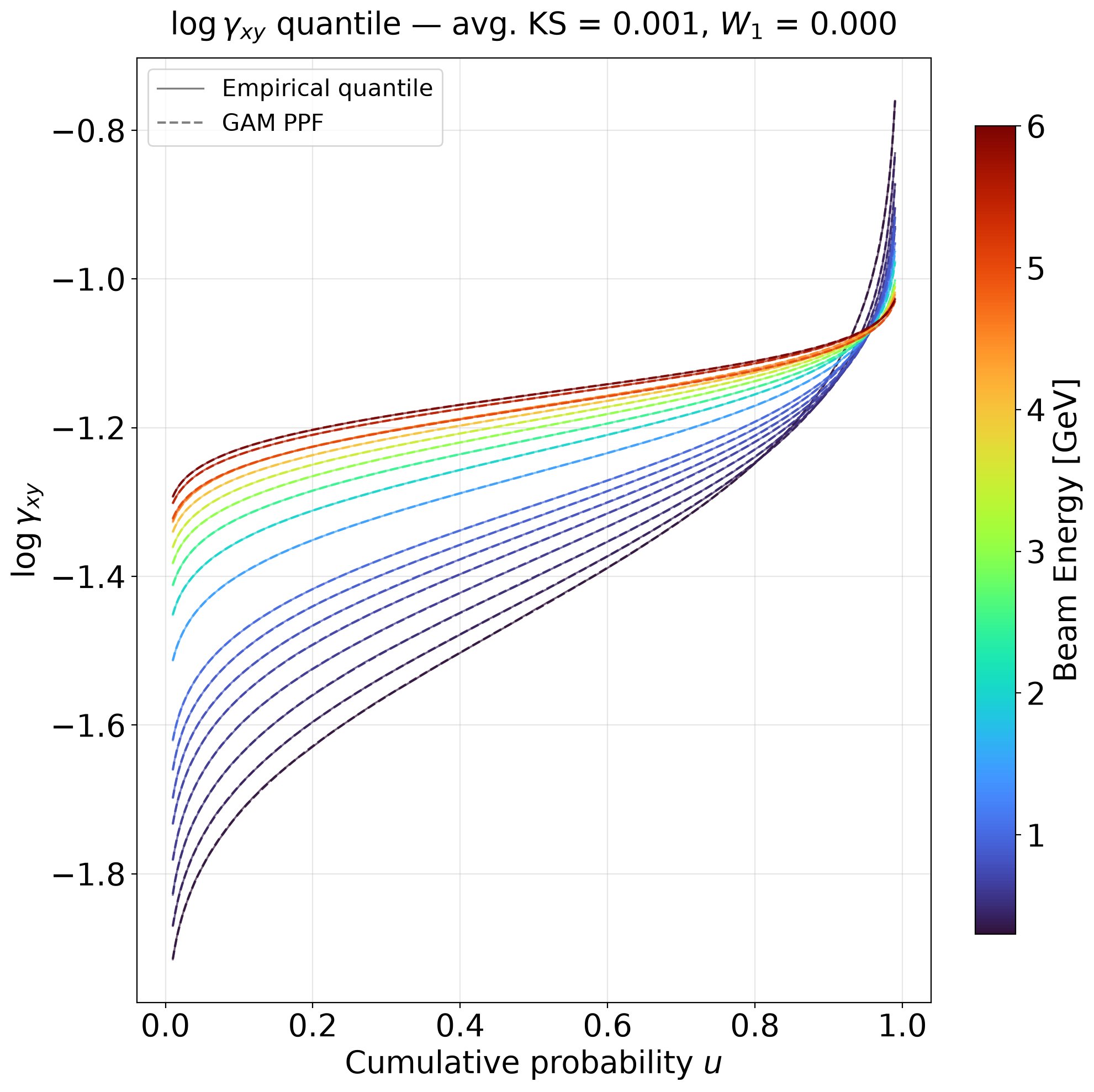}
    \caption{Inverse ECDF of the $\log{\gamma_{xy}}$ variable. Else, as in Fig.~\ref{fig:fig_cdf_evolution_gamma_xy}.}
    \label{fig:fig_ppf_evolution_gamma_xy}
    \end{subfigure}
    \vfill
    \begin{subfigure}{0.4\linewidth}
    \includegraphics[width=\linewidth]{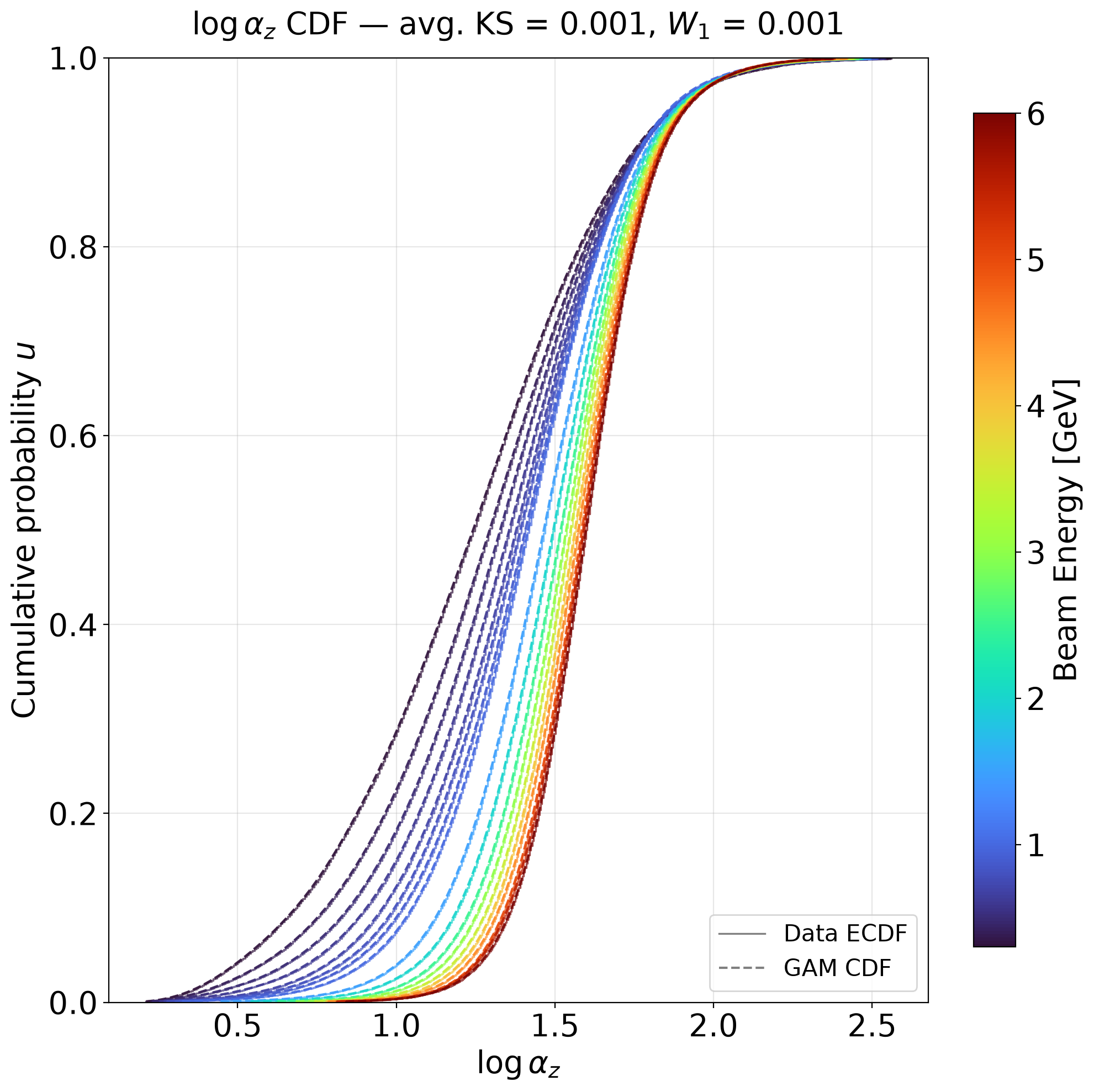}
    \caption{ECDF of the $\log{\alpha_{z}}$ variable. Else, as in Fig.~\ref{fig:fig_cdf_evolution_gamma_xy}.}
    \label{fig:fig_cdf_evolution_alpha_z}
    \end{subfigure}
    \hfill
    \begin{subfigure}{0.4\linewidth}
    \includegraphics[width=\linewidth]{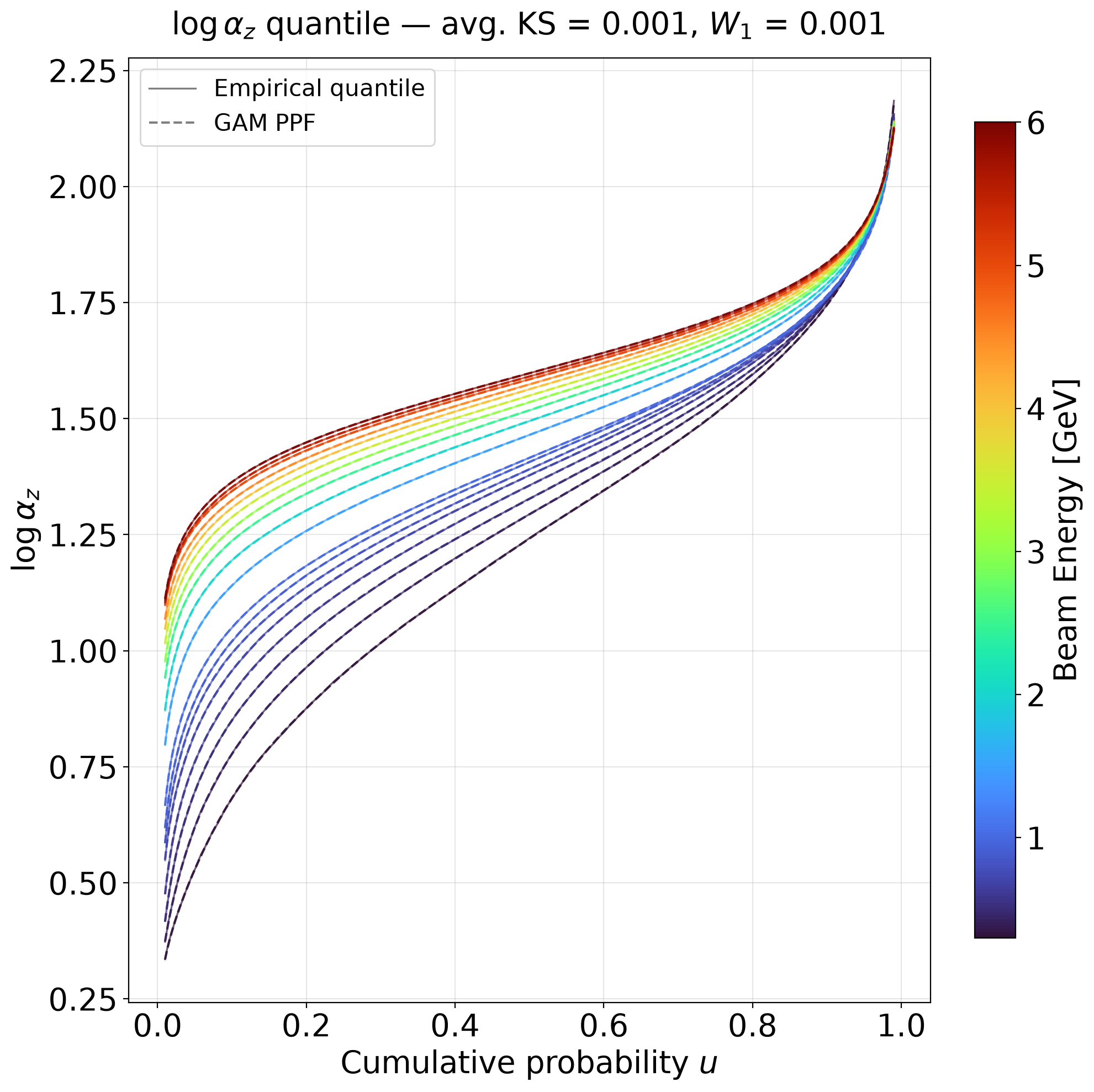}
    \caption{Inverse ECDF of the $\log{\alpha_{z}}$ variable. Else, as in Fig.~\ref{fig:fig_cdf_evolution_gamma_xy}.}
    \label{fig:fig_ppf_evolution_alpha_z}
    \end{subfigure}
    \vfill
    \begin{subfigure}{0.4\linewidth}
    \includegraphics[width=\linewidth]{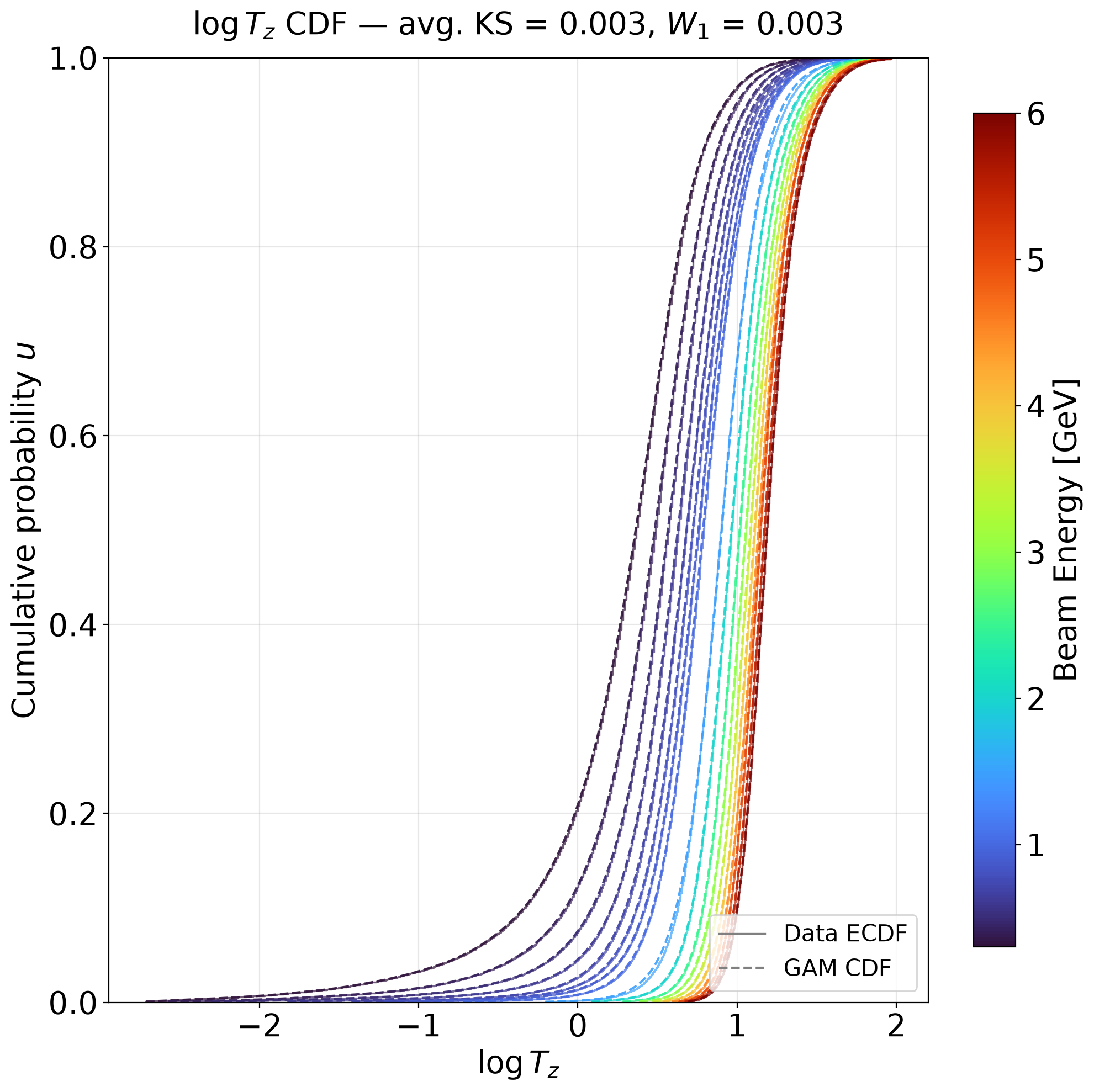}
    \caption{ECDF of the $\log{T_{z}}$ variable. Else, as in Fig.~\ref{fig:fig_cdf_evolution_gamma_xy}.}
    \label{fig:fig_cdf_evolution_T_z}
    \end{subfigure}
    \hfill
    \begin{subfigure}{0.4\linewidth}
    \includegraphics[width=\linewidth]{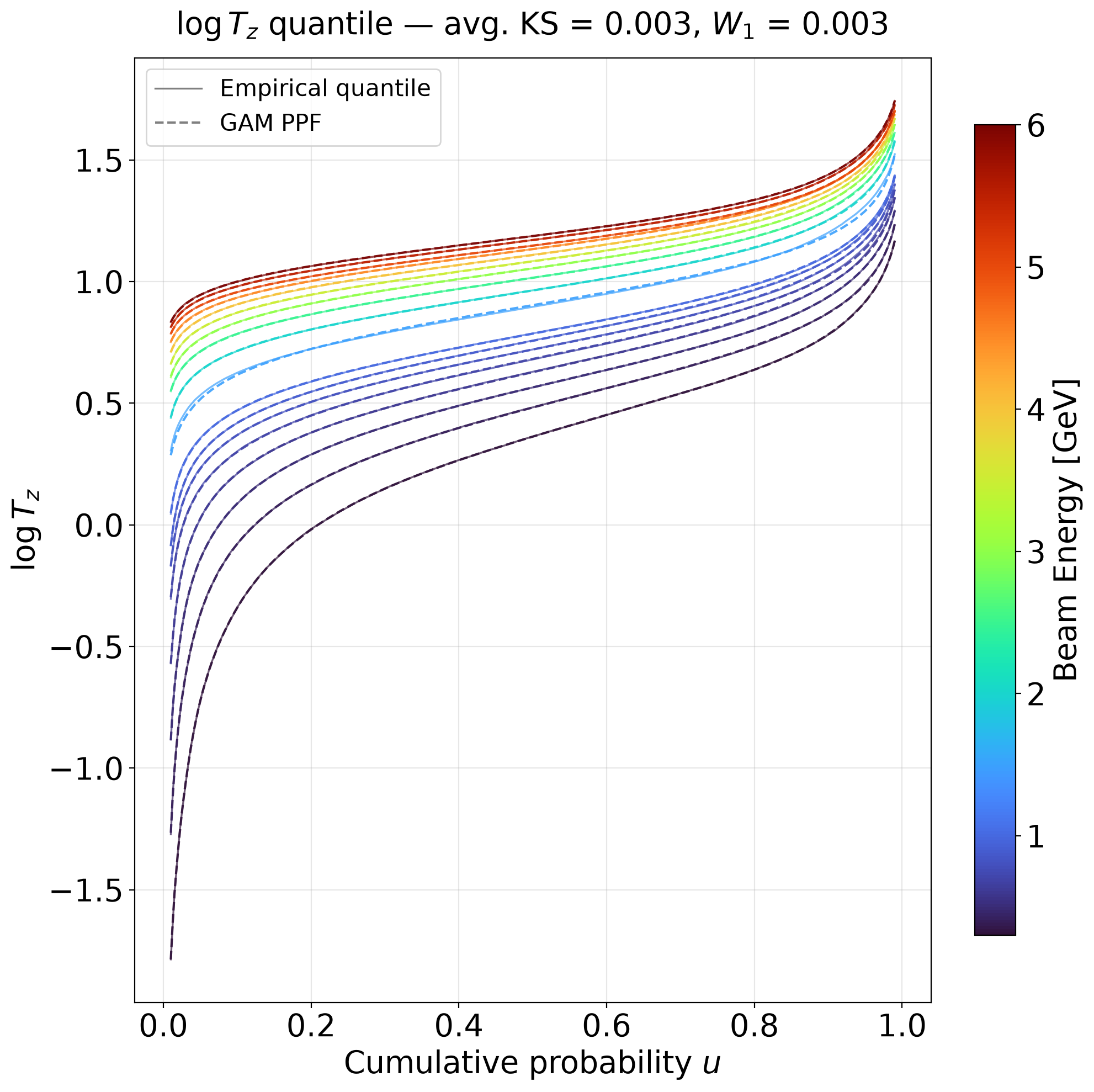}
    \caption{Inverse ECDF of the $\log{T_{z}}$ variable. Else, as in Fig.~\ref{fig:fig_cdf_evolution_gamma_xy}.}
    \label{fig:fig_ppf_evolution_T_z}
    \end{subfigure}

    \caption{ECDFs (left) and Inverse ECDFs (right) of the log variables $\log{\gamma_{xy}}$ (top), $\log{\alpha_{z}}$ (middle),  $\log{T_{z}}$ (bottom). The solid line indicates the true ECDF and the dashed line indicates the smoothed spline fit. The color blue-to-red indicates increasing energy. }
    \label{fig:cdf_ppf_evolution}
    
    \label{fig:ECDFs_copula}
\end{figure}

\begin{figure}
    \centering
    \includegraphics[width=0.9\linewidth]{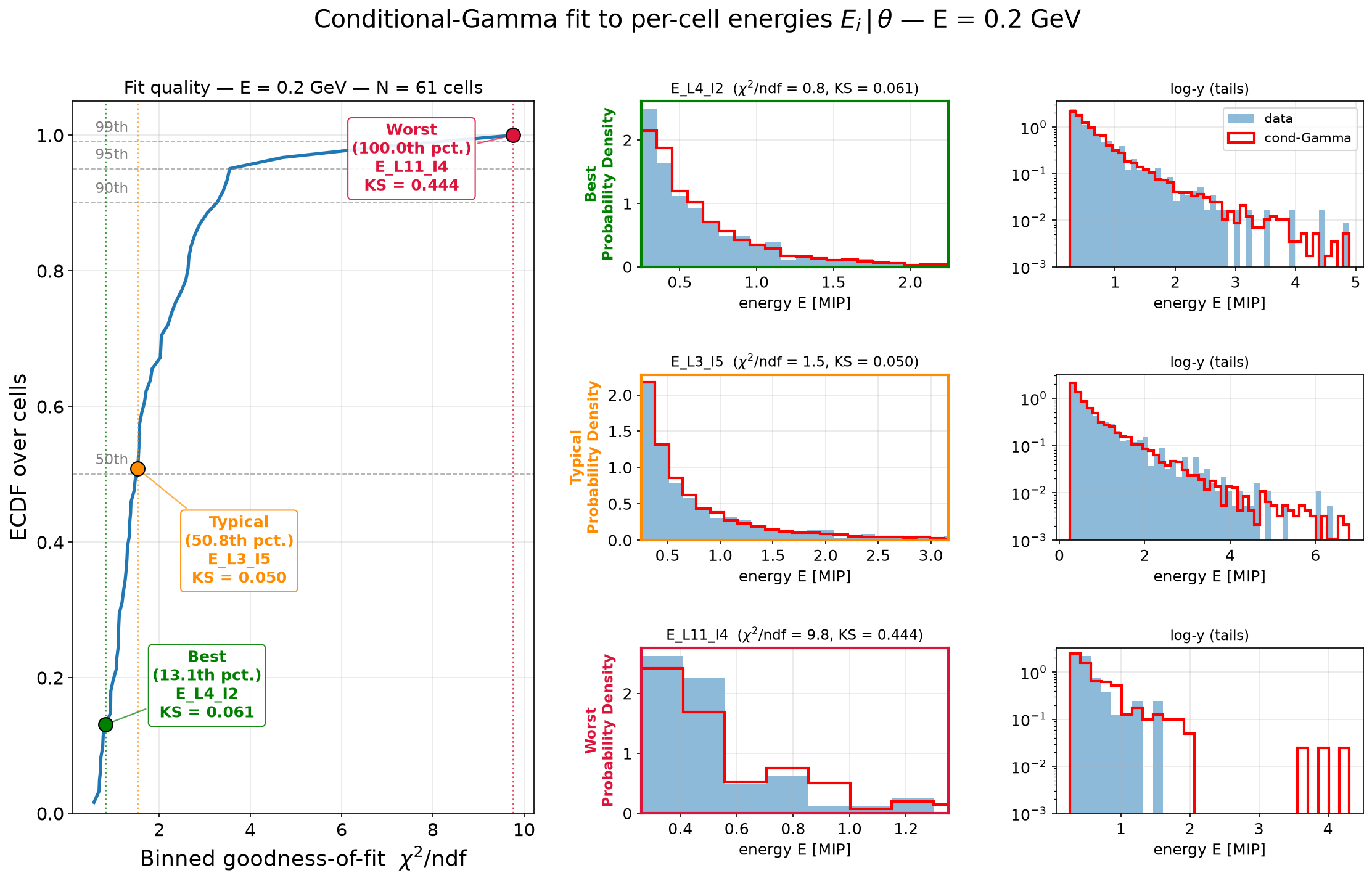}
    \caption{Conditional gamma fits to energy for \qty{0.2}{\giga \electronvolt} electron showers. Else, as in Fig.~\ref{fig:fig_cond_gamma_cdf_fits_1GeV}}
    \label{fig:fig_cond_gamma_cdf_fits_0p2GeV}
\end{figure}

\begin{figure}
    \centering
    \includegraphics[width=0.9\linewidth]{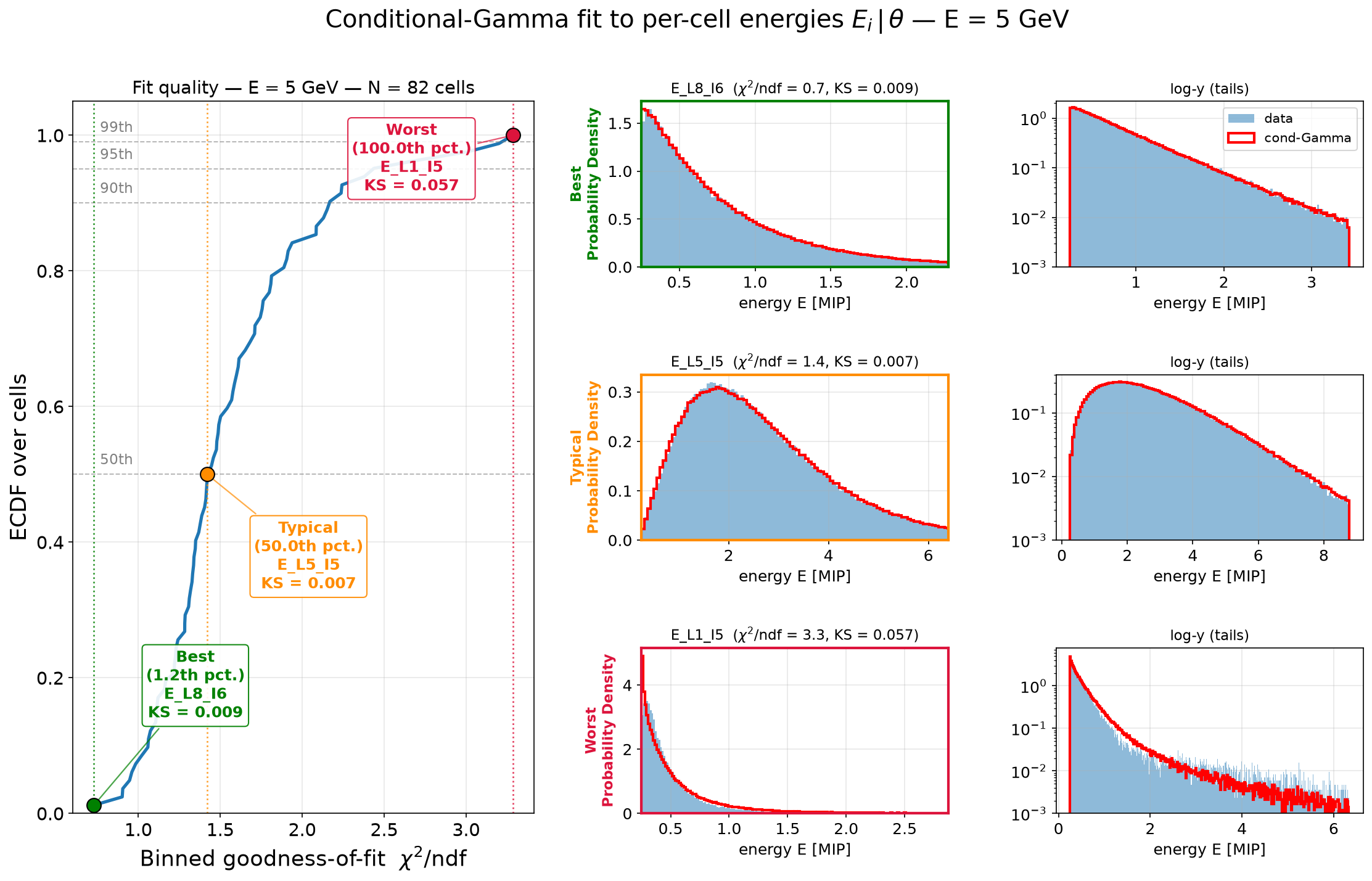}
    \caption{Conditional gamma fits to energy \qty{5}{\giga \electronvolt} electron showers. Else, as in Fig.~\ref{fig:fig_example_fits_profile_1GeV}}
    \label{fig:fig_cond_gamma_cdf_fits_5GeV}
\end{figure}

\begin{figure}
    \centering
    \includegraphics[width=0.9\linewidth]{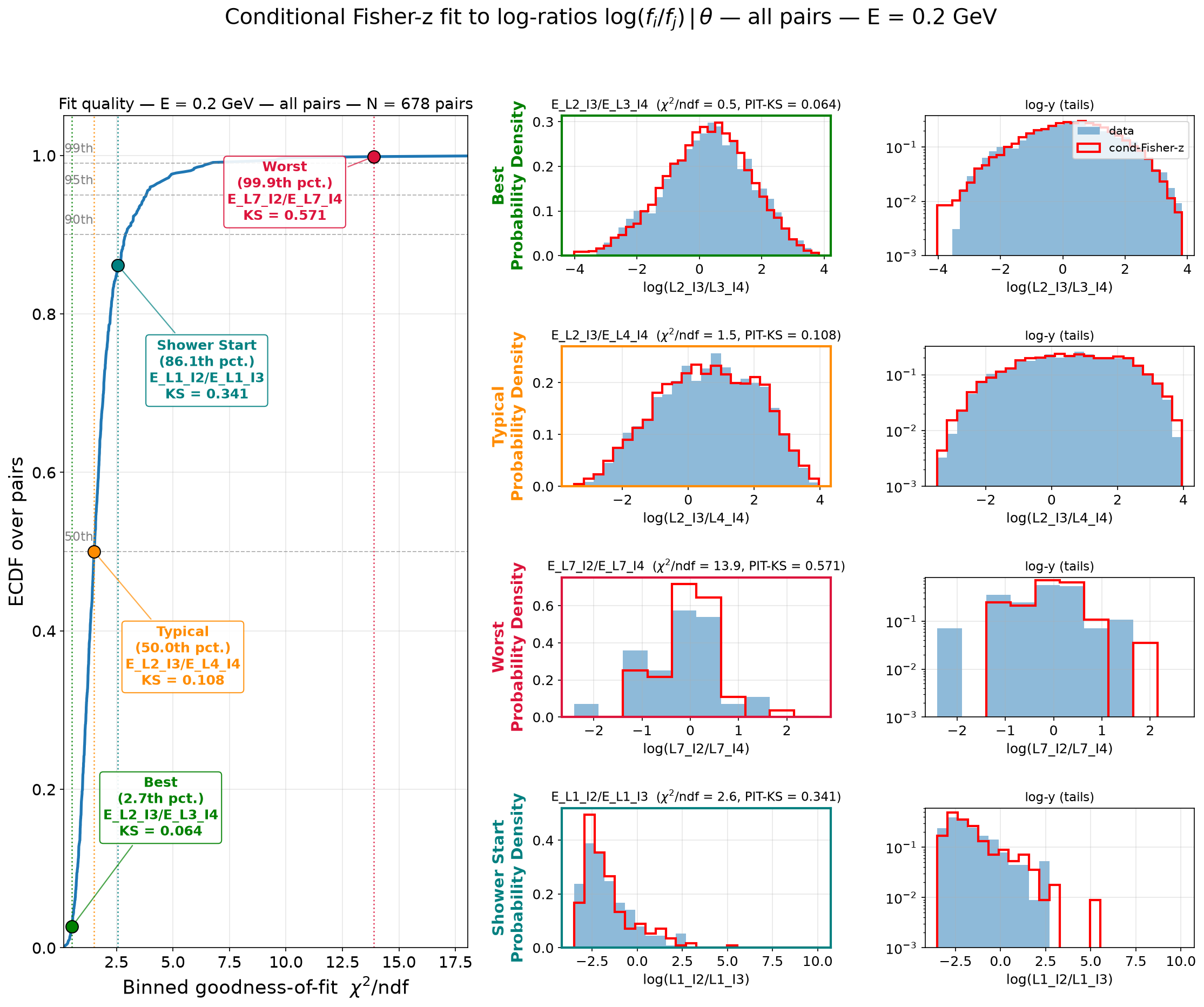}
    \caption{Conditional Fisher's-$z$ fits to energy for \qty{0.2}{\giga \electronvolt} electron showers. Else, as in Fig.~\ref{fig:fig_cond_fisherz_pairs_cdf_fits_1GeV}}
    \label{fig:fig_cond_fisherz_pairs_cdf_fits_0.2GeV}
\end{figure}

\begin{figure}
    \centering
    \includegraphics[width=0.9\linewidth]{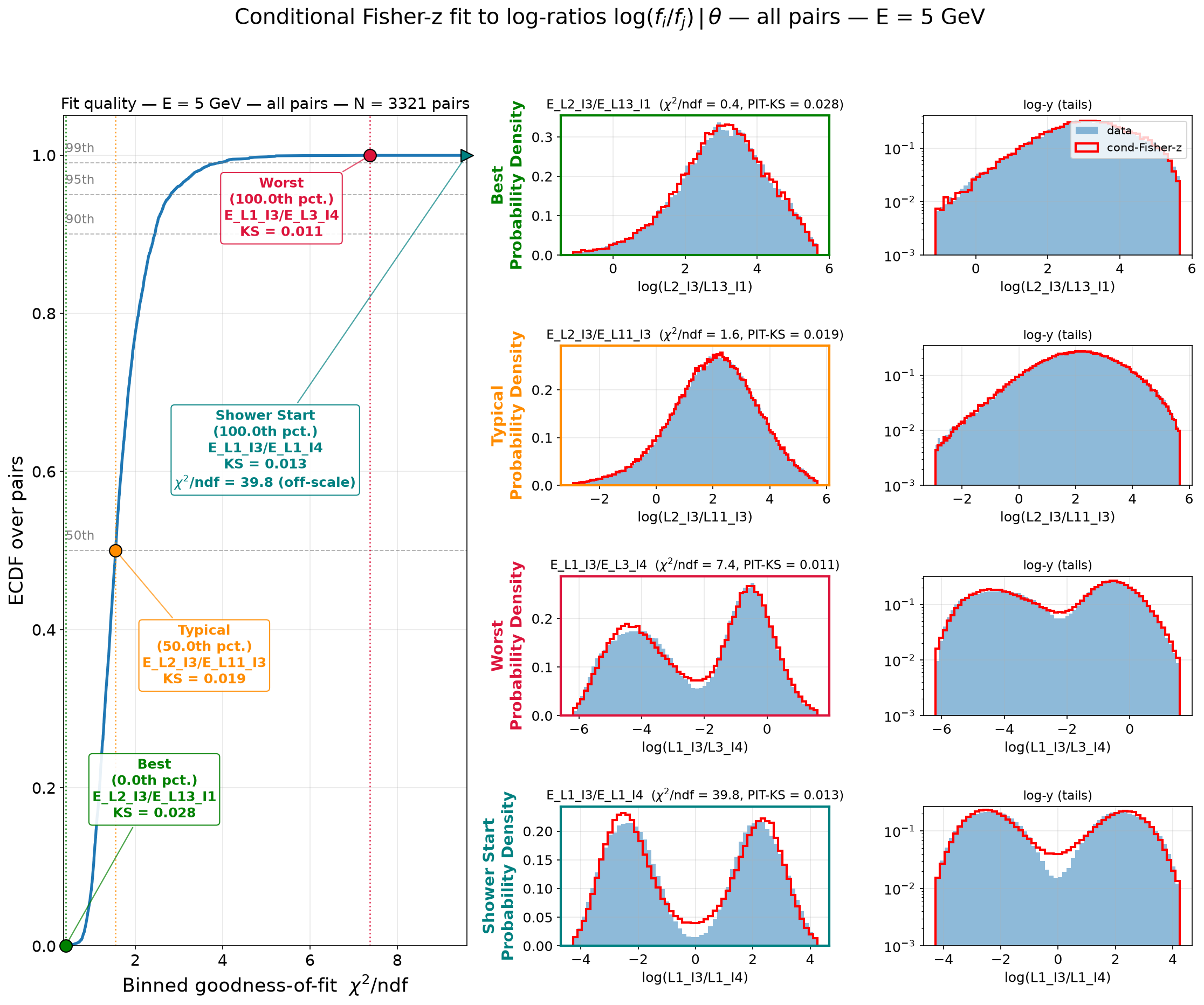}
    \caption{Conditional Fisher's-$z$ fits to energy \qty{5}{\giga \electronvolt} electron showers. Else, as in Fig.~\ref{fig:fig_example_fits_profile_1GeV}}
    \label{fig:fig_cond_fisherz_pairs_cdf_fits_5eV}
\end{figure}

\begin{figure}
    \centering
    \includegraphics[width=\linewidth]{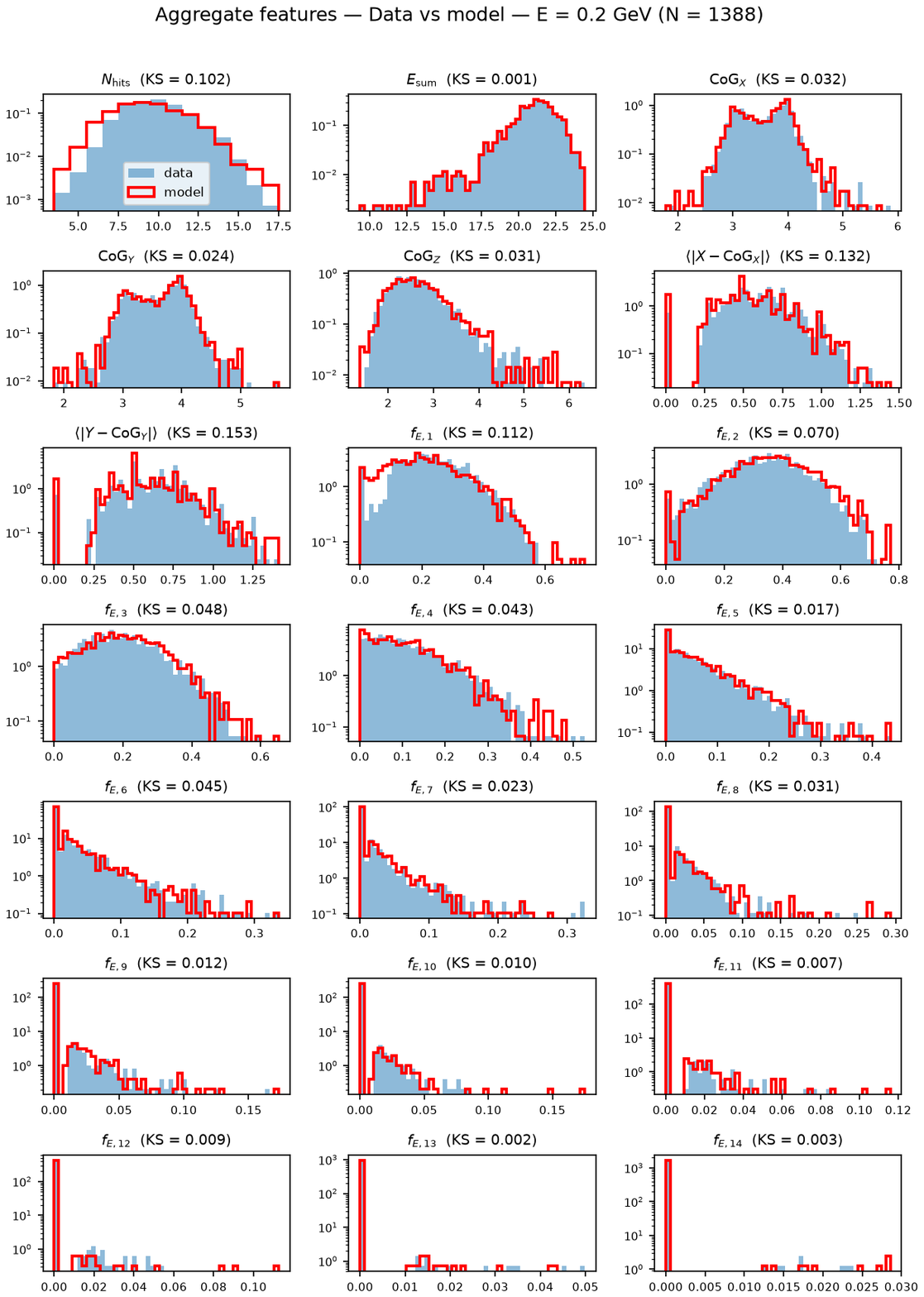}
    \caption{Comparison of the distribution of aggregate variables for the \qty{0.2}{\giga \electronvolt} sample. Else, as in \ref{fig:fig_features_final_model_E1}.}
    \label{fig:fig_features_final_model_E0.2}
\end{figure}

\begin{figure}
    \centering
    \includegraphics[width=\linewidth]{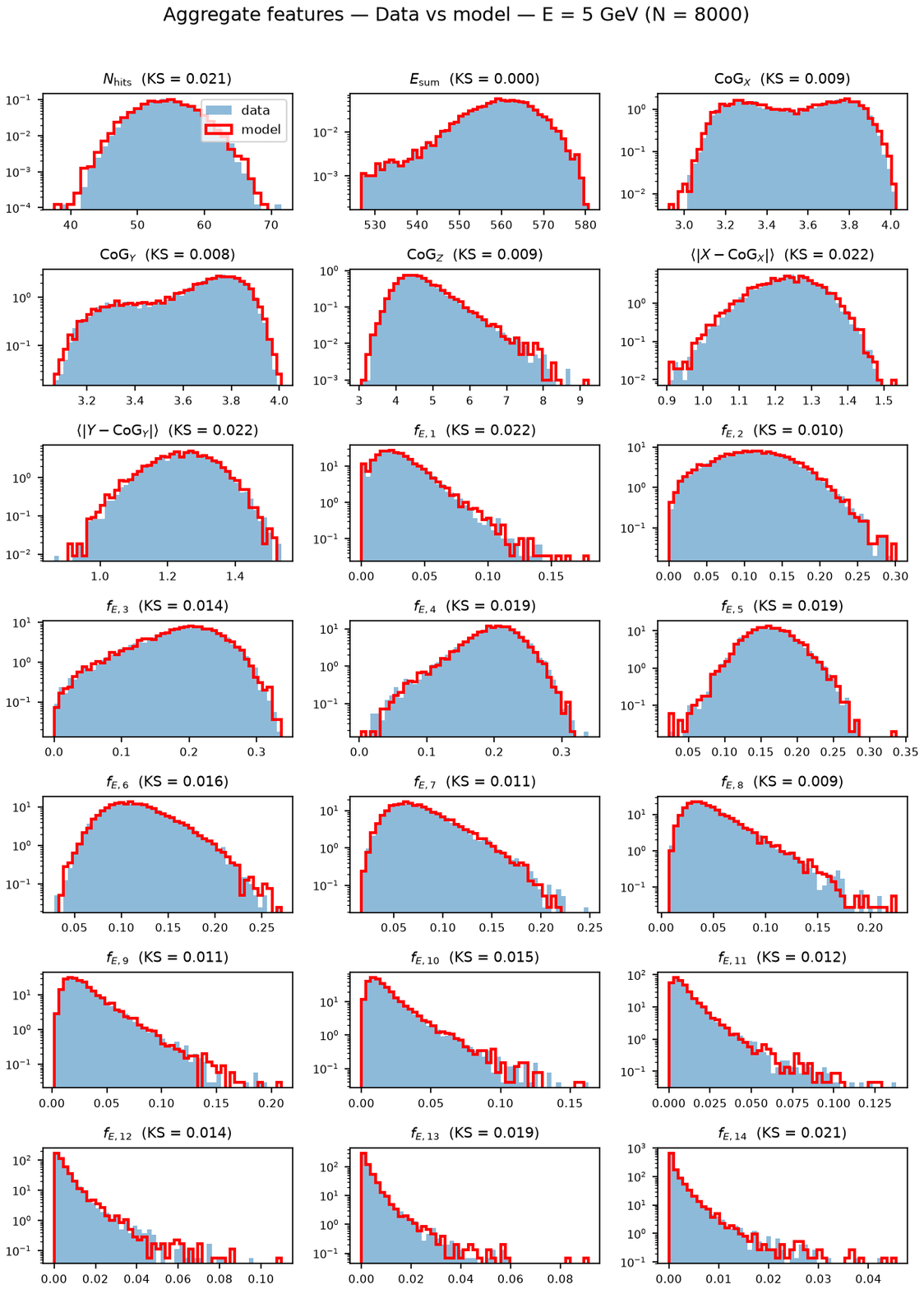}
    \caption{Comparison of the distribution of aggregate variables for the \qty{5}{\giga \electronvolt} sample. Else, as in \ref{fig:fig_features_final_model_E1}.}
    \label{fig:fig_features_final_model_E5}
\end{figure}

\section{Algorithms}

\begin{algorithm}[!t]
\centering
\caption{Sampling $B$ showers from the censored latent-factor Gaussian-copula model
  with zero-inflated LogGamma marginals, conditioned on per-event physical parameters
  $(\eta_i, E_{\mathrm{sum},i})$.}
\label{alg:sample}
\begin{algorithmic}[1]
\Require Conditioners $\eta \in \mathbb{R}^{B \times C}$ (5 shower-shape parameters and $\log E_{\mathrm{beam}}$);
         observed total energies $E_{\mathrm{sum}} \in \mathbb{R}^{B}$;
         censoring cut $E_{\mathrm{cut}} \in \mathbb{R}_{>0}$;
         trained conditional MLPs ($\operatorname{MLP}_{\alpha}$, $\operatorname{MLP}_{\nu}$, $\operatorname{MLP}_{\delta_\nu}$,
         $\operatorname{MLP}_{\pi}$, $\operatorname{MLP}_{\delta_\pi}$, $\operatorname{MLP}_{\Delta}$,
         $\operatorname{MLP}_U$, $\operatorname{MLP}_\Psi$),
         conditional RQS spline $\mathcal{M}(\,\cdot\,, \eta)$;
         learnable per-cell location-cap parameters $\bar\Delta_{\mathrm{max}} \in \mathbb{R}^{D}$;
         latent dimensions $n_{\pi}$, $n_{\nu}$, $n_{c}$.
\Ensure Per-bar log-fractions $\log\phi \in \mathbb{R}^{B \times D}$

\State Draw per-event latents:
\Statex \hspace{1.5em}$\begin{aligned}[t]
   w_\pi &\sim \mathcal{N}(0, I_{n_\pi}), & w_\nu &\sim \mathcal{N}(0, I_{n_\nu}), & w_c &\sim \mathcal{N}(0, I_{n_c}).
\end{aligned}$

\State Compute MLP values conditioned on $(\eta, w_\pi, w_{\nu})$:
\Statex \hspace{1.5em}$\begin{aligned}[t]
    \bar{\alpha} &= \operatorname{MLP}_{\alpha}(\eta), & \bar{\nu} &= \operatorname{MLP}_{\nu}(\eta), & \delta_{\nu} &= \operatorname{MLP}_{\delta_{\nu}}(\eta, w_{\nu}), \\
    \bar{\pi} &= \operatorname{MLP}_{\pi}(\eta), & \delta_{\pi} &= \operatorname{MLP}_{\delta_{\pi}}(\eta, w_{\pi}), & \bar{\Delta} &= \operatorname{MLP}_{\Delta}(\eta), \\
    U &= \operatorname{MLP}_{U}(\eta), & \bar\Psi &= \operatorname{MLP}_{\Psi}(\eta).
\end{aligned}$

\State Parameterise $\alpha$, $\nu$, $\beta$, $\pi$, $\Delta$, $\Psi$:
\Statex \hspace{1.5em}$\begin{aligned}[t]
    \alpha &\gets \alpha_{\min} + \operatorname{Softplus}(\bar{\alpha}), &
    \nu    &\gets \nu_{\min} + \operatorname{Softplus}(\bar{\nu} + \delta_{\nu}), &
    \beta &\gets \alpha\,\nu^{-1}, \\
    \pi    &\gets \operatorname{Sigmoid}(\bar{\pi} + \delta_{\pi}), &
    \Psi   &\gets \epsilon + \operatorname{Softplus}(\bar\Psi), &
    \Delta &\gets \Delta_{\max} \odot \operatorname{Sigmoid}(\bar{\Delta}) \,/ E_{\mathrm{sum}},
\end{aligned}$
\Statex with the learnable per-cell cap
$\Delta_{\max} = \min\!\big(\operatorname{Softplus}(\bar\Delta_{\mathrm{max}} ),\, E_{\mathrm{cut}}\big)$,
and $E_{\mathrm{sum}}$ converting the location from MIP to fraction units.

\State Constrain the factor loadings so each bar's copula correlation satisfies $\rho_i \le \rho_{\max}$:
\Statex \hspace{1.5em}$\begin{aligned}[t]
    U_i &\gets U_i \cdot \min\!\left(1,\; \frac{c_0\,\sqrt{\Psi_i}}{\lVert U_i \rVert}\right), &
    c_0 &= \sqrt{\frac{\rho_{\max}}{1 - \rho_{\max}}}.
\end{aligned}$

\State Evaluate conditional moments of the copula variable $z \mid w_c$:
\Statex \hspace{1.5em}$\begin{aligned}[t]
     \sigma_i &\gets \sqrt{\lVert U_i \rVert^2 + \Psi_i}, &
     m_i &\gets \frac{U_i \cdot w_c}{\sigma_i}, &
     s_i &\gets \frac{\sqrt{\Psi_i}}{\sigma_i}.
\end{aligned}$

\State Sample the copula variable and apply the inverse spline flow:
\Statex \hspace{1.5em}$\begin{aligned}[t]
    z &\gets m + s \odot \varepsilon, \qquad \varepsilon \sim \mathcal{N}(0, I_D), & \zeta &\gets \mathcal{M}^{-1}(z; \eta).
\end{aligned}$

\State Apply the PIT, then the inverse CDF of the zero-inflated LogGamma marginal:
\Statex \hspace{1.5em}$\begin{aligned}[t]
u_i &\gets \Phi(\zeta_i), \qquad \tilde u_i \gets \frac{u_i - (1-\pi_i)}{\pi_i}, \\[4pt]
g_i &\gets \begin{cases}
-\infty \;\;, & u_i \le 1 - \pi_i, \\[4pt]
\log\!\big( F_{\operatorname{Gamma}(\alpha_i,\beta_i)}^{-1}(\tilde u_i) + \Delta_i \big), & u_i > 1 - \pi_i.
\end{cases} 
\end{aligned}$       


\State Define the activity mask in log-fraction space:
\Statex \hspace{1.5em}$\begin{aligned}[t]
    \tilde a_i &= \mathbb{1}\!\left[g_i > \log\phi_{\mathrm{cut}}\right], &
    \log\phi_{\mathrm{cut}} &= \log E_{\mathrm{cut}} - \log E_{\mathrm{sum}}.
\end{aligned}$

\State \textbf{if} $\sum_i \tilde a_i = 0$ for an event \textbf{then} set $\tilde a_{i^\star} = 1$ for $i^\star = \arg\max_i g_i$. (minimum of one hit active)

\State Apply masked log-Softmax of $g$ over the active set to obtain the log-fraction:
\Statex \hspace{1.5em}$\begin{aligned}[t]
    \log\phi_i &= g_i - \log\!\sum_{j:\tilde a_j=1}\exp(g_j).
\end{aligned}$

\State Mask dark cells so that $\exp(\log\phi_i) \to 0$:
\Statex \hspace{1.5em}$\begin{aligned}[t]
    \log\phi_i &\leftarrow \begin{cases} \log\phi_i & \tilde a_i = 1,\\[4pt] -\infty & \tilde a_i = 0. \end{cases}
\end{aligned}$

\State \Return $\log\phi$.
\end{algorithmic}
\end{algorithm}

\begin{algorithm}[ht]
\centering
\caption{Full event sampling using the combined generative methods.}
\label{alg:full_sample}
\begin{algorithmic}[1]
\Require Particle energy $E_{\mathrm{particle}}$;
         calorimeter energy-response model $E_{\mathrm{sum}} \sim f_{\mathrm{calo}}(E_{\mathrm{particle}})$,\,
         censoring cut $E_{\mathrm{cut}} \in \mathbb{R}_{>0}$,\,
         minimally $\mathcal{N}\!\big(E_{\mathrm{particle}},\, a^{2} E_{\mathrm{particle}}\big)$, i.e. $\sigma/E = a/\sqrt{E}$;
         shower-shape conditioner distribution $\theta \sim f_{\theta}(E_{\mathrm{particle}})$,\,
         e.g. $\mathcal{N}_3\!\big(\mu(E_{\mathrm{particle}}),\, 
         \Sigma(E_{\mathrm{particle}})\big)$ over the
         profile parameters $(\log\gamma_{xy}, \log T_z, \log \alpha_z)$;
         the sampler of Alg.~\ref{alg:sample}, $\log\phi \sim f_{\mathrm{sample}}(\eta, E_{\mathrm{sum}})$.
\Ensure Per-bar energies $E \in \mathbb{R}^{B \times D}$.

\State Sample the total measured energy:
\Statex \hspace{1.5em}$\begin{aligned}[t]
    E_{\mathrm{sum}} &\sim f_{\mathrm{calo}}(E_{\mathrm{particle}}).
\end{aligned}$

\State Sample the shower-shape conditioning vector:
\Statex \hspace{1.5em}$\begin{aligned}[t]
    \theta &\sim f_{\theta}(E_{\mathrm{particle}}).
\end{aligned}$

\State Form the model's encoding vector by concatenation:
\Statex \hspace{1.5em}$\begin{aligned}[t]
    \eta &\gets \big[\, \theta \,;\; \log E_{\mathrm{particle}} \,\big] \;\in \mathbb{R}^{C}.
\end{aligned}$

\State Sample the per-bar log energy fractions with Alg.~\ref{alg:sample}:
\Statex \hspace{1.5em}$\begin{aligned}[t]
    \log\phi &\sim f_{\mathrm{sample}}(\eta, E_{\mathrm{sum}}, E_{\mathrm{cut}}).
\end{aligned}$

\State Reconstruct per-bar energies from the fractions:
\Statex \hspace{1.5em}$\begin{aligned}[t]
    E &= E_{\mathrm{sum}} \cdot \exp(\log\phi).
\end{aligned}$

\State \Return $E$.
\end{algorithmic}
\end{algorithm}

\begin{algorithm}[!t]
\centering
\caption{Per-event censored log-likelihood under the latent-factor
  Gaussian-copula model with zero-inflated LogGamma marginals, estimated
  with $K$ scrambled-Sobol (QMC) particles shared across the batch
  (inclusive IWAE objective).
  Lines~\ref{lnll:latent}--\ref{lnll:cormom}
  are identical to Algorithm~\ref{alg:sample} and are reproduced only for
  self-containment.}
\label{alg:nll}
\begin{algorithmic}[1]
\Require Conditioners $\eta \in \mathbb{R}^{B \times C}$ (5 shower-shape parameters and $\log E_{\mathrm{beam}}$);
         observed log-fractions $\log\phi \in \mathbb{R}^{B \times D}$;
         observed total energies $E_{\mathrm{sum}} \in \mathbb{R}^{B}$;
         activity mask $a \in \{0,1\}^{B \times D}$;
         censoring cut $E_{\mathrm{cut}} \in \mathbb{R}_{>0}$;
         trainable parameters (MLP weights, RQS spline parameters,
         per-cell location-cap parameters $\bar\Delta_{\mathrm{max}} \in \mathbb{R}^{D}$);
         conditional RQS spline $\mathcal{M}(\,\cdot\,, \eta)$;
         latent dimensions $n_{\pi}$, $n_{\nu}$, $n_{c}$;
         number of QMC particles $K$.
\Ensure Per-event log-likelihood estimate $\mathcal{L} \in \mathbb{R}^{B}$.
\For{$k = 1, \dots, K$}
    \State Draw latents (identical to Algorithm~\ref{alg:sample}):\label{lnll:latent}
    \Statex \hspace{1.5em}$\begin{aligned}[t]
       w_\pi &\sim \mathcal{N}(0, I_{n_\pi}), & w_\nu &\sim \mathcal{N}(0, I_{n_\nu}), & w_c &\sim \mathcal{N}(0, I_{n_c}).
    \end{aligned}$

    \State Compute MLP values conditioned on $(\eta, w_\pi, w_{\nu})$ (identical to Algorithm~\ref{alg:sample}):
    \Statex \hspace{1.5em}$\begin{aligned}[t]
        \bar{\alpha} &= \operatorname{MLP}_{\alpha}(\eta), & \bar{\nu} &= \operatorname{MLP}_{\nu}(\eta), & \delta_{\nu} &= \operatorname{MLP}_{\delta_{\nu}}(\eta, w_{\nu}), \\
        \bar{\pi} &= \operatorname{MLP}_{\pi}(\eta), & \delta_{\pi} &= \operatorname{MLP}_{\delta_{\pi}}(\eta, w_{\pi}), & \bar{\Delta} &= \operatorname{MLP}_{\Delta}(\eta), \\
        U &= \operatorname{MLP}_{U}(\eta), & \bar\Psi &= \operatorname{MLP}_{\Psi}(\eta).
    \end{aligned}$

    \State Parameterise $\alpha$, $\nu$, $\beta$, $\pi$, $\Delta$, $\Psi$ (identical to Algorithm~\ref{alg:sample}):
    \Statex $\begin{aligned}[t]
        \alpha &\gets \alpha_{\min} + \operatorname{Softplus}(\bar{\alpha}), &
        \nu    &\gets \nu_{\min} + \operatorname{Softplus}(\bar{\nu} + \delta_{\nu}), &
        \beta &\gets \alpha\,\nu^{-1}, \\
        \pi    &\gets \operatorname{Sigmoid}(\bar{\pi} + \delta_{\pi}), &
        \Psi   &\gets \epsilon + \operatorname{Softplus}(\bar\Psi), &
        \Delta &\gets \Delta_{\max} \odot \operatorname{Sigmoid}(\bar{\Delta}) / E_{\mathrm{sum}},
    \end{aligned}$
    \Statex with the learnable per-cell cap
    $\Delta_{\max} = \min\!\big(\operatorname{Softplus}(\bar\Delta_{\mathrm{max}}),\, E_{\mathrm{cut}}\big)$
    and $E_{\mathrm{sum}}$ converting the location from MIP to fraction units.

    \State Constrain the factor loadings so each bar's copula correlation satisfies $\rho_i \le \rho_{\max}$
    (identical to Algorithm~\ref{alg:sample}):
    \Statex \hspace{1.5em}$\begin{aligned}[t]
        U_i &\gets U_i \cdot \min\!\left(1,\; \frac{c_0\,\sqrt{\Psi_i}}{\lVert U_i \rVert}\right), &
        c_0 &= \sqrt{\frac{\rho_{\max}}{1 - \rho_{\max}}}.
    \end{aligned}$

    \State Evaluate conditional moments of the copula variable $z \mid w_c$
    (identical to Algorithm~\ref{alg:sample}):\label{lnll:cormom}
    \Statex \hspace{1.5em}$\begin{aligned}[t]
         \sigma_i &\gets \sqrt{\lVert U_i \rVert^2 + \Psi_i}, &
         m_i &\gets \frac{U_i \cdot w_c}{\sigma_i}, &
         s_i &\gets \frac{\sqrt{\Psi_i}}{\sigma_i}.
    \end{aligned}$

    \State Map the data through the marginal CDF, the inverse PIT, and the forward spline flow
    to obtain the standardised copula scores $z$:
    \Statex \hspace{1.5em}$\begin{aligned}[t]
        u_i &\gets F_{\mathrm{ZILG}}(x_i;\, \alpha_i, \beta_i, \Delta_i, \pi_i), &
        \zeta_i &\gets \Phi^{-1}(u_i), &
        z_i &\gets \mathcal{M}(\zeta_i; \eta),
    \end{aligned}$
    \Statex with $x_i = \log\phi_i$ (dark-cell values replaced by the censoring limit
    before evaluation).

    \State Define the censoring threshold in latent space:
    \Statex \hspace{1.5em}$\begin{aligned}[t]
        z_i^{\mathrm{cut}} &\gets \mathcal{M}\!\Big(\Phi^{-1}\!\big(F_{\mathrm{ZILG}}(\log\phi_{\mathrm{cut}};\, \alpha_i, \beta_i, \Delta_i, \pi_i)\big);\, \eta\Big), &
        \log\phi_{\mathrm{cut}} &= \log E_{\mathrm{cut}} - \log E_{\mathrm{sum}}.
    \end{aligned}$

    \State Censored per-bar log-probability conditioned on the latent draw:
    \Statex $\begin{aligned}[t]
    \log f_i^{(k)} \gets
        \begin{cases}
          \log f_{\mathrm{ZILG}}(x_i;\, \alpha_i, \beta_i, \Delta_i, \pi_i)
          + \log \mathcal{M}'(\zeta_i)
          + \log \varphi(z_i;\, m_i, s_i)
          - \log \varphi(\zeta_i),
           & a_i = 1, \\[6pt]
          \log \Phi\!\left(\dfrac{z_i^{\mathrm{cut}} - m_i}{s_i}\right), & a_i = 0.
        \end{cases}
    \end{aligned}$

    \State Obtain the per-event log-weight conditioned on the latent draw:
    \Statex \hspace{1.5em}$\begin{aligned}[t]
      \log \tilde \omega^{(k)} &\gets \sum_{i=1}^{D} \log f_i^{(k)}.
    \end{aligned}$
\EndFor

\State Estimate the log-marginal-likelihood over the $K$ QMC particles:
\Statex \hspace{1.5em}$\begin{aligned}[t]
  \mathcal{L}(x; \eta) &\gets \log \sum_{k=1}^{K} \exp\!\big(\log \tilde \omega^{(k)}\big) \;-\; \log K.
\end{aligned}$

\State \Return $\mathcal{L}(x; \eta)$.
\end{algorithmic}
\end{algorithm}

\end{appendices}

\clearpage

\printbibliography

\end{document}